\documentclass[sigconf, nonacm]{acmart}

\usepackage{babel}
\usepackage{ifoddpage}
\usepackage{pifont}
\usepackage{siunitx}
\usepackage{multirow}
\usepackage{tikz}

\newcommand*\circled[1]{\tikz[baseline=(char.base)]{
            \node[shape=circle,draw,inner sep=0pt] (char) {#1};}}
\usepackage{amsmath}
\usepackage{enumitem}
\usepackage[most]{tcolorbox}
\usepackage{multicol,lipsum}
\usepackage[utf8]{inputenc}
\usepackage{algorithmicx}
\usepackage{algorithm}
\usepackage[noend]{algpseudocode}
\usepackage{soul}

\usepackage{appendix}
\usepackage{balance}
\usepackage{makecell}
\usepackage[skip=4pt]{caption}
\usepackage{diagbox}
\usepackage{float}
\usepackage{graphicx}
\usepackage{hyperref}
\usepackage{siunitx}
\usepackage{subcaption}
\usepackage{threeparttable}
\usepackage{url}
\usepackage{verbatim} 
\usepackage{xspace}
\usepackage{xcolor}
\usepackage{tikz}
\usepackage{totpages}
\usepackage{mathtools}
\usepackage{physics}
\usepackage{array}
\usepackage{pdfpages}
\usepackage{adjustbox}
\usepackage{lipsum}
\usepackage{multicol}
\usepackage{float}
\usepackage[normalem]{ulem}

\newcommand\vldbdoi{XX.XX/XXX.XX}

\newcommand\vldbavailabilityurl{https://gitlab.com/steamgjk/nezhav2}
\newcommand\vldbpagestyle{plain}

\newenvironment{parafont}{\fontfamily{ptm}\selectfont}{}
\newcommand{\sysname}[0]{Nezha\xspace}
\newcommand{\primname}[0]{DOM\xspace}
\newcommand{\Para}[1]{\vspace{2pt}\noindent\begin{parafont}\textbf{\textit{#1}}\end{parafont}}

\newcommand{\todo}[1]{}
\newcommand{\as}[1]{}
\newcommand{\ag}[1]{}
\newcommand{\rv}[1]{{#1}}

\begin{document}
\title{Nezha: Deployable and High-Performance Consensus\\Using Synchronized Clocks }
\author{[Technical Report]}



\author{Jinkun Geng$^*$, Anirudh Sivaraman$^+$, Balaji Prabhakar$^*$, Mendel Rosenblum$^*$}
\affiliation{%
  \institution{$^*$Stanford University, $^+$New York University}
}

\sloppypar
\begin{abstract}
    This paper presents a high-performance consensus protocol, \sysname, which can be deployed by cloud tenants without any support from their cloud provider. \sysname bridges the gap between protocols such as Multi-Paxos and Raft, which can be readily deployed and protocols such as NOPaxos and Speculative Paxos, that provide better performance, but require access to technologies such as programmable switches and in-network prioritization, which cloud tenants do not have.
    
    \sysname uses a new multicast primitive called deadline-ordered multicast (DOM). DOM uses high-accuracy software clock synchronization to synchronize sender and receiver clocks. Senders tag messages with deadlines in synchronized time; receivers process messages in deadline order, on or after their deadline.
    
    \as{Are the throughput and latency benefits with or without proxy in Nezha?}
    \as{The term open-loop doesn't exist in abstract. We shouldn't use it without defining it.}
    We compare \sysname with Multi-Paxos, Fast Paxos, Raft, a NOPaxos version we optimized for the cloud, and 2 recent protocols, Domino and TOQ-EPaxos, that use synchronized clocks. In throughput, \sysname outperforms all baselines by a median of 5.4$\times$ (range: 1.9--20.9$\times$). In latency, \sysname outperforms five baselines by a median of 2.3$\times$ (range: 1.3--4.0$\times$), with one exception: it sacrifices 33\% latency compared with our optimized NOPaxos in one test. We also prototype two applications, a key-value store and a fair-access stock exchange, on top of \sysname to show that \sysname only modestly reduces their performance relative to an unreplicated system. \sysname is available at {\color{blue} \url{https://github.com/Steamgjk/Nezha}}.
\end{abstract}




\maketitle

\pagestyle{\vldbpagestyle}




\section{Introduction}
\label{sec:introduction}

Our goal in this paper is to build a high-performance consensus protocol which can be deployed by cloud tenants with no help from their cloud provider. We are motivated by the fact that the cloud hosts a number of applications that need both high performance (i.e., low latency and high throughput) and fault tolerance. We provide both current and futuristic examples motivating our work below.

First, modern databases (e.g., Cosmos DB, TiKV and CockroachDB) would like to provide high throughput and strong consistency (linearizability) over all their data. Yet, they often need to split their data into multiple instances because a single instance's throughput is limited by the consensus protocol~\cite{cosmosdb,TiKv-sharding,CockroachDB-multiRaft}, thereby losing consistency guarantees over the whole data. Second, microsecond-scale applications are pushing the limits of computing~\cite{osdi20-rdma-consensus,nsdi19-shinjuku, hotos19-granular-computing,nsdi21-millisort,nanosort}. Such applications often have stateful components that must be made fault-tolerant (e.g., the matching engine within a fair-access cloud stock exchange~\cite{hotos21-cloudex}, details in \S\ref{sec:app}). To effectively support such applications on the public cloud, we need the consensus protocol to provide low latency and high throughput.

Despite significant improvements in consensus protocols over the years, the status quo falls short in 2 ways. First, protocols such as Multi-Paxos~\cite{paxos} and Raft~\cite{atc14-raft} can be (and are) widely deployed without help from the cloud provider. However, they only provide modest performance: latency in the millisecond range and throughput in the 10K requests/second range~\cite{etcd-benchmark}. Second, high-performance alternatives such as NOPaxos~\cite{osdi16-nopaxos}, Speculative Paxos~\cite{nsdi15-specpaxos}, NetChain~\cite{nsdi18-netchain}, NetPaxos~\cite{nsdi18-netchain}, and Mu~\cite{osdi20-rdma-consensus}, require technologies such as programmable switches, switch multicast, RDMA, priority scheduling, and control over routing---most of which are out of reach for the cloud tenant.\footnote{We note that many of these technologies are available to \emph{cloud providers}, but in most cases they are not exposed to tenants of the cloud. RDMA instances~\cite{rdma-instances} are an exception, but such instances are expensive.}

Here, we develop a protocol, \sysname, that provides high performance for cloud tenants without access to such technologies. Our starting point in designing \sysname is to observe that a common approach to improve consensus protocols is through \emph{optimism}: in an optimistic protocol, there is a common-case fast path that provides low client latency, and there is a fallback slow path that suffers from a higher latency. Examples of this approach include Fast Paxos~\cite{fastpaxos}, EPaxos~\cite{sosp13-epaxos}, Speculative Paxos~\cite{nsdi15-specpaxos}, and NOPaxos~\cite{osdi16-nopaxos}.

For optimism to succeed, however, the fast path must indeed be the common case, i.e., the fraction of client requests that take the fast path should be high. For a sequence of client requests to take the fast path, these requests must arrive in the same order at all servers involved in the consensus protocol. In the public cloud, however, cloud tenants have no control over paths from clients to these servers. As we empirically demonstrate in \S\ref{sec:motivation}, this leads to frequent cases of \emph{reordering}: client requests arrive at servers in different orders. Thus, for an optimistic protocol to improve performance in the public cloud, reordering must be reduced. This observation influenced the design of \sysname, which has 3 key ideas.

\Para{Deadline-ordered multicast.} \sysname uses a new network primitive, called deadline-ordered multicast (DOM), designed to reduce the rate of reordering in the public cloud. DOM is a type of multicast~\footnote{Unless otherwise specified, ``multicast'' in this paper refers to application-based multicast, because switch-based multicast is not supported in cloud environment.} that works as follows. The senders' and receivers' clocks are synchronized to each other to produce a global time shared by the sender and all receivers. The sender attaches a deadline in global time to its message and multicasts the message to all its receivers. Receivers process a message on or after its deadline, and process multiple messages in the increasing order of deadline. Because the deadline is a message property and common across all receivers of a message, ordering by deadline provides the same order of processing at all receivers and undoes the reordering effect. DOM is best effort: messages arriving after their deadlines or lost messages are no longer DOM's responsibility. Thus, for DOM to be effective, the deadline should be set so that most messages arrive before their deadlines---despite variable network delays and despite clock synchronization errors. However, if messages arrive after their deadlines, correctness is still maintained because \sysname falls back to the slow path. Here, DOM follows Liskov's suggestion of using accurate clock synchronization for performance improvements, but not as a necessity for correctness~\cite{podc91-clock-sync-remark}.

\Para{Speculative execution.} DOM combats reordering and increases the fraction of client requests that take the fast path. Our next idea reduces client latency of \sysname in the slow path, by decoupling execution of a request from committing the request. Consensus protocols like Multi-Paxos and Raft wait until the request is committed at a quorum of servers before executing the request at one of them (typically the leader). However, the leader in \sysname executes the request before it is committed and sends the execution result to the client. The client then accepts the leader's execution result only if it also gets a quorum of replies from other servers that indicate commitment; otherwise, the client just retries the request. Thus a leader's execution is \emph{speculative} in that the execution result might not actually be accepted by a client because (1) the leader was deposed after sending its execution result and (2) the new leader executed a different request instead.


\Para{Proxy for deployability.} Performing quorum checks, multicasting, and clock synchronization at the client creates additional work for a \sysname client relative to a typical client of a protocol like Multi-Paxos or Raft. To address this, \sysname uses a proxy (or a fleet of proxies if higher throughput is needed), which multicasts requests, checks the quorum sizes, and performs clock synchronization---on the client's behalf. Because \sysname's proxy is stateless, it is easy to scale with the number of clients and it is easier to make fault tolerant.

\Para{Evaluation.} We compare \sysname to six baselines in the public cloud: Multi-Paxos, Fast Paxos, our optimized version of NOPaxos, Raft, Domino and TOQ-EPaxos under closed-loop and open-loop workloads. In a closed-loop workload, commonly used in the literature~\cite{sosp13-epaxos,nsdi15-specpaxos,osdi16-nopaxos, vldb21-compartmentalized-paxos}, a client only sends a new request after receiving the reply for the previous one. In open-loop workloads, recently suggested as a more realistic benchmark~\cite{nsdi21-epaxos}, clients submit new requests according to a Poisson process, without waiting for replies for previous ones. We find:
\vspace{-0.05cm}
\begin{enumerate}[wide, labelwidth=!,nosep,label=(\arabic*)]
\item In closed-loop tests, \sysname (with proxies) outperforms all the baselines by 1.9--20.9$\times$ in throughput, and by 1.3--4.0$\times$ in latency at close to their saturation throughputs. 

\item In open-loop tests, \sysname (with proxies) outperforms all the baselines by 2.5--9.0$\times$ in throughput. It also outperform five baselines by 1.3--3.8$\times$ in latency at close to their saturation throughputs. The only exception is that, it sacrifices 33\% of latency compared with our optimized version of NOPaxos.

\item \sysname can achieve better latency without a proxy, if clients perform multicasts and quorum checks. In open-loop tests, \sysname (without proxies) outperforms all the baselines by 1.3--6.5$\times$ in latency at close to their respective saturation throughputs. In closed-loop tests, \sysname (without proxies) outperforms them by 1.5--6.1$\times$.

\item We also use \sysname to replicate two applications (Redis and a prototype financial exchange, CloudEx~\cite{hotos21-cloudex}) and show that \sysname can provide fault tolerance with only a modest performance degradation: compared with the unreplicated system, \sysname sacrifices 5.9\% throughput for Redis; it saturates the processing capacity of CloudEx and prolongs the order processing latency by 4.7\%. 
\end{enumerate}

\sysname is open-sourced at {\color{blue} \url{https://github.com/Steamgjk/Nezha}}.

\section{Background}
\label{sec:background}
\subsection{Clock Synchronization}
In a distributed system, each node (server or VM) may report a different time due to clock frequency variations. Clock synchronization aims to bring clocks close to each other, by periodically correcting each node's current clock offset and/or frequency. Given two nodes $i$ and $j$, their clock times are denoted as $c_i(t)$ and $c_j(t)$ at a certain real time $t$. We consider their clocks synchronized if 
\begin{equation*}
    |c_i(t) - c_j(t)| \leq \epsilon
\end{equation*}
where $\epsilon$ is the clock error bound, indicating how closely the clocks are synchronized at $t$. 

Guaranteeing an $\epsilon$ is difficult because clock synchronization may fail occasionally, and the error bound can grow arbitrarily in such cases. Therefore, \sysname does not depend on clock synchronization for correctness. However, well synchronized clocks can improve the performance of \sysname by increasing the fraction of requests that can be committed in the fast path of \sysname using the DOM primitive. While \sysname is compatible with any clock synchronization algorithm, in our implementation, we chose to build \sysname on Huygens~\cite{huygens}, because Huygens is a high-accuracy software-based system that can be easily deployed in the public cloud.  

\subsection{Consensus Protocols}
\label{sec-back-protocols}
In this section, we overview consensus protocols most closely related to \sysname, namely, Multi-Paxos~\cite{paxos}/Raft~\cite{atc14-raft}, Fast Paxos~\cite{fastpaxos}, Speculative Paxos~\cite{nsdi15-specpaxos}, and NOPaxos~\cite{osdi16-nopaxos}. Table~\ref{tab:cmp} summarizes the basic properties of them and \sysname. \S\ref{sec:related-work} provides a more detailed comparison to related work.  


\newcolumntype{x}[1]{>{\centering\arraybackslash\hspace{0pt}}p{#1}}
\begin{table*}[t]
\begin{threeparttable}
  \centering
\caption{Basic Properties of Typical Consensus Protocols (F:fast path; S:slow path)}
\begin{tabular}{|x{2.4cm}|x{1.5cm}|x{1.5cm}|x{2.3cm}|x{1.5cm}|x{1.8cm}|x{3cm}|}
\hline 
{\small \makecell{Protocols}} &
{\small \makecell{Message \\ Delays 
~\tnote{\color{green!80!black} 1}}} & { \small \makecell{Load on \\ Leader~\tnote{\color{green!80!black} 2}} } & {\small \makecell{Quorum\\ Size} } & {\small \makecell{Quorum \\ Check} }  & {\small \makecell{Inconsistency \\ Penalty} }   & {\small \makecell{Deployment \\ Requirement} } \\ \hline
{\small \makecell{Multi-Paxos/Raft}} & {\small \makecell{4}} &
{\small \makecell{$2(2f+1)$}} &
{\small \makecell{$f+1$}} &
{\small \makecell{Leader}} &
{\small \makecell{Low}} &
{\small \makecell{No special requirement}}
\\ \hline  

{\small \makecell{Fast Paxos}} & {\small \makecell{F:3\quad S:5}} &
{\small \makecell{$2f+2$}} &
{\small \makecell{F: $f+\lceil f/2 \rceil+1$ \\ S: $f+1$}} &
{\small \makecell{F/S: Leader}} &
{\small \makecell{Medium}} &
{\small \makecell{No special requirement}}
\\ \hline  

{\small \makecell{Speculative Paxos}} & {\small \makecell{F:2\quad S:6}} &
{\small \makecell{$2$}} &
{\small \makecell{F: $f+\lceil f/2 \rceil+1$ \\ S: $f+1$}} &
{\small \makecell{F: Client\\S: Leader}} &
{\small \makecell{High\\ (rollback)}} &
{\small \makecell{Priority scheduling,\\SDN control, etc}}
\\ \hline  

{\small \makecell{NOPaxos}} & {\small \makecell{F: 2 or 3~\tnote{\color{green!80!black} 3}\\ S: 4 or 5}} &
{\small \makecell{$2$}} &
{\small \makecell{F: $f+1$ \\ S: $f+1$}} &
{\small \makecell{F: Client\\S: Leader}} &
{\small \makecell{Medium}} &
{\small \makecell{Priority scheduling,\\programmable switch~\tnote{\color{green!80!black} 4},\\SDN control, etc}}
\\ \hline  

{\small \makecell{Mencius}} & {\small \makecell{F:4\quad S:6}} &
{\small \makecell{$2(2f+1)/l$~\tnote{\color{green!80!black} 5}}} &
{\small \makecell{F: $f+1$\\ S: $f+1$ }} &
{\small \makecell{F: Leaders\\ S: Leaders}} &
{\small \makecell{Medium}} &
{\small \makecell{No special requirement}}
\\ \hline  

{\small \makecell{EPaxos}} & {\small \makecell{F:4~\tnote{\color{green!80!black} 6}\quad S:6}} &
{\small \makecell{$2(2f+1)/l$}} &
{\small \makecell{F: $f+\lfloor (f+1)/2 \rfloor$ \\ S: $f+1$}} &
{\small \makecell{F: Leaders\\S: Leaders}} &
{\small \makecell{Medium}} &
{\small \makecell{No special requirement}}
\\ \hline  

{\small \makecell{CURP}} & {\small \makecell{F:2\quad S:6}} &
{\small \makecell{$2f+2$}} &
{\small \makecell{F: $f+1$ \\ S: $f+1$}} &
{\small \makecell{F: Client\\S: Leader}} &
{\small \makecell{Low}} &
{\small \makecell{NVM, a non-faulty \\ configuration manager}}
\\ \hline  

{\small \makecell{Nezha (No proxy)}} & {\small \makecell{F: 2\quad S: 3}} &
\multirow{2}{*}{\small \makecell{$2+2f/m$~\tnote{\color{green!80!black} 7}}} &
\multirow{2}{*}{\small \makecell{F: $f+\lceil f/2 \rceil+1$ \\ S: $f+1$}} &
{\small \makecell{F/S: Client}} &
\multirow{2}{*}{\small \makecell{Low}} &
\multirow{2}{*}{\small \makecell{Clock \\ synchronization}} 
\\  
\cline{1-2} \cline{5-5}

{\small \makecell{(With proxy)}} & {\small \makecell{F: 4\quad S: 5}} &
 &
 &
{\small \makecell{F/S: Proxy}} & 
&

\\ \hline  

\end{tabular}
    \begin{tablenotes}
      \small
      \item[1] We illustrate the number of message delays for each protocol in Appendix~\ref{sec-protocol-message-delay}.
      \item[2] Load on Leader indicates the total number of messages the leader needs to send or receive in order to commit one client request. 
      \item[3] When NOPaxos uses a switch-based sequencer, the message flow client$\xrightarrow{}$sequencer$\xrightarrow{}$replica incurs one message delay because the switch is on path; hence, the overall latency is 2 message delays. With a software sequencer, both client$\xrightarrow{}$sequencer and sequencer$\xrightarrow{}$replica incur one message delay; hence, the overall latency is 3 message delays. The same holds for the slow path.
      \item[4] Programmable switch serves as the hardware sequencer, and it is unnecessary when NOPaxos uses software sequencer. However, software sequencers can reduce throughput.
     \item[5] Mencius and EPaxos use multiple leaders to amortize the load. Given the number of leaders as $l$, the load on one leader becomes $2(2f+1)/l$.
    \item[6] Here, we consider EPaxos in a single datacenter environment. When EPaxos is deployed in WAN environments, and clients are co-located with some replica, the 4 message delays of EPaxos include 2 LAN message delays and 2 WAN message delays.
     \item[7] 2 of the messages are the request and the reply. The other $2f$ messages are \emph{sync} messages with much smaller size, and can be batched, so the load is amortized by a factor of $m$ ($m$ is the batch size).
    \end{tablenotes}
\label{tab:cmp}
\end{threeparttable}
\end{table*}

\Para{Deployable but low-performance.} Multi-Paxos/Raft and Fast Paxos are generally deployable, but have lower performance. It takes 4 message delays (i.e., 2 RTTs) for Multi-Paxos/Raft to commit one request and the leader can become a throughput bottleneck. Fast Paxos can save 1 message delay if the request is committed in the fast path. However, when there is no in-network functionality that increases the frequency of the fast path (e.g., the MOM primitive~\cite{nsdi15-specpaxos} or the OUM primitive~\cite{osdi16-nopaxos}), Fast Paxos performs much worse (\S\ref{sec-latency-throughput}) because most requests can only be committed in the slow path, but its slow path (2.5 RTTs) is slower than Multi-Paxos/Raft and causes even more work for its leader because the leader needs to reconcile the inconsistent sequences of requests among followers.

\Para{High-performance but needs in-network functionality.} Speculative Paxos~\cite{nsdi15-specpaxos} and NOPaxos~\cite{osdi16-nopaxos} can achieve high performance. However, both require considerable in-network functionality. Speculative Paxos requires that most requests arrive at replicas in the same order to commit them in 1 RTT. Speculative Paxos achieves this by (1) controlling routing to ensure the same path length for all client-to-replica requests and (2) using in-network priorities to ensure that these requests encounter low queues. When reordering occurs, the request can only be committed via the slow path (3 RTTs); the slow path also requires application-specific rollback. Further, Speculative Paxos is very sensitive to packet reordering. Its throughput drops by $10\times$ with a 1\% reordering rate~\cite{nsdi15-specpaxos}[Figure 9]; such rates can easily occur in public cloud, where routing is out of a client's control and packets can take different paths. We include a micro-benchmark in \S\ref{sec:motivation}, which shows the reordering rate can exceed 20\%.  NOPaxos requires a programmable switch as the sequencer to achieve its optimal latency (1 RTT). When such a switch is unavailable, NOPaxos uses a server as a software sequencer, which adds 1 additional message delay to its fast path. Besides, as we show (\S\ref{sec-latency-throughput}), NOPaxos also loses throughput when using a software sequencer in the public cloud. In particular, it is not resistant to bursts in our open-loop tests, which further increases packet reordering/drop and trigger its slow path, causing distinct performance degradation.

\section{Motivating \sysname: Packet reordering in the cloud}
\label{sec:motivation}

Consensus protocols are often used to provide the abstraction of a replicated state machine (RSM)~\cite{schneider_rsm}, where multiple servers cooperate to present a fault-tolerant service to clients. In the RSM setting, the goal of consensus protocols is to get multiple servers/replicas to reach agreement on the contents of an ordered log, which represents a sequence of operations issued to the RSM. This amounts to 2 requirements, one for the order of the log and one for the contents of the log. We state these 2 requirements below.

For any two replicas $R_1$ and $R_2$:
\begin{itemize}[wide, labelwidth=!,nosep]
    \item \textbf{Consistent ordering.} If $R_1$ processes request $a$ before request $b$, then $R_2$ should also process request $a$ before request $b$, if $R_2$ received both $a$ and $b$.
    \item \textbf{Set equality.} If $R_1$ processes request $a$, then $R_2$ also processes request $a$.
\end{itemize}

Many \emph{optimistic} protocols leverage the fact that the ordering of messages from clients to replicas is usually consistent at different locations: they employ a fast path during times of consistent ordering and fall back to a slow path when ordering is not consistent~\cite{fastpaxos, nsdi15-specpaxos, osdi16-nopaxos,  conext20-domino}. However, for an optimistic protocol to actually improve performance, the fast path should indeed be the common case. If not, such protocols can potentially hurt performance~\cite{fastpaxos, nsdi15-specpaxos} relative to a protocol that doesn't optimize for the common case like Raft or Multi-Paxos.

\begin{figure*}[!t]
\begin{minipage}{0.3\textwidth}
    \includegraphics[width=1\linewidth]{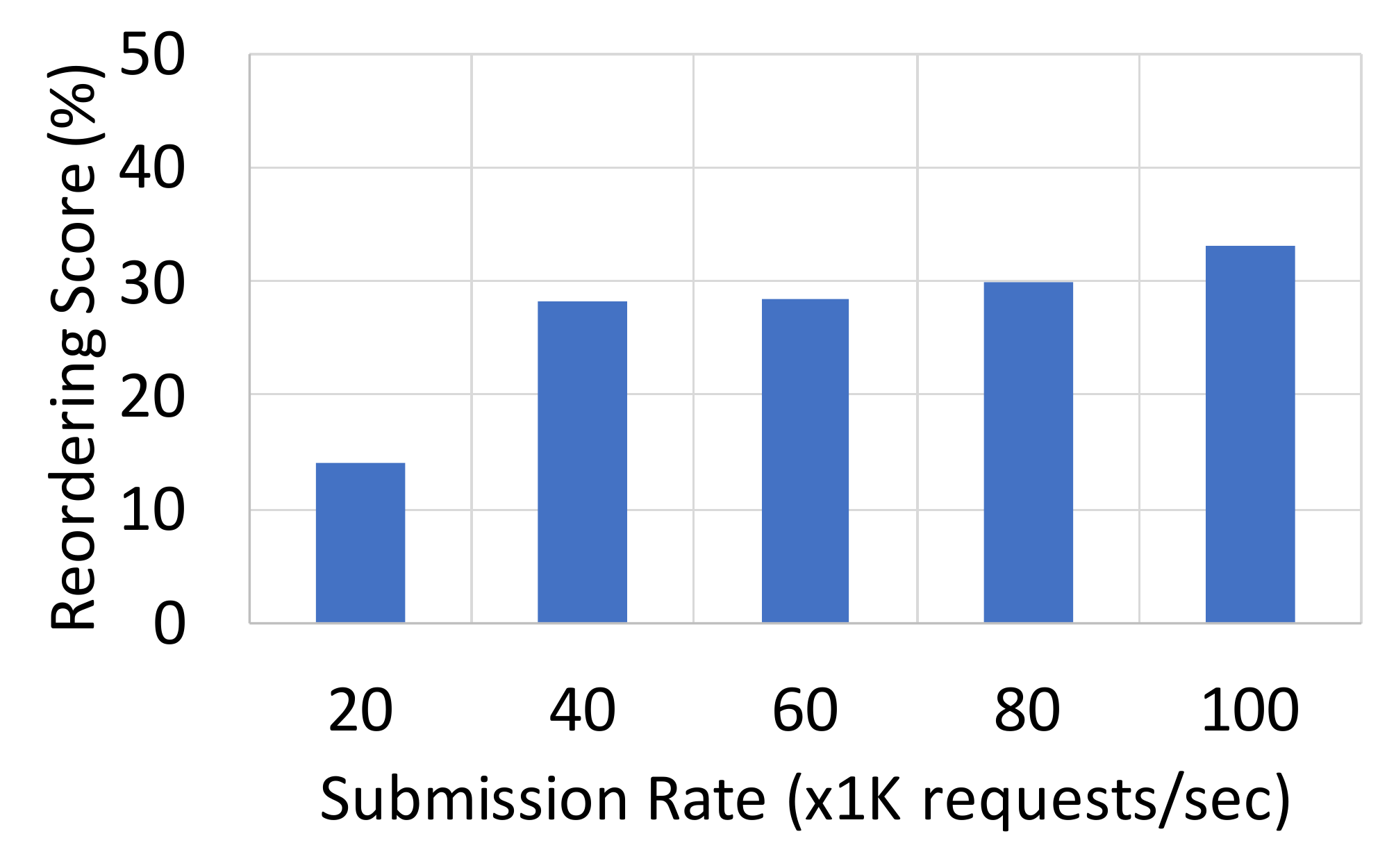}
   \caption{Packet reordering vs. submission rate on Google Cloud}
   \label{fig-micro-vary-rate}
\end{minipage}\hspace{0.5cm}
\begin{minipage}{0.3\textwidth}
    \includegraphics[width=1\linewidth]{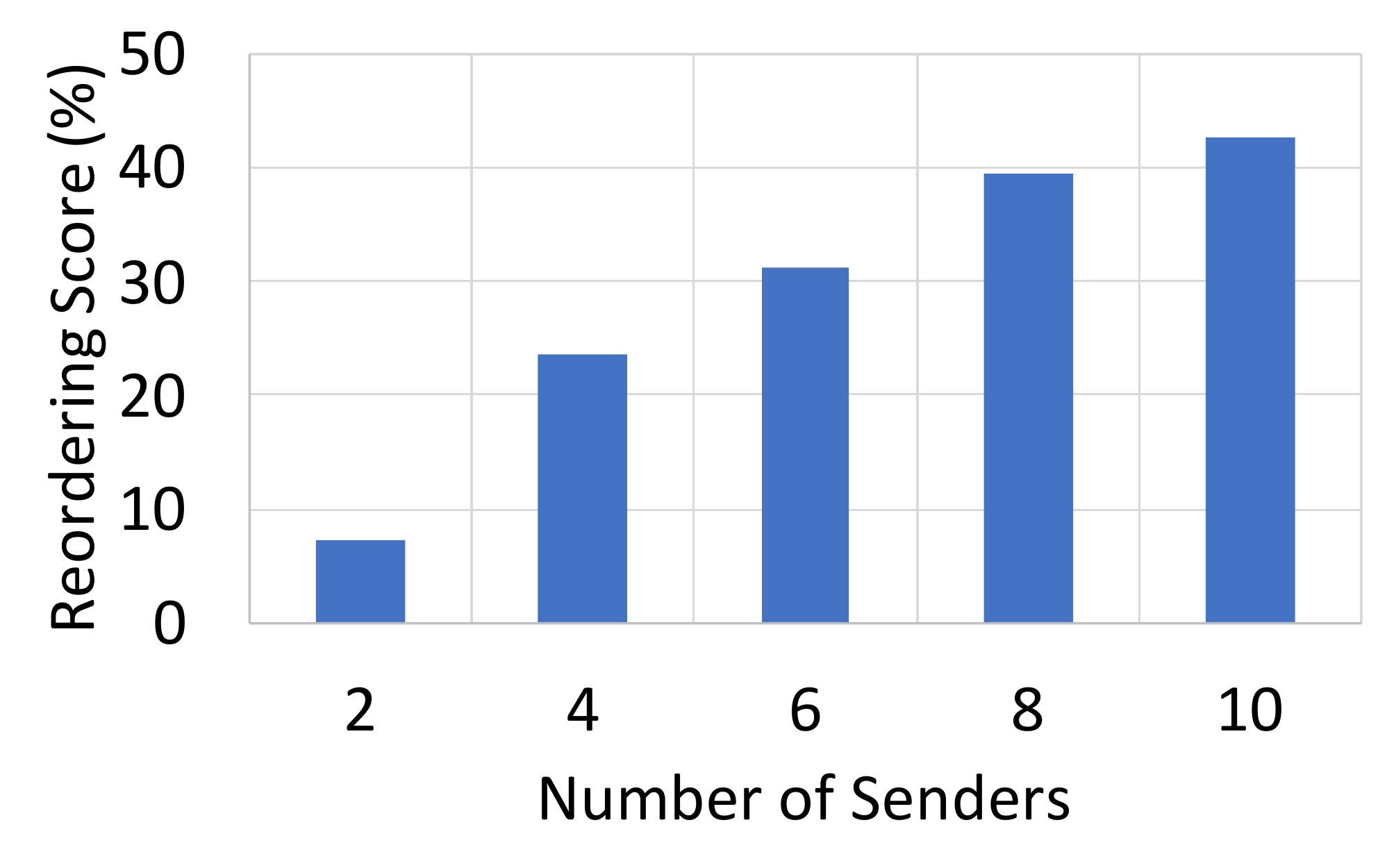}
    \caption{Packet reordering vs. number of senders on Google Cloud}
    \label{fig-micro-vary-client}
\end{minipage}\hspace{0.5cm}
\begin{minipage}{0.3\textwidth}
     \includegraphics[width=0.98\linewidth]{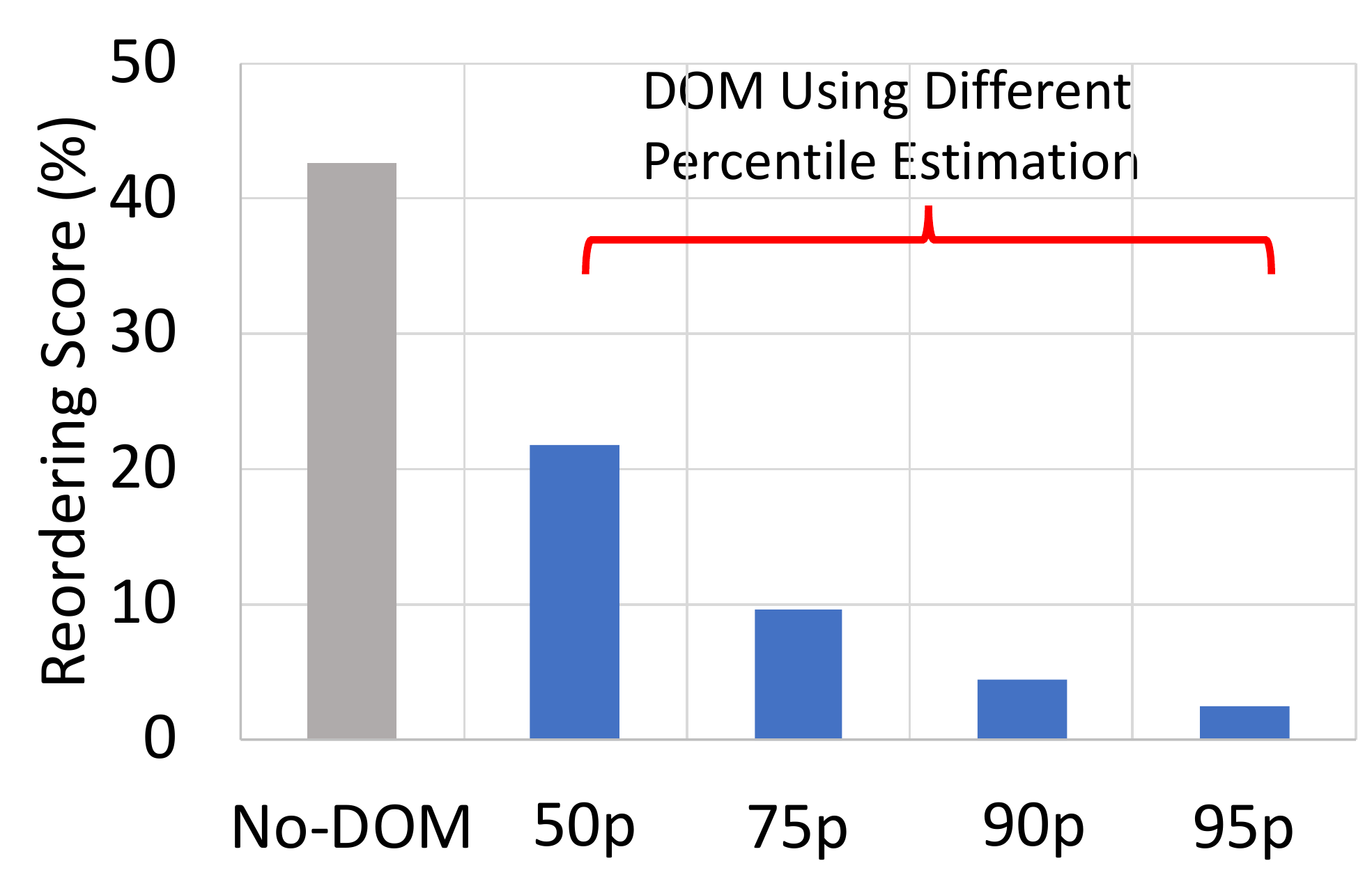}
      \caption{Effectiveness of DOM on packet reordering on Google Cloud}
      \label{fig-micro-dom-effectiveness}
\end{minipage}
\end{figure*}

Consistent ordering is violated if messages arrive in different orders at different receivers. This situation is especially common in the public cloud where there is frequent reordering: messages from one or more senders to different receivers take different network paths and arrive in different orders at the receivers.

We measure the reordering with a simple experiment on Google Cloud. We use two receiver VMs, denoted as $R_1$ and $R_2$. We use a variable number of sender VMs to multicast messages to $R_1$ and $R_2$. We vary the rate of a Poisson process used by each sender to generate multicast messages (Figure~\ref{fig-micro-vary-rate}) or vary the number of multicasting senders (Figure~\ref{fig-micro-vary-client}). After the experiment, $R_1$ receives a sequence of messages, which serves as the ground truth: each message is assigned a sequence number based on its arrival order at $R_1$. Based on these sequence numbers, we calculate a metric called \emph{reordering score}  to check how reordered $R_2$ is. We calculate the length of the longest increasing subsequence (LIS)~\cite{lis-ref,lis-slides} in $R_2$'s sequence, and the \emph{reordering score} is calculated as: 
\begin{equation*}
    \text{reordering score}=(1-\frac{\text{Length of $R_2$'s LIS}}{\text{Total Length of $R_2$'s Sequence }})\times 100\%
\end{equation*}
A higher reordering score indicates more reordering occurring in the public cloud. Figure~\ref{fig-micro-vary-rate} shows that when we vary the submission rate, keeping the number of senders fixed at 2, the reordering score quickly exceeds 28\%. Further in Figure~\ref{fig-micro-vary-client}, when we vary the number of senders, keeping the submission rate fixed at 10K messages/second, the reordering score increases rapidly up to 43\% with the number of senders, because more senders lead to more concurrency in request submission but the network of public cloud has no way to serialize the concurrent requests into a commonly-agreed sequence. 


In the public cloud, with such high reordering rates, optimistic protocols are forced to take the slow path often, which reduces their performance (\S\ref{sec-latency-throughput}). In order to design a high-performance protocol, we need to reduce the rate of reordering. This motivates us to design the \emph{deadline-ordered multicast} (DOM) primitive to guarantee consistent ordering among replicas. DOM
does not guarantee set equality. This is intentional and is also why we need a consensus protocol, \sysname, to go along with DOM because guaranteeing both requirements has been shown to be as hard as consensus itself~\cite{weakestfd}.

\section{Deadline-Ordered Multicast}
\label{sec-dom}



Informally, deadline-Ordered Multicast (DOM) is designed to reduce the rate of reordering by (1) waiting to process a message at a receiver until the message's deadline is reached and (2) delivering messages to the receiver in deadline order. This gives other messages with a lower deadline the ability to ``catch up'' and reach the receiver before a message with a later deadline is processed.

Formally, in DOM, a sender wishes to send a message $M$ to multiple receivers $R_1$, $R_2$, ..., $R_n$. The sender attaches a deadline $D(M)$ to the message, where $D(M)$ is specified in a global time that is shared by senders and receivers because their clocks are synchronized. Then the DOM primitive attempts to deliver $M$ to receivers within $D(M)$. Receivers (1) can only process $M$ on or after $D(M)$ and (2) must process messages in the order of $D(M)$s regardless of $M$'s sender.

We stress that DOM is a \emph{best-effort} primitive: a sequence of messages is processed in order at a receiver \emph{if} they all arrive before their deadline, but DOM does not guarantee that messages arrive \emph{reliably} at all receivers either before the deadline or at all. There are two situations that cause DOM messages to arrive late or be lost.

The first is network variability: messages may not reach some receivers or reach them so late that other messages with larger deadlines have been processed. The second is a temporary loss of clock synchronization. If clocks are poorly synchronized, the deadline on a message might be set much earlier in time than the actual time at which the receiver receives the message.

While DOM is a general primitive, we comment briefly on its specific use for consensus as in \sysname. When DOM is used for consensus, because DOM makes no guarantees on reliable or timely delivery, it is up to the slow path of the consensus protocol to handle lost or late messages. If client requests are lost because of drops in the network and haven't been received by a quorum of replicas, it is up to clients to retry the requests. These weaker guarantees in DOM are important because providing both reliable delivery and ordering of multicast messages is just as hard as solving consensus~\cite{weakestfd}. The use of clock synchronization for performance (i.e., increasing the frequency of the fast path) rather than correctness (i.e., linearizability) is also in line with Liskov's suggestion on how synchronized clocks should be used~\cite{podc91-clock-sync-remark}.

\Para{Setting DOM deadlines.}
Setting deadlines is a trade-off between avoiding message reordering and adding too much waiting time to a message before it can be processed. In the public cloud, where VM-to-VM latencies can be variable and reordering is common, these deadlines should be set adaptively based on recent measurements of one-way delays (OWDs), which are also enabled by clock synchronization. We pick the deadline for a message by taking the maximum among the estimated OWDs from all receivers and adding it to the sending time of the message. The estimation of OWD is formalized as below.

\begin{equation*}
    \widetilde{OWD} = 
    \begin{cases}
        P + \beta (\sigma_{S}+\sigma_{R}), \quad 0<\widetilde{OWD}<W \\
        W
    \end{cases}
\end{equation*}
To track the varying OWDs, each receiver maintains a sliding window for each sender, and records the OWD samples by subtracting the message's sending time from its receiving time. Then the receiver picks a percentile value from the samples in the window as $P$. We previously tried moving average but found that just a few outliers (i.e. the tail latency samples) can inflate the estimated value. Therefore, we use percentiles for robust estimation. The percentile is a DOM parameter set by the user of DOM. 

Besides $P$, DOM also obtains from the clock synchronization algorithm, Huygens~\cite{huygens} the standard deviation  for the sending time and receiving time, denoted as $\sigma_{S}$ and $\sigma_{R}$.~\footnote{$\sigma$ values are calculated based on the method in ~\cite{yilong-spn}[Appendix A]. $\beta=3$ in our setting, i.e., a 3$\sigma$ confidence interval for the sending/receiving time provided by Huygens.} $\sigma_{S}$ and $\sigma_{R}$ provide an \emph{approximate} error bound for the synchronized clock time, so we add the error bound with a factor $\beta$ to $P$ and obtain the final estimated OWD. The involvement of $\beta (\sigma_{S}+\sigma_{R})$ enables an adaptive increase of the estimated value, leading to a graceful degradation of \sysname as the clock synchronization performs worse. Moreover, in case that clock synchronization goes wrong and provides invalid OWD values (i.e. very large or even negative OWDs), we further adopt a clamping operation: If the estimated OWD goes out of a predefined scope $[0, W]$, we will use $W$ as the estimated OWD. The estimated OWD will be sent back to the sender to decide the deadlines of subsequent requests.

To illustrate DOM's benefits, we redo our experiments from \S\ref{sec:motivation} with 10 Poisson senders, each submitting 10K requests/sec to 2 receivers. Figure~\ref{fig-micro-dom-effectiveness} shows different percentiles (i.e., 50\textsuperscript{th}, 75\textsuperscript{th}, 90\textsuperscript{th}, and 95\textsuperscript{th}) for DOM to decide its deadlines. We can see that a higher percentile leads to more reduction of reordering. However, a higher percentile also means a longer holding delay for messages in DOM, which in turn undermines the latency benefit of \sysname protocol.

\section{\sysname Overview}
\label{sec:design}
\begin{figure*}
\begin{minipage}[t]{0.48\textwidth}
    \vspace{0pt}
    \includegraphics[width=8cm]{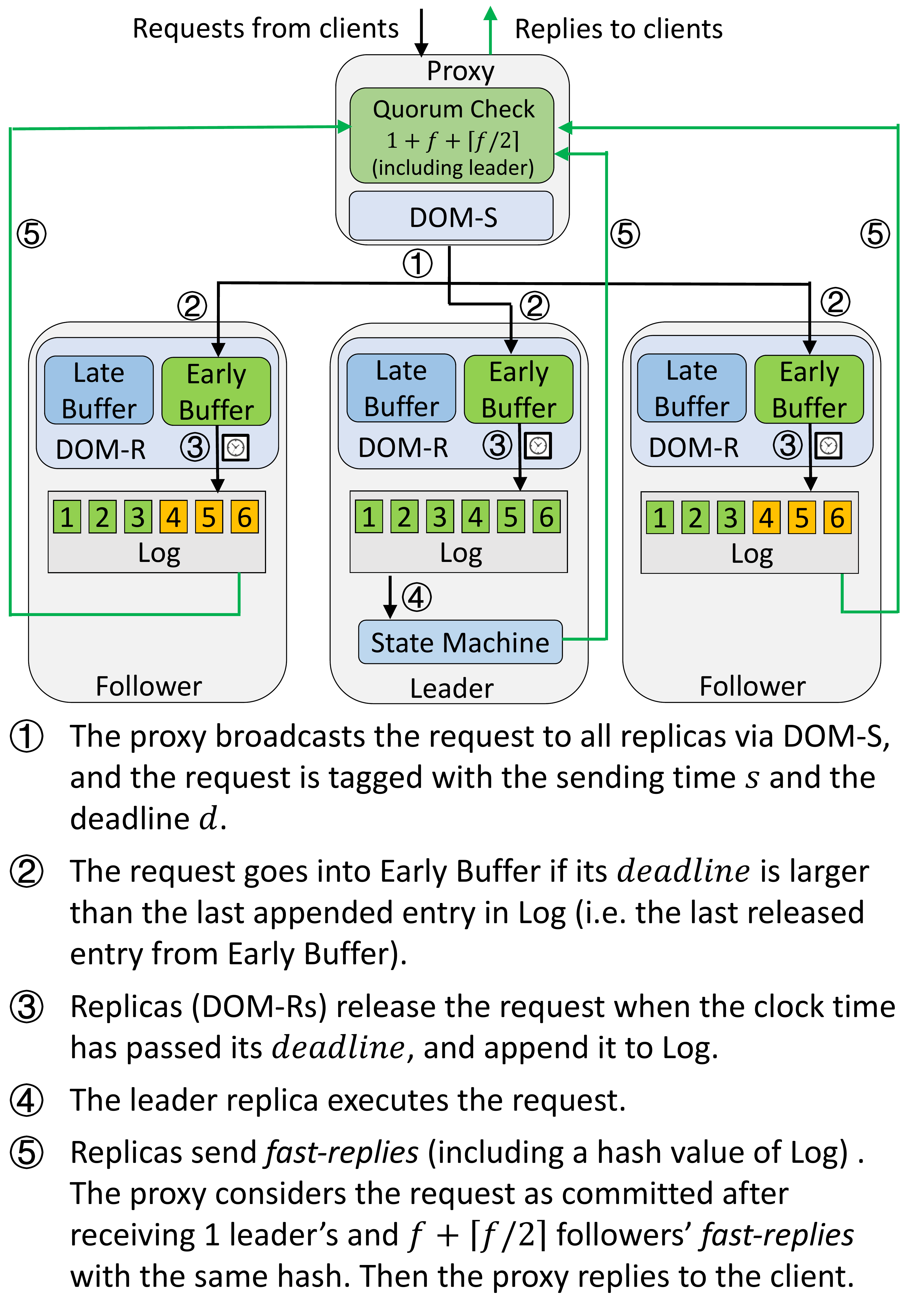}
    \caption{Fast path of \sysname}
    \label{fig-workflow-fast}
\end{minipage}\hspace{0.6cm}
\begin{minipage}[t]{0.48\textwidth}
    \vspace{0pt}
    \includegraphics[width=8cm]{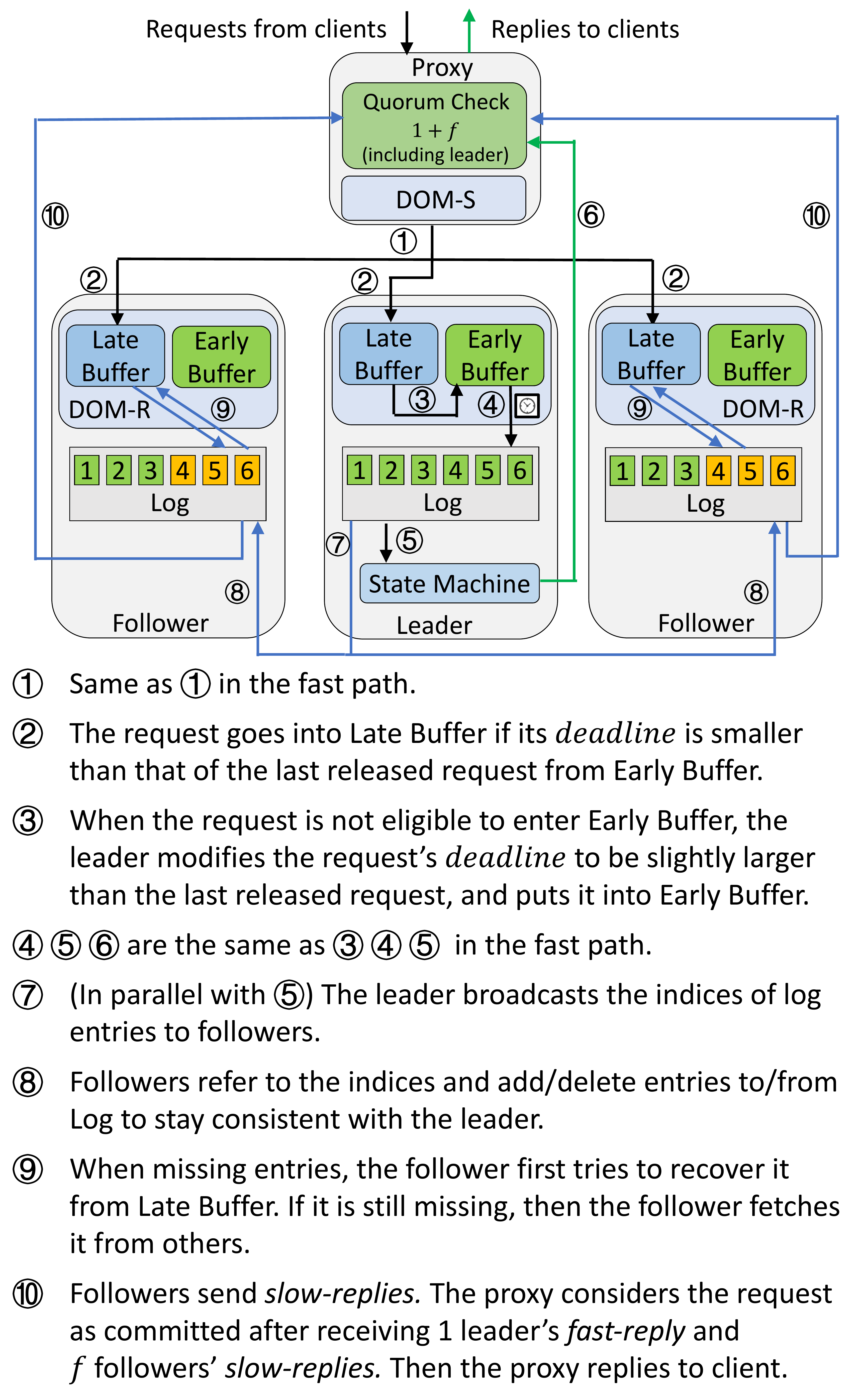}
    \caption{Slow path of \sysname}
    \label{fig-workflow-slow}
\end{minipage}
\end{figure*}

We use DOM as a building block to develop a consensus protocol, called \sysname, atop DOM. We overview the protocol here and describe it in detail in subsequent sections. Recall that DOM maintains consistent ordering across replicas by ordering messages based on their deadlines. This allows \sysname to use a fast path that assumes consistent ordering across replicas. When DOM fails to deliver a message to enough replicas before the message's deadline (either because of delays or drops), \sysname uses a slow path instead.

\Para{Model and assumptions.} \sysname assumes a fail-stop model and does not handle Byzantine failures. It uses $2f+1$ replicas: 1 leader and $2f$ followers, where at most $f$ can be faulty and crash. \sysname guarantees safety (linearizability) at all times and liveness under the same assumptions as Multi-Paxos/Raft (i.e.,  ``the majority of servers are up and communicating with reasonable timeliness''~\cite{john-paxos-slides}). However, \sysname's performance is improved by DOM, whose effectiveness depends on accurate clock synchronization among VMs and the variance of OWDs between proxies and replicas. Here ``accurate'' means the clocks among  proxies and replicas are synchronized with a small error bound \emph{in most cases}. But Nezha does not assume the existence of a worst-case clock error bound that is never violated because clock synchronization can also fail~\cite{podc84-clock-sync,nancy-clock-bound,podc91-clock-sync-remark}.

\Para{\sysname architecture.}
Nezha uses a stateless proxy/proxies (Figures~\ref{fig-workflow-fast} and \ref{fig-workflow-slow}) interposed between clients and replicas to relieve clients' computational burden of quorum checks and multicasts. Using a stateless proxy also makes Nezha a drop-in replacement for Raft/Multi-Paxos because the client just communicates with a Nezha proxy like it would with a Raft leader. This proxy serves as the DOM sender, while the replicas serve as DOM receivers. The DOM deadline is set to be the sum of (1) the request sending time from the proxy and (2) the maximum of a sliding window median (50\textsuperscript{th} percentile) of OWD estimates between the proxy and each replica; these deadlines also take into account the current estimate of clock synchronization errors (\S\ref{sec-dom}). Another benefit of a proxy is that it is sufficient if the proxy's clock is synchronized with the receivers; the client can remain unsynchronized.

\Para{\sysname fast path.} We very briefly describe Nezha's fast path and slow path, leaving details to later sections. Figure~\ref{fig-workflow-fast} shows the fast path. The request is multicast from the proxy \circled{1}. If the request's deadline is larger than the last request released from the \emph{early-buffer}, the request enters the \emph{early-buffer} \circled{2}. It will be released from the \emph{early-buffer}s at the deadline, so that replicas can append the request to their \emph{log}s \circled{3}. The \emph{log} of requests is ordered by request deadline. After that, followers immediately send a reply to the proxy without executing the request \circled{5}, whereas the leader first executes the request \circled{4} and sends a reply including the execution result. The proxy considers the request as committed after receiving replies from the leader and $f+\lceil f/2 \rceil$ followers. The proxy also obtains the execution result from the leader's reply, and then replies with the execution result to its client. The fast path requires a super quorum ($f+\lceil f/2 \rceil+1$) rather than a simple quorum ($f+1$) for the same reason as Fast Paxos~\cite{fastpaxos}: Without leader-follower communication, a simple quorum cannot persist sufficient information for a new leader to always distinguish committed requests from uncommitted requests (details in \S\ref{sec-fast-quorum-check}).

\Para{\sysname slow path.} Figure~\ref{fig-workflow-slow} shows the more involved slow path: when a multicasted \circled{1} request goes to the \emph{late-buffer} because of its small deadline \circled{2}, followers do not handle it. However, the leader must pick it out of its \emph{late-buffer} eventually for liveness. So the leader modifies the request's deadline to make it eligible to enter the \emph{early-buffer} \circled{3}. After releasing and appending this request to the log \circled{4}, the leader broadcasts this request's \emph{unique} identifier (a 3-tuple consisting of \emph{client-id}, \emph{request-id}, and request deadline) to followers \circled{7}, to force followers to keep consistent logs with the leader. On hearing this broadcast \circled{8}, the followers add/modify entries from their log to stay consistent with the leader: as an optimization, followers can retrieve missing requests from their \emph{late-buffer}s without having to ask the leader for these entries \circled{9}. After this, followers send replies to the proxy \circled{10}. Meanwhile, the leader has executed the request \circled{5} and replied to the proxy \circled{6}. After collecting $f+1$ replies (including the leader's reply), the proxy considers the request as committed. Notably, \sysname differs from other optimistic protocols (e.g.~\cite{fastpaxos, nsdi15-specpaxos,osdi16-nopaxos}): in the slow path, to improve performance, \sysname decouples request execution (at the leader) from commit. The request is immediately executed by the leader and included in the leader's reply. The request is only committed by the proxy if a quorum exists, i.e., if the leader's speculation is safe.


\section{The \sysname Protocol}
\label{sec-protocol}

\begin{figure}[!t]
    \centering
    \includegraphics[width=9cm]{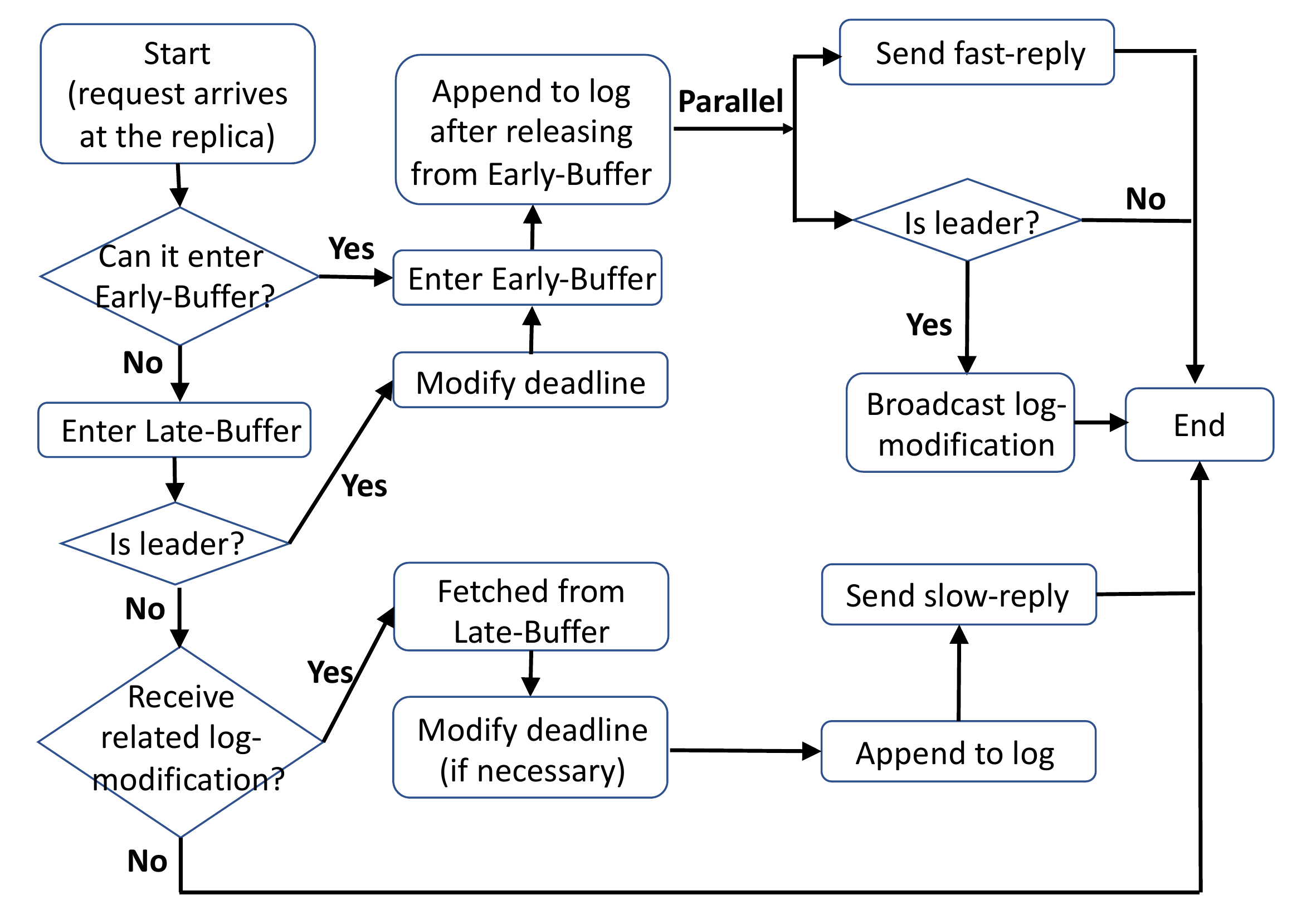}
    \caption{Life cycle of one request on the replica}
    \label{fig:branch}
\end{figure}

We first describe the state maintained by \sysname,  \sysname's message formats and \sysname's fast and slow path. In addition, Figure~\ref{fig:branch} illustrate the life cycle of a request after arriving at a replica. Algorithms~\ref{algo-replica} and ~\ref{algo-proxy} describe the replica and proxy algorithms with object-oriented and event-driven methods.

\algdef{SE}[EVENT]{Event}{EndEvent}[1]{\textbf{upon}\ \textsc{\small #1}\ \algorithmicdo}{\algorithmicend\ \textbf{event}}%
\algtext*{EndEvent}
\algdef{SE}[DOWHILE]{Do}{doWhile}{\algorithmicdo}[1]{\algorithmicwhile\ #1}%
\newcommand{\tsc}[1]{{\textsc{\small #1}}}
\makeatletter
\NewDocumentCommand{\LeftComment}{s m}{%
  \Statex \IfBooleanF{#1}{\hspace*{\ALG@thistlm}}\(\triangleright\) #2}
\makeatother

\begin{algorithm}[!t]

\small 
\begingroup


\caption{Replica Actions}
\label{algo-replica}
\hspace*{\algorithmicindent} \textbf{Member Variables}: \Comment{} \\
\hspace*{\algorithmicindent} \emph{eb},\Comment{\emph{early-buffer}} \\
\hspace*{\algorithmicindent} \emph{lb},\Comment{\emph{late-buffer}} \\
\hspace*{\algorithmicindent} \emph{synced-log}, \Comment{the replica's \emph{log} which has been synced with leader} \\
\hspace*{\algorithmicindent} \emph{unsynced-log}, \Comment{the replica's \emph{log} which hasn't been synced with leader} \\
\hspace*{\algorithmicindent}  \emph{replica-id}, \emph{view-id}, \emph{status}, \emph{f} \Comment{Other state variables} \\

\begin{algorithmic}[1]
  \Event{Receive $request$} 
    \State $lastReq\xleftarrow{}$ the last released request from $eb$. 
    \If{$request.deadline>lastReq.deadline$\footnotemark} 
    \State $eb$.\textbf{insert}($request$)
    \Comment{$eb$ is a priority queue}
    \Else
    \State $lb$.\textbf{insert}($request$) \Comment{$lb$ is a map}
    \If{$\emph{replica-id}=\emph{view-id}\  \% \ (2f+1)$} \Comment{It is leader}
    \State $newDdl=\textbf{max}(\textsc{clockTime}(), lastReq.deadline+1)$
    \State $request.deadline=newDdl$ \Comment{Modify its deadline}
    \State $eb.$\textbf{insert}$(request)$ \Comment{Can enter $eb$ with the new deadline}
    \EndIf
    \EndIf
  \EndEvent
  
  \Event{$eb.\mathbf{empty}$()=$\mathbf{false}$ $\mathbf{and}$ $eb.\mathbf{top}$().$deadline$ $\leq$ clockTime() }
  \State $request=$$eb$.\textbf{top}() \Comment{The request to be released from $eb$}
  \State $eb$.\textbf{erase}(request)
  \If{$\emph{replica-id}=\emph{view-id}\  \% \ (2f+1)$} \Comment{Replica is leader}
  \State $result=$\textsc{execute}($request$) \Comment{Only leader executes request}
  \LeftComment{Leader directly appends $request$ and $result$ to \emph{synced-log}}
  \State \emph{synced-log}.\textbf{append}(\{$request$, $result$\})
  \State \emph{hash}=\textsc{calcIncrementHash}(\emph{synced-log}) 
  \State \textsc{sendFastReply}(\emph{result}, \emph{hash}) 
  \LeftComment{In parallel with sending \emph{fast-reply}, leader conducts broadcast}
  \State \textsc{broadcastLogModification}(\emph{request}) 
  \Else
    \LeftComment{Follower appends $request$ to \emph{unsynced-log} without execution}
    \State \emph{unsynced-log}.\textbf{append}(\{$request$, $null$\})
    \State \emph{synced-hash}=\textsc{calcIncrementHash}(\emph{synced-log}) 
    \State \emph{unsynced-hash}=\textsc{calcIncrementHash}(\emph{unsynced-log}) 
    \LeftComment{Follower's hash is generated by concatenating the two parts.}
    \State \textsc{sendFastReply}(\emph{null}, \emph{synced-hash} \textbf{XOR} \emph{unsynced-hash}) 
  \EndIf
  \EndEvent
  
  \LeftComment{Only followers will receive $log\emph{-}modification$ messages}
  \Event{receive $log\emph{-}modification$}
  \LeftComment{Both synced part and unsynced part will be modified, details at \S\ref{sec-slow-quorum-check}}
    \State \textsc{ModifySynced}(\emph{synced-log}, \emph{synced-hash})
    \State \textsc{ModifyUnSynced}(\emph{unsynced-log}, \emph{unsynced-hash})
    \State \textsc{sendSlowReply}() \Comment{Only followers send \emph{slow-reply}}
  \EndEvent
\end{algorithmic}
\endgroup
\end{algorithm}

\begin{algorithm}[!htbp]
\small 
\begingroup
\caption{Proxy Actions}
\label{algo-proxy}
\hspace*{\algorithmicindent} \textbf{Member Variables}: \Comment{} \\
\hspace*{\algorithmicindent} \emph{f},\Comment{the number of replicas is \emph{2f+1}}\\
\hspace*{\algorithmicindent} \emph{replySet},\Comment{the set of replies received from replicas} \\
\begin{algorithmic}[1]

\Event{receive $request\  \mathbf{from}\ \mathbf{client}$}
\State Tag $request$ with the sending time and the deadline
\For{$r$ $\leftarrow$0 to $2f$}
     \State send $request$ to replica $r$ 
\EndFor
\EndEvent

  \Event{receive $reply\ \mathbf{from}\  \mathbf{replica}$}
    \If{$reply$ $\mathbf{is}$ $\mathbf{duplicate}$ $\mathbf{or}$ $\mathbf{from\ previous}\ view$ }
    \State \Return
    \EndIf
    \If{$reply$ $\mathbf{is\ from\ new}\ view$ }
    \LeftComment{Replicas experienced view change, all previous replies are stale}
    \State \emph{replySet}.\textbf{clear}() 
    \EndIf
    
    \State \emph{replySet}.\textbf{insert}(\emph{reply})
    \State $committedReply$= \textsc{checkCommitted}($reply$)
    \If{$committedReply\neq null$}
    \State \textsc{replyToClient}($committedReply$)
    \EndIf
  \EndEvent

 \Function{checkCommitted}{\emph{reply}}
 \LeftComment{If the proper quorum is established, return the leader's reply because it contains the exeuction result}
 \State  $quorum=\{ msg \in replySet: msg.\emph{view-id}=reply.\emph{view-id}$ $\And$ $\emph{msg.client-id}=\emph{reply.client-id}\And\emph{msg.request-id}=\emph{reply.request-id}\}$
    \State $\emph{leader-id}=\emph{reply}.\emph{view-id}\ \%\  \emph{(2f+1)}$

    \If{\emph{quorum} \textbf{not} \textbf{contains} \emph{replica} \emph{leader-id}'s \emph{fast-reply} }
    \State \Return \emph{null} \Comment{Leader's \emph{fast-reply} must be included}
    \EndIf
    \State $\emph{fast-reply-num, slow-reply-num}=0,0$
    \For{$r$ $\leftarrow$0 to $2f$}
        \If{\emph{quorum} \textbf{contains} \emph{replica} \emph{r}'s \emph{slow-reply}}
        \State $\emph{slow-reply-num++}$
        \LeftComment{\emph{slow-reply} can serve as \emph{fast-reply}, but not the opposite}
        \State $\emph{fast-reply-num++}$ 
        \ElsIf{\emph{quorum} \textbf{contains} \emph{replica} \emph{r}'s \emph{fast-reply} $\mathbf{and}$ \emph{replica} \emph{r}'s \emph{fast-reply.hash=} \emph{replica} \emph{leader-id}'s \emph{fast-reply.hash}  }
        \State    $\emph{fast-reply-num++}$ 
        \EndIf
    \EndFor
    \If{$\emph{fast-reply-num}\geq 1+f+\lceil f/2 \rceil$}
    \State \Return \emph{replica} \emph{leader-id}'s \emph{fast-reply} \Comment{Committed in fast path}
    \EndIf 
    \If{$\emph{slow-reply-num}\geq f$} \Comment{We also have leader's \emph{fast-reply}}
    \State \Return \emph{replica} \emph{leader-id}'s \emph{fast-reply} \Comment{Committed in slow path}
    \EndIf 
\EndFunction
\end{algorithmic}
\endgroup
\end{algorithm}

\setlength{\textfloatsep}{0.1cm}
\setlength{\floatsep}{0.2cm}

\subsection{Replica State}
\label{sec-request-state}
\begin{figure}[!t]
    \centering
    \begin{tcolorbox}[colback=white!10,
                      width=8cm,
                      arc=1mm, auto outer arc,
                      boxrule=0.5pt, 
                     ]
                     
    \begin{itemize}[leftmargin=*,nosep]
        \item \emph{replica-id}--- replica identifier ($0,1,\cdots, 2f$).
        \item \emph{view-id}--- the view identifier, initialized as 0 and increased by 1 after every view change.
        \item \emph{status}--- one of \textsc{normal}, \textsc{viewchange}, or \textsc{recovering}.
        \item \emph{early-buffer}--- the priority queue provided by \primname, which sorts and releases the requests according to their deadlines. 
        \item \emph{late-buffer}-- the map provided by \primname, which is searchable by \emph{<client-id, request-id>} of the request.
        \item \emph{log}--- a list of requests, which are appended in the  order of their deadlines.  
        \item \emph{sync-point}--- the log position indicating this replica's \emph{log} is consistent with the leader up to this point. 
        \item \emph{commit-point}--- the log position indicating the replica has checkpointed the state up to this point.
    \end{itemize}
    \end{tcolorbox}
    \caption{Local state of \sysname replicas}
    \label{fig:replica-state}
\end{figure}

 Figure~\ref{fig:replica-state} summarizes the state variables maintained by each replica. We omit some variables (e.g., \emph{crash-vector}) related to \sysname's recovery (\S\ref{sec:recovery}). Below we describe them in detail.
 
 \textbf{replica-id}:  Each replica is assigned with a unique \emph{replica-id}, ranging from 0 to $2f$. The \emph{replica-id} is provided to the replica during the initial launch of the replica process, and is then persisted to \emph{stable storage}, so that the replica can get its \emph{replica-id} after crash and relaunch.
 
 \textbf{view-id}: Replicas leverage a view-based approach~\cite{viewstamp}: each view is indicated by a \emph{view-id}, which is initialized to 0 and incremented by one after every view change. Given a \emph{view-id}, this view's leader's \emph{replica-id} is $\emph{view-id}\%(2f+1)$. 
 
 \textbf{status}: Replicas switch between three different \emph{status}es: \textsc{normal}, \textsc{viewchange}, and \textsc{recovering}. Replicas are initially launched in \textsc{normal} status. When the leader is suspected of failure, followers switch from \textsc{normal} to \textsc{viewchange} and initiate the view change process. They will switch back to \textsc{normal} after completing the view change. For a failed replica to rejoin the system, it starts from \textsc{recovering} status and will switch to \textsc{normal} after recovering its state from the other replicas. 
 
 \textbf{early-buffer}: \emph{early-buffer} is implemented as a priority queue, sorted by requests' deadlines. \emph{early-buffer} is responsible for (1) conducting eligibility checks of incoming requests: if the incoming request's deadline is larger than the last released one from \emph{early-buffer}, then the incoming request can enter the \emph{early-buffer}; and (2) release its accepted requests in the order of their deadlines, thus maintaining DOM's consistent ordering across replicas.

  \footnotetext{When deadlines are equal, the tie is broken by \emph{<client-id},\emph{request-id>}.}
  
 \textbf{late-buffer}: \emph{late-buffer} is implemented as a map using the \emph{<client-id, request-id>} as the key. It is used to hold those requests which are not eligible to enter the \emph{early-buffer}. Replicas maintain such a buffer because those requests may later be needed in the slow path (\S\ref{sec-slow-quorum-check}). In that case, replicas can directly fetch those requests locally instead of asking remote replicas.

 \textbf{log}: Requests released from the \emph{early-buffer} will be appended to the \emph{log} of replicas. The requests then become the entries in the \emph{log}. The \emph{log} is ordered by request deadline.

 \textbf{sync-point}: Followers modify their \emph{log}s to stay consistent with the leader (\S\ref{sec-slow-quorum-check}). \emph{sync-point} indicates the log position up to which this replica's \emph{log} is consistent with the leader. Specially, the leader always advances its \emph{sync-point} after appending a request.
 
 \textbf{commit-point}: Requests (log entries) up to \emph{commit-point} are considered as committed/stable, so that every replica can execute requests up to \emph{commit-point} and checkpoint its state up to this position. \emph{commit-point} is used in an optional optimization (\S\ref{sec-checkpoint}).

\subsection{Message Formats}
\label{sec-message-formats}
There are five types of messages closely related to \sysname. We explain their formats below. Since \sysname uses a view-based approach for leader change, we omit the description of messages related to leader changes; these messages have been defined in Viewstamped Replication~\cite{viewstamp}.

\textbf{request}: \emph{request} is generated by the client and submitted to the proxy. The proxy will attach some necessary attributes and then submit \emph{request} to replicas. \emph{request} is represented as a 5-tuple: 
\begin{equation*}
    \emph{request=<}\emph{client-id}, \emph{request-id}, command, s, d\emph{>}
\end{equation*}
\emph{client-id} represents the client identifier and \emph{request-id} is assigned by the client to uniquely identify its own request. On one replica, \emph{client-id} and \emph{request-id} combine to uniquely identify the request. $command$ represents the content of the request, which will be executed by the leader. $s$ and $d$ are tagged by proxies. $s$ is the sending time of the \emph{request} and $d$ is the estimated deadline that the request is expected to arrive at all replicas. When the request arrives at the replica, the replica can also derive the proxy-replica OWD by subtracting $s$ from its receiving time.

\textbf{fast-reply}: \emph{fast-reply} is sent by every replica after they have appended or executed the request, and it is used for quorum checks in the fast path. \emph{fast-reply} is represented as a 6-tuple:
\begin{equation*}
\emph{fast-reply} = \emph{<}\emph{view-id}, \emph{replica-id}, \emph{client-id}, \emph{request-id}, result, hash\emph{>}
\end{equation*}
\emph{view-id} and \emph{replica-id} are from the replica state variables (see \S\ref{sec-request-state}). \emph{client-id} and \emph{request-id} are from the appended request that lead to this reply. \emph{result} is only valid in the leader's \emph{fast-reply}, and is \emph{null} in followers' \emph{fast-replies}. The proxy can recognize the leader's reply by checking whether its \emph{replica-id} is \emph{view-id}$\%$($2f+1$). \emph{hash} captures a hash of the replica's \emph{log} (the hash calculation is explained in \S\ref{sec-incremental-hash}). Proxies can check the \emph{hash} values to know whether the replicas involved in a quorum have consistent \emph{log}s. 

\textbf{log-modification}: \emph{log-modification} message is broadcast by the leader to convey the \emph{log} entry's \emph{deadline}, \emph{client-id} and \emph{request-id} to followers, making the followers modify their \emph{log}s to keep consistent with the leader. Meanwhile, \emph{log-modification} also doubles as the leader's heartbeat. \emph{log-modification} is represented as a 5-tuple: 
\begin{equation*}
    \emph{log-modification=}\emph{<}\emph{view-id}, \emph{log-id}, \emph{client-id}, \emph{request-id}, \emph{deadline}\emph{>}
\end{equation*} 
\emph{view-id} is from the replica state. \emph{log-id} indicates the position of this log entry (request) in the leader's \emph{log}. \emph{client-id}
and \emph{request-id} uniquely identify the request. \emph{deadline} is the request's deadline shown in the leader's \emph{log}, which is either assigned by proxies on the fast path (i.e., \circled{1} in Figure~\ref{fig-workflow-fast}) or overwritten by the leader on the slow path (i.e., \circled{3} in Figure~\ref{fig-workflow-slow}). \emph{log-modification} messages can be batched under high throughput to reduce the leader's burden of broadcast.

\textbf{slow-reply}: \emph{slow-reply} is sent by followers after all the entries in their \emph{log}s have become the same as the leader's \emph{log} entries up to this request. It is used by the client to establish the quorum in the slow path. \emph{slow-reply} is represented as a 4-tuple:
\begin{equation*}
        \emph{slow-reply} = \emph{<}\emph{view-id}, \emph{replica-id}, \emph{client-id}, \emph{request-id}\emph{>}
\end{equation*}
The four fields have the same meaning as in the \emph{fast-reply}. 

\textbf{log-status}: \emph{log-status} is periodically sent from the followers to the leader, reporting the \emph{sync-point} of the follower's \emph{log}, so that the leader can know which requests have been committed and update its \emph{commit-point}. \emph{log-status} is represented as a 3-tuple.
\begin{equation*}
    \emph{log-status=}\emph{<}\emph{view-id},
    \emph{replica-id}, \emph{sync-point} \emph{>}
\end{equation*} 
The three fields come from the followers' replica state variables.


\subsection{Fast Path}
\label{sec-fast-quorum-check}
\sysname relies on DOM to increase the frequency of its fast path. As shown earlier, the percentile at which DOM estimates OWDs is a parameter set by the DOM user. A lower percentile will set smaller deadlines, which improve fast path latency, but reduce the frequency of the fast path. Higher percentiles have the opposite problem. For \sysname, we use the 50\textsuperscript{th} percentile to strike a balance between the two. This does reduce the fast path frequency compared with using a higher percentile; hence, \sysname compensates for this by optimizing its slow path for low client latency as well.

To commit the request in the fast path (Figure~\ref{fig-workflow-fast}), the proxy needs to get the \emph{fast-reply} messages from both the leader and $f+\lceil f/2 \rceil$ followers. (1) It must include the leader's \emph{fast-reply} because only the leader's reply contains the execution result. (2) It also requires the $f+\lceil f/2 \rceil+1$ replicas have matching \emph{view-id}s and the same \emph{log} (requests). In \S\ref{sec-incremental-hash} we will show how to efficiently conduct the quorum check by using the \emph{hash} field included in \emph{fast-reply}. If both (1) and (2) are satisfied, the proxy can commit the request in 1 RTT.

As briefly explained in the sketch of the fast path (\S\ref{sec:design}), the fast path requires a super quorum ($f+\lceil f/2 \rceil+1$) rather than a simple quorum ($f+1$), because a simple quorum is insufficient to guarantee the correctness of \sysname's fast path. Consider what would happen if we had used a simple majority ($f+1$) in the fast path. Suppose there are two requests \emph{request-1} and \emph{request-2}, and \emph{request-1} has a larger  \emph{deadline}. \emph{request-1} is accepted by the leader and $f$ followers. They send \emph{fast-replies} to the proxy, and then the proxy considers \emph{request-1} as committed and delivers the execution result to the client application. Meanwhile, \emph{request-2} is accepted by the other $f$ followers. After that, the leader fails, leaving $f$ followers with \emph{request-1} accepted and the other $f$ followers with \emph{request-2} accepted. Now, the new leader cannot tell which of \emph{request-1} or \emph{request-2} is committed at a provided log position. If the new leader adds \emph{request-2} into the recovered \emph{log}, it will be appended and executed ahead of \emph{request-1} due to \emph{request-2}'s smaller \emph{deadline}. This violates linearizability~\cite{linearizability}: the client sees \emph{request-1} executed before \emph{request-2} with the old leader but sees the reverse with the new leader.  
\subsection{Slow Path}
\label{sec-slow-quorum-check}

The proxy is not always able to establish a super quorum to commit the request in the fast path. When requests are dropped or are placed into the \emph{late-buffers} on some replicas, there will not be sufficient replicas sending fast replies. Thus, we need the slow path to resolve the inconsistency among replicas and commit the request. We explain the details of the slow path (Figure~\ref{fig-workflow-slow}) below in temporal order starting with the request arriving at the leader.

\Para{Leader processes request.} After the leader receives a \emph{request}, the leader ensures it can enter the \emph{early-buffer}: if it is not eligible due to its small deadline~\circled{2}, the leader will modify its deadline to make it eligible~\circled{3}. Specifically, the modified deadline should be larger than the last released request's deadline. Therefore, we choose the max between (a) the replica's current clock time and (b) the last released request's $\emph{deadline}+\SI{1}{\micro\second}$. The leader then conducts the same operations as in the fast path ( i.e., appending the request~\circled{4}, applying it to the state machine~\circled{5}, and sending \emph{fast-reply}~\circled{6}).

\Para{Leader broadcasts log-modification.}  In parallel with \circled{5}-\circled{6}, the leader also broadcasts a \emph{log-modification} message to followers~\circled{7} after appending each request. Every time a follower receives a \emph{log-modification} message \circled{8}, it checks its log entry at the position \emph{log-id} included in the \emph{log-modification} message. (1) If the entry has the same 3-tuple \emph{<}\emph{client-id}, \emph{request-id}, \emph{deadline}\emph{>} as that included in the \emph{log-modification} message, it means the follower has the same log entry as the leader at this position. (2) If only the 2-tuple \emph{<}\emph{client-id}, \emph{request-id}\emph{>} is matched with that in the \emph{log-modification} message, it means the leader has modified the deadline, so the follower also needs to replace the deadline in its entry with the deadline from the \emph{log-modification} message. (3) Otherwise, the entry has different \emph{<}\emph{client-id}, \emph{request-id}\emph{>}, which means the follower has placed a wrong entry at this position. In this third case, the follower removes the wrong entry and tries to put the right one. It first searches its \emph{late-buffer} for the right entry with matching \emph{<}\emph{client-id}, \emph{request-id}\emph{>}.  As a rare case, when the entry does not exist on this replica because the request was dropped or delayed, the follower fetches it from other replicas and puts it at the position.

\Para{Follower sends {slow-reply}.} After the follower has processed the \emph{log-modification} message, and has ensured the requests in its \emph{log} are the same as the leader's log entries, the follower updates its \emph{sync-point}, indicating it has the same \emph{log} entries as the leader up to the log position represented by the \emph{sync-point}. The leader itself can directly advance its \emph{sync-point} after appending the request to \emph{log}. Then, the follower sends a  \emph{slow-reply} message for every synced request \circled{10}. The \emph{slow-reply} will be used to establish the quorum in the slow path. Specially, a \emph{slow-reply} can be used in place of the same follower's \emph{fast-reply} in the fast path's super quorum, because it indicates the follower's \emph{log} is consistent with the leader. By contrast, the follower's \emph{fast-reply} cannot replace its \emph{slow-reply} for the quorum check in the slow path. 

\Para{Proxy conducts quorum check.} The proxy considers the request as committed when it receives the \emph{fast-reply} from the leader and the \emph{slow-replies} from $f$ followers. The execution result is obtained from the leader's \emph{fast-reply}. Decoupling (leader's) execution from commit enables the proxy to know whether the request is committed even earlier than the leader replica. Meanwhile, replicas can continue to process subsequent requests and are \emph{not blocked by the quorum check in the slow path}, which proves to be an advantage compared to other opportunistic protocols like NOPaxos (see \S\ref{sec-latency-throughput}). Unlike the quorum check of the fast path (\S\ref{sec-fast-quorum-check}), the slow path does not need a super quorum ($1+f+\lceil f/2 \rceil$). This is because, before sending \emph{slow-replies}, the followers have updated their \emph{sync-point}s and ensured that all the requests (log entries) are consistent with the leader up to the \emph{sync-point}s. A simple majority ($f+1$) is sufficient for the \emph{sync-point} to survive the crash. All requests before \emph{sync-point} are committed requests, whose log positions have all been fixed. During the recovery (\S\ref{sec:recovery}), they are directly copied to the new leader’s \emph{log}.

\Para{In the background: followers report {sync-statuses}.} In response to \emph{log-modification} messages, followers send back \emph{log-status} messages to the leader to report their \emph{sync-point}s. The leader can know which requests have been committed by collecting the \emph{sync-point}s from $f+1$ replicas including itself: the requests up to the smallest \emph{sync-point} among the $f+1$ ones are definitely committed. Therefore, the leader can update is \emph{commit-point} and checkpoint its state at the \emph{commit-point}. It can also broadcast the \emph{commit-point} to followers, which enables \emph{followers} to checkpoint their states for acceleration of recovery (\S\ref{sec-checkpoint}). Note that the followers' reporting \emph{sync-status} is not on the critical part of the client's latency on the slow path; it happens in the background. Therefore, the slow path only needs three message delays (1.5 RTTs) for the proxy to commit the request. Besides, the \emph{log-status} messages only serve for an optional optimization to accelerate recovery (details in \S\ref{sec-checkpoint}). The correctness and performance of \sysname will not be affected even if all \emph{log-status} messages are lost.

\subsection{Other Concerns}
\label{sec-timeout}
\Para{Proxy Failure.} Proxy failures do not hurt \sysname's correctness: proxy failures cause the same effect as packet drops, which is already handled by consensus protocols because consensus protocols do not assume reliable communication~\cite{paxos-original, sigmod21-pigpaxos}.


\Para{Client Timeout and Retry.} The client starts a timer while waiting for the proxy's reply. If the timeout is triggered (due to packet drop or proxy failure), the client eventually retries the request with the same or different proxy (if the previous proxy is suspected of failure), and the proxy resubmits the request with a different sending time and (possibly) a different latency bound. As in traditional distributed systems, replicas maintain \emph{at-most-once} semantics. When receiving a request with duplicate \emph{<client-id, request-id>}, the replica resends the previous reply without re-execution.

\section{Recovery}
\label{sec:recovery}

\Para{Assumptions.} We assume replica processes can fail because of process crashes or a reboot of the replica's server. When a replica process fails, it will be relaunched on the same server. However, we assume that there is some stable storage (e.g., disk) that survives process crashes or server reboots. A more general case, which we do not handle, is to relaunch the replica process from a different server with a new disk where the stable storage assumption no longer holds. We also do not handle the case of changing \sysname's $f$ parameter by adding or removing replicas from the system. Both cases are handled by the literature on reconfigurable consensus~\cite{matchmaker_paxos, viewstamp}, which we believe can be adapted to \sysname as well.

\Para{Recovery protocol.} \sysname's recovery protocol consists of two components: replica rejoin and leader change. After a replica fails, it can only rejoin as a follower. If the failed replica happens to be the leader, then the remaining followers will stop processing requests after failing to receive the leader's heartbeat for a threshold of time. Then, they will initiate a view change to elect a new leader before resuming service. We describe the recovery protocol in  pseudo-code in Appendix~\S\ref{recovery-algo}, and include a model-checked TLA+ specification in Appendix~\S\ref{sec-nezha-tla}. We also include the correctness proof in Appendix~\ref{append-correctness}. Here, we only sketch the major steps for the new leader to recover its state ($\emph{log}$). 

After the new leader is elected via the view change protocol, it contacts the other $f$ survived replicas\as{What if there are more than f? Can it contact any f then? yes}, acquiring their \emph{log}s, \emph{sync-point}s and \emph{last-normal-view}s (i.e., the last view in which the replica's status is \textsc{normal}, defined in \S\ref{sec-request-state}). Then, it recovers the \emph{log} by aggregating the \emph{log}s of those replicas with the largest \emph{last-normal-view}. The aggregation involves two key steps. 
\begin{enumerate}[wide, labelwidth=!,nosep,label={(\arabic*)}]
    \item The new leader chooses the largest \emph{sync-point} from the qualified replicas (i.e., the replicas with the largest \emph{last-normal-view}). Then the leader directly copies all the \emph{log} entries up to the \emph{sync-point} from that replica. 
    \item For the remaining \emph{log} entries which have not been copied to the new leader, the new leader checks each of them as follows: if the entry has larger \emph{deadline} than the one located at \emph{sync-point}, the leader checks whether this entry exists on $\lceil f/2 \rceil +1$ out of the qualified replicas. If so, the entry will also be added to the leader's \emph{log}. All the entries are sorted by their \emph{deadline}s.
\end{enumerate}

After the leader rebuilds its \emph{log}, it executes the entries in their \emph{deadline} order. It then switches to \textsc{normal} \emph{status}. After that, the leader distributes its rebuilt \emph{log} to followers. Followers replace their original \emph{log}s with the new ones, and also switch to \textsc{normal}. 


In some cases, the leader change can happen not only because of a process crash but also because of a network partition, where followers fail to hear from the leader for a long time and start a view change to elect the new leader. When the deposed leader notices the existence of a higher view, it needs to abandon its current state, because its current state may have diverged from the state of the new leader. In other words, the state of the deposed leader may include the execution of some uncommitted requests, which do not exist in the new view. To maintain the correct state, the deposed leader obtains the correct new state from another replica in the fresh view. 

\as{Paragraph below could be improved.}
\Para{Avoiding disk writes during normal processing.} While designing the recovery protocol, we aim to avoid disk writes as much as possible. This is because disk writes add significant delays (0.5ms$\sim$20ms per write), significantly increasing client latency. At the same time, we also want to preserve the correctness of our protocol from \emph{stray messages}~\cite{tdsn21-recovery}, which have caused bugs to multiple diskless protocols (e.g., \cite{viewstamp}\cite{nsdi15-specpaxos}\cite{osdi16-nopaxos}\cite{conext20-domino}). \sysname adopts the \emph{crash-vector} technique invented by Michael et al.~\cite{disc17-dcr,disc17-dcr-techreport}, to develop \sysname's recovery protocol. While \sysname still uses stable storage to distinguish whether it is the first launch or reboot,~\footnote{During the startup of the replica, we pass the \emph{replica-id} to the replica. Then the replica tries reading \emph{replica-id} from the stable storage, if the \emph{replica-id} cannot be read from the stable storage, the replica notices it is the first launch and persists its \emph{replica-id} to stable storage, and then it enters the system with \textsc{normal} status. Otherwise, the replica also fetches the same value of \emph{replica-id} from the stable storage, then the replica knows this is a reboot, so it will first go through the recovery process with \textsc{recovering} status, and switch back to \textsc{normal} status after recovery.} it does not use disk writes during normal processing and preserves its correctness from the \emph{stray message} effect. In Appendix~\S\ref{recovery-algo} we describe how to use \emph{crash-vector} to prevent \emph{stray-message}s for \sysname. 

\Para{Correctness.} In Appendix~\S\ref{append-correctness}, we have proved \sysname's three correctness properties. The three properties have also been model-checked in our TLA+ specification (Appendix \S\ref{sec-nezha-tla}).

\begin{itemize}[leftmargin=*]
    \item \textbf{Durability:} if a client considers a request as committed, the request survives replica crashes.
    \item \textbf{Consistency:} if a client considers a request as committed, the execution result of this request remains unchanged after the replica's crash and recovery.
    \item \textbf{Linearizability:} A request appears to be executed exactly once between start and completion. The definition of linearizability can also be reworded as: if the execution of a request is observed by the issuing client or other clients, no contrary observation can occur afterwards (i.e., it should not appear to revert or be reordered). 
\end{itemize}

\section{Optimizations in \sysname}
\label{sec-optimization}

\subsection{Incremental Hash}
\label{sec-incremental-hash}
In \sysname's fast path, \emph{fast-replies} from replicas can form a super quorum only if these replies indicate that the replicas' ordered logs are identical. This is because---unlike the slow path---replicas do not communicate amongst themselves first before replying to the client. One impractical way to check that the ordered logs are identical is to ship the logs back with the reply. A better approach is to perform a hash over the sequence corresponding to the ordered log, and update the hash every time the log grows. However, if the log is ever modified in place (like we need to in the slow path), such an approach will require the hash to be recomputed from scratch, starting from the first log entry.

Instead, we use a more efficient approach by decomposing the equality check of two ordered logs into two components: checking the contents of the 2 logs and checking the order of the 2 logs. Because logs are always ordered by deadline at all our replicas, it suffices for us to check the contents of the 2 logs. The contents of the logs can be checked by checking equality of the 2 sets corresponding to the entries of the 2 logs: this requires only a hash over a set rather than a hash over a sequence.

To compute this hash over a set, we maintain a running hash value for the set. Every time an entry is added or removed from this set, we compute a hash of this entry (using SHA-1) and XOR this hash with the running hash value. This allows us to rapidly update the hash every time a log entry is appended (an addition to the set) or modified (a deletion followed by an addition to the set). The proxy checks for equality of this set hash across all replicas, knowing that equality of the set of log entries guarantees equality of the ordered logs because logs are always ordered by deadlines.

To be more specific, when the replica sends the \emph{fast-reply} for its $n$\textsuperscript{th} request, the hash, denoted as $H_n$, represents the set of all previously appended entries using an incremental hash function~\cite{increment-hash}: 
\begin{equation*}
H_n =  \bigoplus_{1\leq i \leq n} h(request_i)
\end{equation*}
Here, $h(*)$ is a standard hash function (we use SHA1) and  $\oplus$ is the XOR operation. To calculate $h(request_i)$,  we concatenate the values of the request's \emph{deadline}, \emph{client-id}, \emph{request-id} into a bitvector, and then transform it into a hash value. 

To avoid the \emph{stray message} effect~\cite{disc17-dcr,disc17-dcr-techreport}, we also XOR $H_n$ with the hash of the \emph{crash-vector} to get the final hash value:
\begin{equation*}
hash_n = H_n \oplus h(\emph{crash-vector})
\end{equation*}
Here, $h(\emph{crash-vector})$ is calculated by concatenating every integer in the vector (recall that \emph{crash-vector} consists of $2f+1$ integers) and transforming it into a hash value. The inclusion of \emph{crash-vector} is necessary for the correctness of \sysname and we explain this in Appendix~\S\ref{sec-crash-vector-stray-message}.

The replica includes $hash_n$ while sending the \emph{fast-reply} for the $n$\textsuperscript{th} request in its \emph{log}. Assuming no hash collisions, $hash_n$ represents the replica state when replicas reply to the proxy. By comparing the \emph{hash} in the \emph{fast-replies} from different replicas, the proxy can check whether they have the same requests. 

\subsection{Commutativity Optimization}
\label{sec-commutativity-optimization}
To enable a high fast commit ratio without a long holding delay of DOM, we employ a commutativity optimization in \sysname. As an example, commutative requests refer to those requests operating on different keys in a key-value store, so that the execution order among them does not matter~\cite{sosp13-commutativity-rule, nsdi19-curp}. The commutativity optimization enables us to choose a modest percentile (50\textsuperscript{th} percentile) while still achieving a high fast commit ratio, because it eases the fast path in two aspects. 


First, it relaxes the eligibility check condition of the \emph{early-buffer}. Without commutativity, DOM prevents the incoming request from entering the \emph{early-buffer} if its deadline is smaller than the last request released from the \emph{early-buffer} (\S\ref{sec-dom}). Otherwise, DOM's consistent ordering property is violated. However, the execution results of commutative requests are not affected by their order~\cite{nsdi19-curp}. Hence, consistent ordering is only required among \emph{non-commutative} requests, which enables the relaxation of the \emph{early-buffer}'s entrance check: the request can enter the \emph{early-buffer} if its deadline is larger than the last released request, \emph{which is not commutative with the incoming request}.

Second, it refines the hash computation, making hash consistency among replicas becomes easier. Since read requests do not modify replica state, the \emph{hash} field in the \emph{fast-reply} does not need to encode read requests. Besides, when encoding previous write requests, the \emph{hash} field only considers those that are not commutative to the current request. To do that, \sysname maintains a table of per-key hashes for the write requests. For every newly appended write request, the replica will XOR its hash to update the corresponding per-key hash in the table according to its key. While sending the \emph{fast-reply} for a specific request, the replica only includes the hash of the same key. For compound requests, which write (and hence do not commute with) multiple keys (e.g., ``move 10 from $x$ to $y$ and return $x$ and $y$''), the replica fetches the hashes of all relevant keys (e.g. $x$ and $y$), and includes the XORed hash value (e.g. $hash_x \oplus hash_y$) in the \emph{fast-reply}.

We also evaluate across a range of workloads in Appendix \S\ref{sec:extensive-eval}, with different read/write ratios and skew factors. The result shows that the commutativity optimization helps reduce the latency by \rv{\SI{7.7}{\percent}-\SI{28.9}{\percent}}.

\subsection{Periodic Checkpoints}
\label{sec-checkpoint}
To (1) accelerate the recovery process after leader failure and (2) enable a deposed leader to quickly catch up with the fresh state, we integrate a periodic checkpoint mechanism in \sysname.

Since \sysname only allows the leader to execute requests during normal processing, it can lead to inefficiency during leader change, either caused by leader's failure or network partitions. This is because the new leader is elected from followers, and it has to execute all requests from scratch after it becomes the leader. To optimize this, we adopt an idea from NOPaxos~\cite{osdi16-nopaxos} and conduct synchronization of the replica state in the background.

Periodically, the followers report their \emph{sync-point}s to the leader, and the leader chooses the smallest \emph{sync-point} among the $f+1$ replicas as the \emph{commit-point}, and broadcasts the \emph{commit-point} to all replicas. Both the leader and followers checkpoint state at their \emph{commit-point}s. The periodic checkpoints bring acceleration benefit in two aspects: (1) When the leader fails, the new leader only needs to recover and execute the requests from its \emph{commit-point} onwards. (2) When network partitions happen, the leader is deposed and it later notices the existence of a higher view. Instead of abandoning its complete state (as what we described in \S\ref{sec:recovery}), it can start from its latest checkpoint state to stable storage, and only retrieve from another replica (in the fresh view) requests beyond its \emph{commit-point}. 

\section{Evaluation}
\label{sec:evaluation}

We answer the following questions during the evaluation:
\begin{enumerate}[wide, labelwidth=!,nosep,label=(\arabic*)]
\item How does \sysname compare to the baselines (Multi-Paxos, Fast Paxos, NOPaxos) in the public cloud? (\S\ref{sec-latency-throughput})
\item How does \sysname compare to the recent protocols which also use clock synchronization (i.e., Domino and TOQ-EPaxos)? (\S\ref{sec-eval-domino-toq})
\item How effective are the proxies in saving clients' CPU cost, especially when there is a large number of replicas? (\S\ref{sec:proxy})
\item How fast can \sysname recover from the leader failure? (\S\ref{eval-failure-recovery})
\item How does \sysname compare to Raft when both are equipped with log persistence to stable storage? (\S\ref{sec-nezha-vs-raft})
\item Does \sysname provide sufficient performance for replicated applications? (\S\ref{sec:app})
\end{enumerate}

\subsection{Settings}
\Para{Testbed.} We run experiments in Google Cloud. We employ \texttt{n1-standard-4} VMs for clients, \texttt{n1-standard-16} VMs for replicas and the NOPaxos sequencer, and \texttt{n1-standard-32} VMs for \sysname proxies. All VMs are in a single cloud zone (except \S\ref{sec:wan-comp}). Huygens is installed on all VMs and has an average 99\textsuperscript{th} percentile clock offset of \SI{49.6}{\nano\second}. 
\as{49.6 ns does seem very good at the 99th percentile. I thought Huygens could only do 100s of ns in the public cloud.}

\Para{Baselines.} We compare with Multi-Paxos, Fast Paxos and NOPaxos. For the 3 baselines, we use the implementation from the NOPaxos repository~\cite{nopaxos-code} with necessary modification: (1) we change switch multicast into multiple unicasts because switch multicast is unavailable in cloud. (2) we use a software sequencer with multi-threading for NOPaxos because tenant-programmable switches are not yet available in cloud. We also added two recently proposed protocols that leverage synchronized clocks for comparison, i.e., Domino~\cite{conext20-domino} and TOQ-EPaxos~\cite{nsdi21-epaxos}. We choose to compare them with \sysname because they also use clock synchronization to accelerate consensus. Domino is previously tested with clocks synchronized by Network Time Protocol (NTP) in \cite{conext20-domino}, but in our test, we also provide Huygens synchronization for Domino to give it more favorable conditions because Huygens has higher accuracy than NTP~\cite{huygens}. However, since both Domino and TOQ-EPaxos target WAN settings, we do not expect them to perform better than \sysname or our other baselines in LAN settings, which is verified by our experiments (\S\ref{sec-latency-throughput}). We intended to compare with Derecho~\cite{tocs19-derecho}. However, its performance degrades significantly in the public cloud due to a lack of RDMA (see Appendix~\S\ref{sec-derecho-issue}). We think the comparison is not fair to Derecho and do not include it in the main body.


\Para{Metrics.} We measure execution latency: the time between when a client submits a request to the system and receives an execution result from it along with a confirmation that the request is committed. We also measure throughput. To measure latency, we use median latency because it is more robust to heavy tails. We attempted to measure tail latency at the 99th and 99.9th percentile. But we find it hard to reliably measure these tails because tail latencies within a cloud zone can exceed a millisecond~\cite{nsdi13-bobtail,hotos15-tail-latency,silo} despite a median of only hundreads of microseconds. This is unlike the WAN setting where tails can be more reliably estimated~\cite{nsdi21-epaxos}. We run each experiment 5 times and average values before plotting.

\Para{Evaluation method.} We follow the method of NOPaxos~\cite{osdi16-nopaxos} and run a \emph{null application} with no execution logic. Traditional evaluation of consensus protocols~\cite{sosp13-epaxos,nsdi15-specpaxos, osdi16-nopaxos,osdi16-janus,podc19-paxos-raft,vldb21-compartmentalized-paxos} use closed-loop clients, which issue a continuous stream of back-to-back requests, with exactly one outstanding request at all times. However, the recent work~\cite{nsdi21-epaxos} suggests a more realistic open-loop test with a Poisson process where the client can have multiple outstanding requests (sometimes in bursts). We use both closed-loop and open-loop tests. While comparing the latency and throughput in \S\ref{sec-latency-throughput}, we use 3 replicas. For the closed-loop test, we increase load by adding more clients until saturation.~\footnote{Specially, when the system is saturated, the throughput can drop instead of continuously increasing~\cite{osreview01-seda,osdi20-overload-control}, as shown in Figure~\ref{fig-latency-tp}.} For the open-loop test, we use 10 clients and increase load by increasing the Poisson rate until saturation.

\Para{Workloads.} Since the three baselines (Multi-Paxos, Fast Paxos and NOPaxos) are oblivious to the read/write type and commutativity of requests, and the \emph{null application} does not involve any execution logic, we simply measure their latency and throughput under one type of workload, with a read ratio of \SI{50}{\percent} and a skew factor~\cite{sigmod94_zipfian} of 0.5. We also evaluate \sysname under various read ratios and skew factors in Appendix~\S\ref{sec:extensive-eval}, which verifies the robustness of its performance.

\subsection{Comparison with Multi-Paxos, Fast Paxos and NOPaxos} 
\label{sec-latency-throughput}

\begin{figure*}[!t]
    \centering
    \begin{minipage}{0.6\textwidth}
    \subcaptionbox{Closed-loop workload\label{closed-loop-latency-throughput-part} }
      {
      \centering
      \includegraphics[width=1\linewidth]{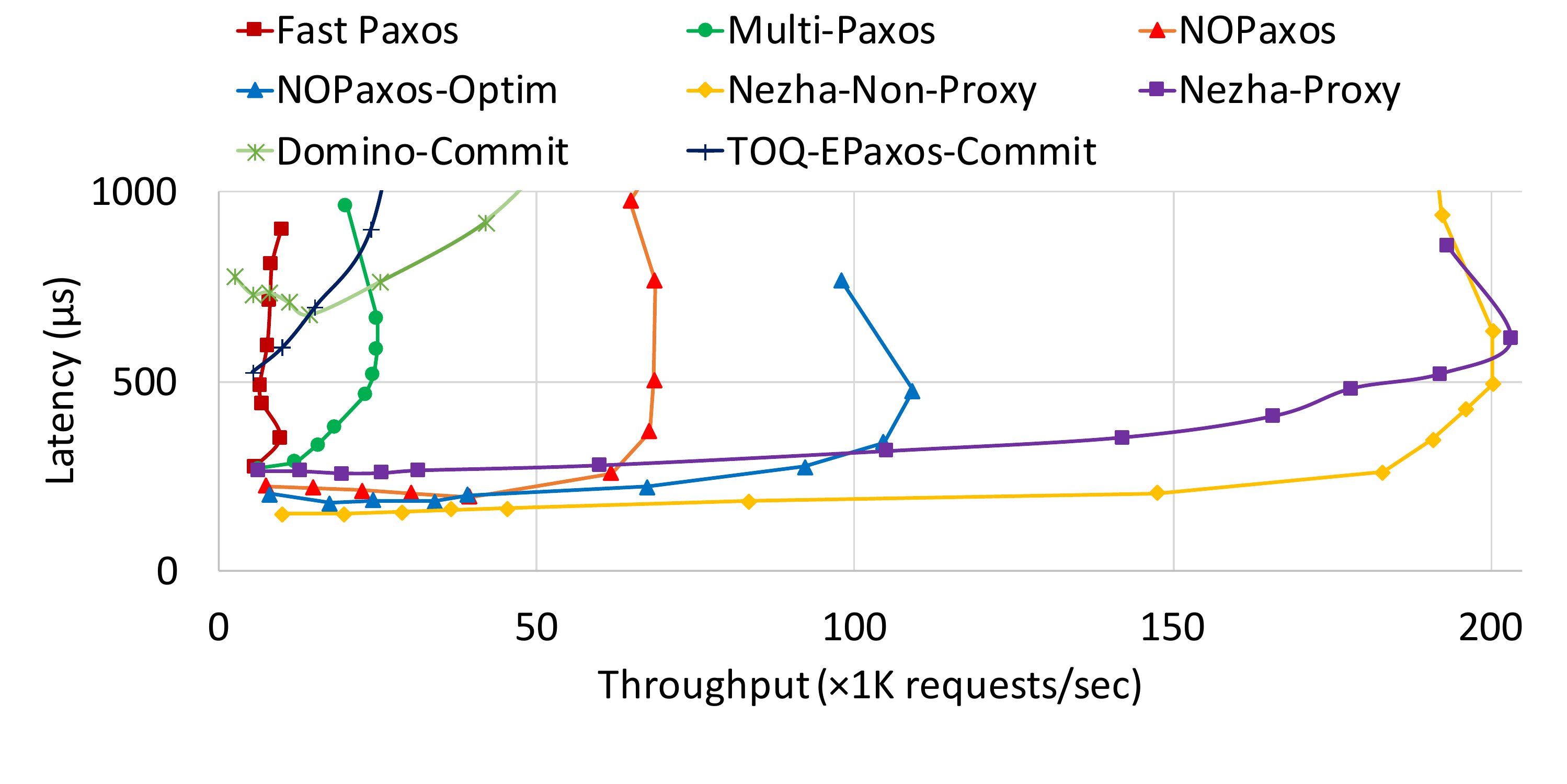}
      }
      \end{minipage}
      
     \begin{minipage}{0.6\textwidth}
    \subcaptionbox{Open-loop workload\label{open-loop-latency-throughput-part}}
      {
      \centering
      \includegraphics[width=1\linewidth]{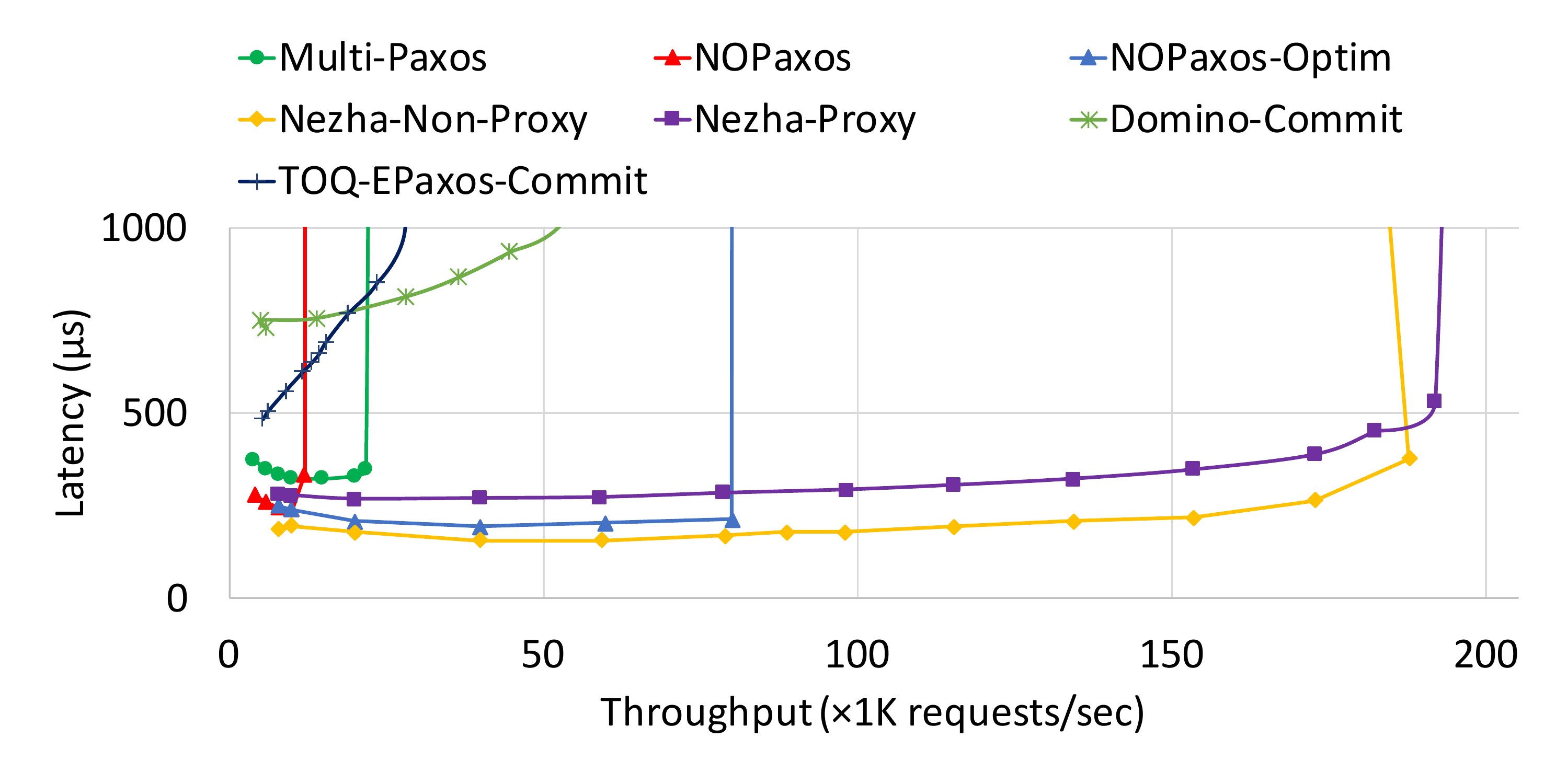}
     }
   \caption{Latency vs.throughput}
   \label{fig-latency-tp}
   \end{minipage}
\end{figure*}


The closed-loop and open-loop evaluation results are shown in Figure~\ref{fig-latency-tp}. We plot two versions of \sysname. \sysname-Proxy uses standalone proxies whereas \sysname-Non-Proxy lets clients undertake proxies' work. Below we discuss three main takeaways.


First, all baselines yield poorer latency and throughput in the public cloud, in comparison with their published numbers from highly-engineered networks~\cite{osdi16-nopaxos}. This is not suprising since the public cloud environment without network support is not the main focus of~\cite{nsdi15-specpaxos, osdi16-nopaxos}.  Fast Paxos suffers the most and reaches only 4.0K requests/second at \SI{425}{\micro\second} in open-loop test (not shown in Figure~\ref{open-loop-latency-throughput-part}). When clients send at a higher rate, Fast Paxos suffers from heavy reordering, and the reordered requests force Fast Paxos into its slow path, which is even more costly than Multi-Paxos.

Second, NOPaxos performs unexpectedly poorly in the open-loop test, because it performs \emph{gap handling} and \emph{normal request processing} in one thread. NOPaxos \emph{early binds} the sequential number with the request at the sequencer. When request reordering/drop inevitably happens from the sequencer to replicas, the replicas trigger much gap handling and consume significant CPU cycles. We realized this issue and developed an optimized version (NOPaxos-Optim in Figure~\ref{fig-latency-tp}) by using separate threads for the two tasks. NOPaxos-Optim outperforms all the other baselines because it offloads request serialization to the sequencer and quorum check (fast path) to clients. But it still loses significant throughput in the open-loop test compared with the closed-loop test. This is because open-loop tests create more bursts of requests, and cause packet reordering/drop more easily. When the gap occurs, NOPaxos needs at least one RTT for the leader to coordinate with followers to fetch the missing request or mark no-op at the gap position. During the gap handling process, all the incoming requests have to be pending and can no longer be processed (i.e., the normal request processing logic is \emph{blocked}). Thus, all these follow-up requests will make the gap handling cost as part of their latencies, and they can also continue to cause more gaps. Meanwhile, the system's overall throughput is also degraded because no more requests are processed until the gap handling is completed.

Third, \sysname achieves much higher throughput than all the baselines, and \sysname-Non-Proxy also achieves the lowest latency because of co-locating proxies with clients. Even equipped with standalone proxies, \sysname-Proxy still outperforms all baselines at their saturation throughputs, except NOPaxos-Optim (open-loop). \sysname's improved throughput and latency come from three design aspects: (1) \primname helps create consistent ordering for the replication protocol, and makes it easier for replicas to achieve consistency. (2) \sysname separates request execution and quorum check, letting clients/proxies undertake quorum check instead of the leader, which effectively relieves leader's burden and enables better pipelining (i.e., avoid the blocking problem in NOPaxos). (3) The use of commutativity further reduces the latency by allowing more requests to be committed in fast path. To verify the benefit of each component, we further conduct an ablation study in \S\ref{sec-ablation-study-1}.

\subsection{Comparison with Domino and TOQ-EPaxos}
\label{sec-eval-domino-toq}
For a fair comparison of Domino and TOQ-EPaxos with \sysname, we originally wanted to plot their execution latencies in Figure~\ref{fig-latency-tp}.  However, both Domino and TOQ-EPaxos decouple commit from execution, and execution happens much later than commit, which causes high execution latencies. In our experiments, we found that Domino's execution latency exceeds \SI{10}{\milli\second} and TOQ-EPaxos' execution latency ranges from \SI{1.3} to \SI{3.3}{\milli\second}, which are significantly larger than \sysname (as well as our other baselines). This makes it hard to show them in our figure, and hence we plot their commit latencies instead. 

When comparing the commit latency of Domino and TOQ-EPaxos to \sysname's execution latency, we still find that \sysname performs better. We believe this is because of differences in implementation: Domino and TOQ-EPaxos are implemented in Golang with gRPC~\cite{domino-repo,toq-repo} whereas \sysname and the other baselines are implemented in C++ with UDP~\cite{nopaxos-code}). These differences in implementation likely arise from the fact that the additional latency incurred by Golang+gRPC is tolerable for wide-area use cases that typically have higher latencies. We also compare \sysname with Domino and TOQ-EPaxos below from a design perspective.

\as{Try to shrink Nezha with Domino a little bit.}

\subsubsection{\sysname compared with Domino.} 

\sysname and Domino both use synchronized clocks with deadlines attached to messages. They differ in 3 ways. 

(1) Unlike Domino, \sysname does not decouple commit from execution, which makes it easier for \sysname to be a drop-in replacement for Paxos/Raft, where applications can directly get the execution result from the commit reply. 

(2) \sysname's correctness is independent of clock skew, whereas Domino can violate durability when the clock values are not monotonically increasing. When a request arrives later than its deadline, Domino replicas are expected to reject it. But if clock skew occurs at this moment, the replicas will accept it and make the client consider it as committed, even though this request may later be replaced by a no-op. We explain with error traces in Appendix~\S\ref{error-trace-domino}. 

(3) \sysname guarantees the linearizability, whereas Domino does not specify the consistency model it uses. We suspect Domino uses a weak consistency model, i.e., eventual consistency. However, the durability violation in (2) also causes the eventual consistency not guaranteed (details in Appendix~\S\ref{error-trace-domino}).

\subsubsection{\sysname compared with TOQ-EPaxos.} 

(1) TOQ-EPaxos only synchronizes replicas to reduce reordering of messages between replicas, so that replicas are more likely to process non-commutative requests in the same order (refer to \S 4 in~\cite{nsdi21-epaxos}). \sysname synchronizes replicas and proxies to reduce the reordering from proxies to replicas. Compared with TOQ-EPaxos, \sysname controls more paths in the consensus workflow, which makes \sysname's acceleration more effective: when clients and replicas are located in different zones, TOQ provides little benefit, whereas \sysname can still reduce latency (\S\ref{sec:wan-comp}). 

(2) TOQ does not guarantee \emph{consistent ordering} but DOM does. In TOQ, when one replica multicasts the requests with a \texttt{ProcessAt} time, if some requests arrive at some replicas very late, then different replicas can still have different message orders. By contrast, DOM prioritizes \emph{consistent ordering} over \emph{set equality}, which means, any two replicas can never release the same requests in different order. DOM adopts such design because our \sysname protocol can rapidly fix set inequality based on its single leader design, unlike EPaxos which involves multiple leaders in its design.

(3) TOQ does not improve EPaxos performance in a LAN. As shown in~\cite{sigmod19-paxi}, even implemented under the same framework,  EPaxos is less performant than Multi-Paxos in LAN. By contrast, \sysname is a generally high-performance protocol which yields good performance in both LAN (Figure~\ref{fig-latency-tp}) and WAN (Figure~\ref{fig:wan-comp}) settings.

\subsection{Ablation Study}
\label{sec-ablation-study-1}
\begin{figure}[!t]
    \centering
    \includegraphics[width=0.5\textwidth]{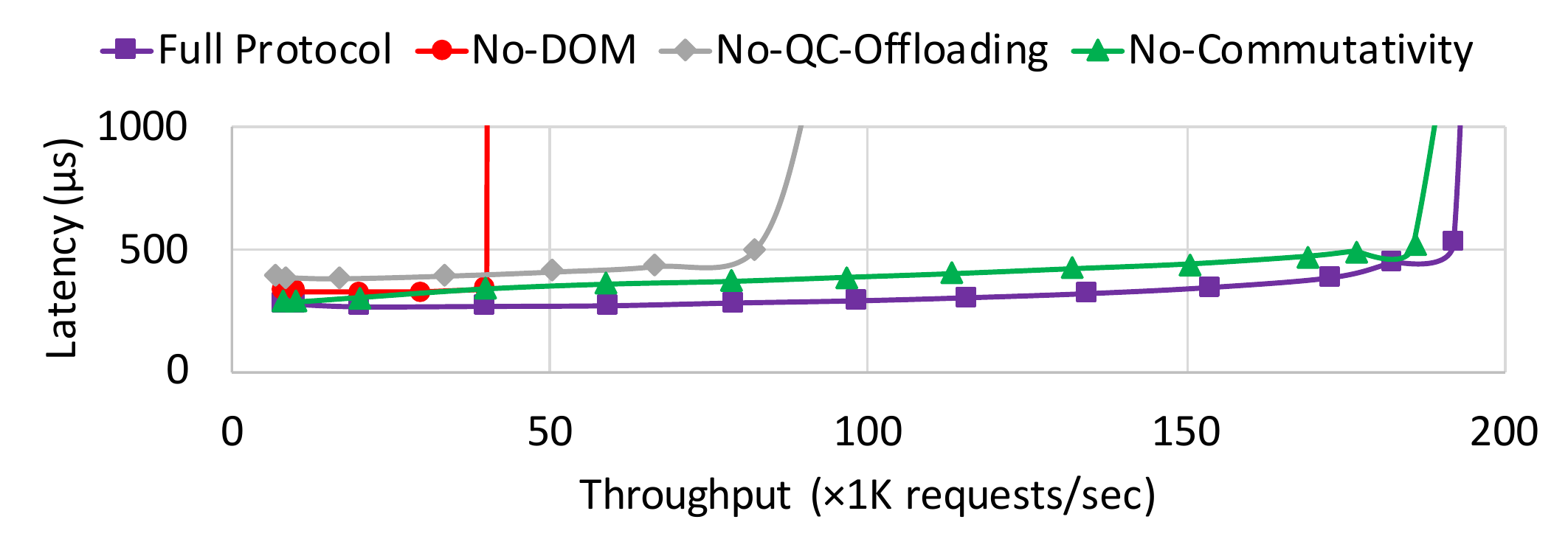}
    \caption{Ablation study of \sysname}
    \label{fig-ablation-1}
\end{figure}

During the ablation study of \sysname, we remove one component from the full protocol of \sysname each time to yield three variants, shown as No-DOM, No-QC-Offloading, No-Commutativity in Figure~\ref{fig-ablation-1}. No-DOM variant removes the DOM primitive from \sysname. No-QC-Offloading variant relies on the leader replica to do the quorum check, and it still relies on DOM for consistent ordering (the proxies still perform request multicast). No-Commutativity variant disables \sysname's commutativity optimization. We run all protocols under the same setting as Figure~\ref{open-loop-latency-throughput-part}.

Figure~\ref{fig-ablation-1} shows that, removing any of the three components degrades performance (i.e., throughput and/or latency). 

\begin{enumerate}[wide,label=(\arabic*)]
    \item The No-DOM variant makes the fast path meaningless: requests are no longer ordered by their deadlines without DOM, so consistent ordering is not guaranteed and set equality (i.e. reply messages with consistent hash) no longer indicates the state consistency among replicas. In this case, the No-DOM variant actually becomes the Multi-Paxos protocol with quorum check offloading, and the leader replica still takes the responsibility of ordering and request multicast,  which makes No-DOM variant yield a much lower throughput and higher latency. 

\item The No-QC-Offloading variant still uses DOM for ordering and request multicast, but it relies on the leader to do quorum check for every request. Therefore, the leader's burden becomes much heavier than the full protocol, and the heavy bottleneck at the leader replica degrades the throughput and latency performance.

\item The No-Commutativity variant degrades the fast commit ratio and causes more requests to commit via the slow path. It does not cause a distinct impact on the throughput. However, compared with the full protocol,  the lack of commutativity optimization degrades the latency performance by up to \SI{24.2}{\percent}. 
\end{enumerate}


\subsection{DOM's Trade-Off at Different Percentiles}
\label{sec:dom-tradeoff}

\begin{figure}[!t] 
\begin{minipage}{0.5\textwidth}
\centering
    \subcaptionbox{Without Commutativity\label{dom-tradeoff-no-comm} }
      {
      \includegraphics[width=0.6\linewidth]{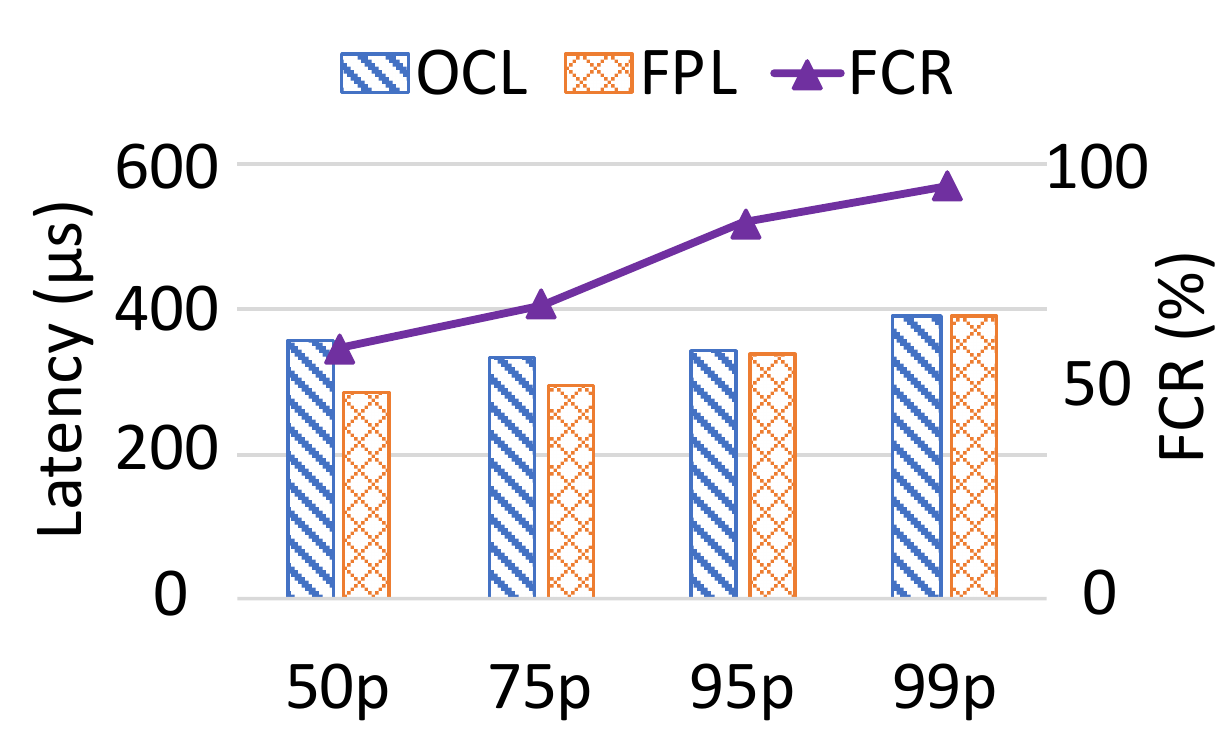}
      }
      \hspace{-0.2cm}
    \subcaptionbox{With Commutativity\label{dom-tradeoff-with-comm}}
      {
      \includegraphics[width=0.6\linewidth]{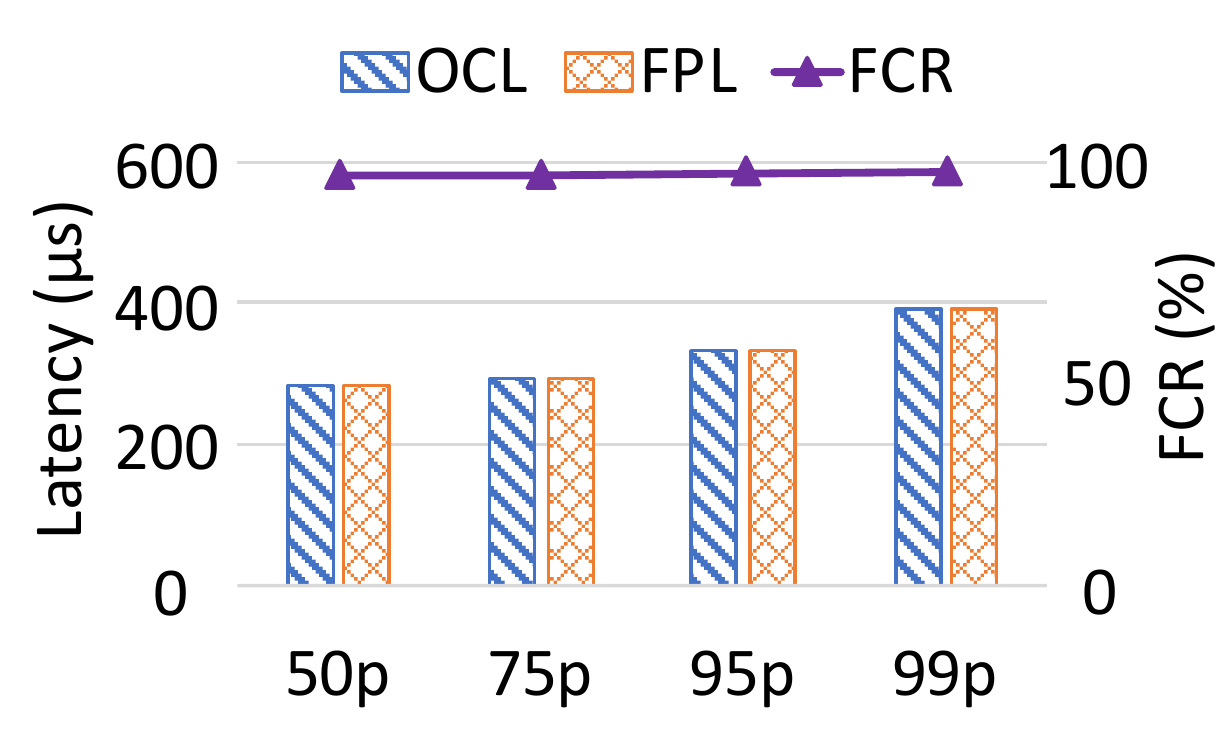}
     }
  \caption{Trade-off of using different percentiles in DOM}
  \label{fig:dom-trade-off}
\end{minipage}
\end{figure}


At the level of DOM primitive, the percentile used by DOM makes a trade-off between reordering rate and latency (Figure~\ref{fig-micro-dom-effectiveness}). When it comes to \sysname, the percentile makes a trade-off between \emph{how fast the request can be committed via the fast path}  and \emph{how frequently the request can be committed via the fast path}. We measure these two aspects with \emph{fast path latency} (FPL) and \emph{fast commit ratio} (FCR) respectively. FPL is the median latency for requests to commit in fast path; FCR is the ratio of requests committed in fast path. A larger percentile leads to higher (better) FCR but longer (worse) FPL, but both FCR and FPL affect the the overall commit latency (OCL) \emph{of all requests}. 

To measure this, in Figure~\ref{fig:dom-trade-off} we run \sysname with/without commutativity optimization in the open-loop test (20K requests/second).

(1) Without commutativity optimization (Figure~\ref{dom-tradeoff-no-comm}), as we use larger percentiles (from 50p to 75p), the improvement of FCR outweighs the increase of FPL, thus leading to lower OCL. However, as we continue to use larger percentiles (e.g., 95p and 99p), though FCR keeps growing, FPL also becomes longer due to the increasing holding delay in \emph{early-buffer}, which undermines the benefit of fast path, and no longer helps reduce OCL.

(2) After adding the commutativity optimization (Figure~\ref{dom-tradeoff-with-comm}), \sysname already reaches a high FCR using 50p. Therefore, using larger percentiles brings little FCR improvement, but only increases FPL, and further FPL. Therefore, we choose 50p in DOM and have verified its robustness under different workloads (details in \cite{nezha-tech-report}).

\subsection{Scalability}
\label{sec-scalability}

\begin{figure*} 
\centering
\begin{minipage}{0.8\textwidth}
    \subcaptionbox{Closed-loop\label{closed-loop-tp-scalability-part} }
      {
      \includegraphics[width=0.5\linewidth]{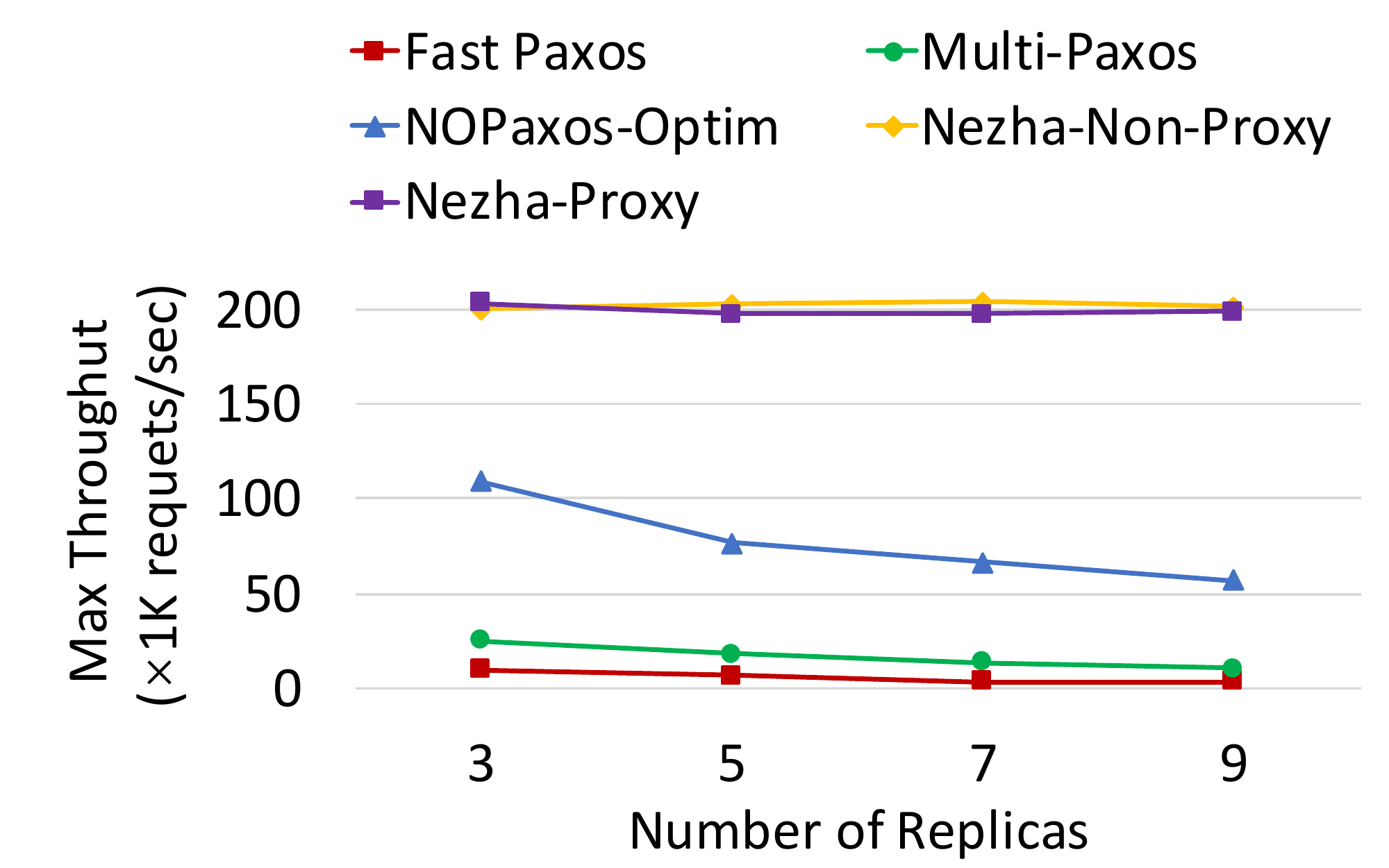}
      }
    \subcaptionbox{Open-loop\label{open-loop-tp-scalability-part}}
      {
      
      \includegraphics[width=0.4\linewidth]{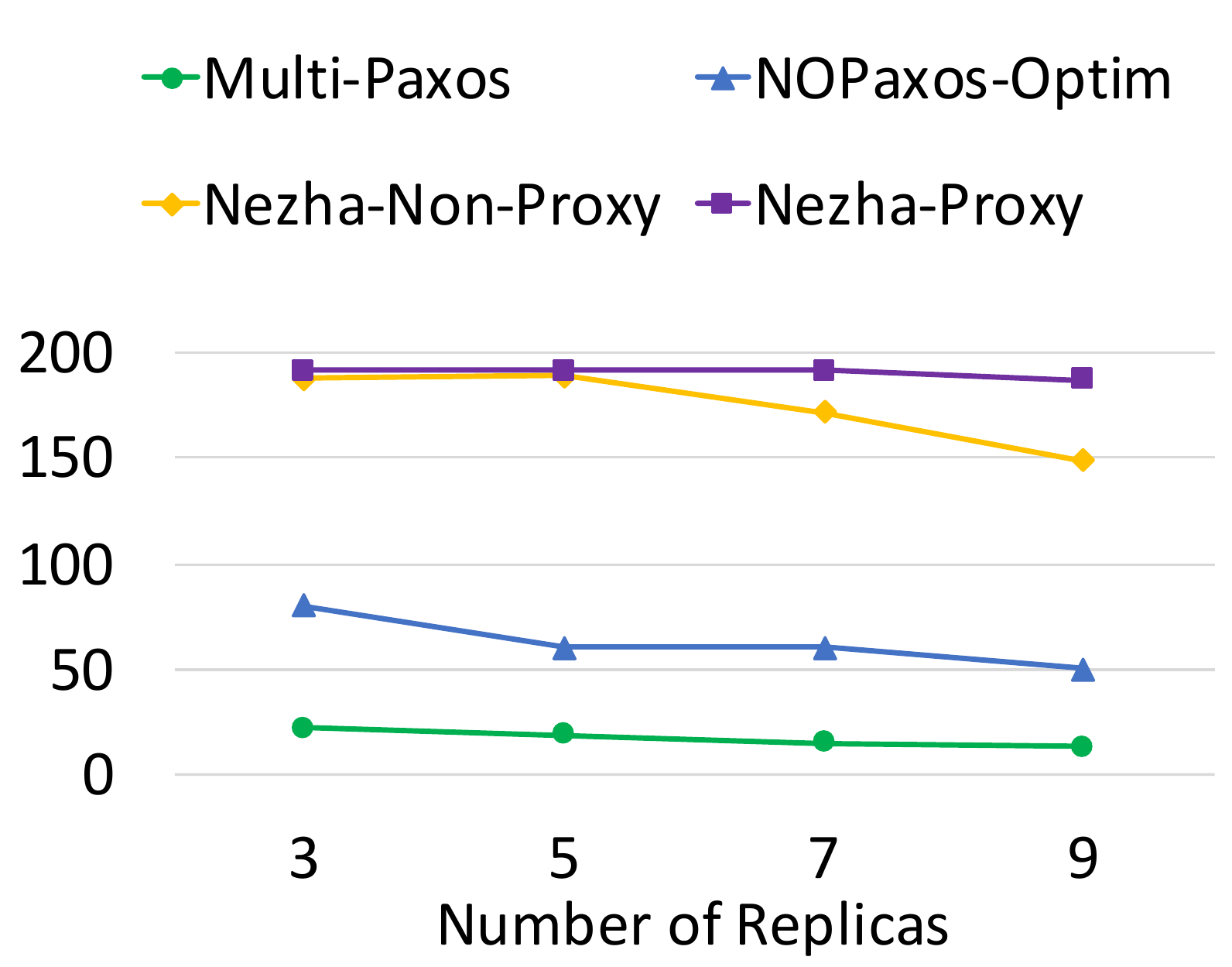}
      
     }
  \caption{Max throughput vs. number of replicas}
  \label{tp-scalability-part}
\end{minipage}
\end{figure*}

\begin{figure*}[!t]
    \begin{minipage}{1\linewidth}
    \centering
    \subcaptionbox{Latency vs. throughput \label{proxy-latency-tp}}
      {
      \includegraphics[width=0.32\linewidth]{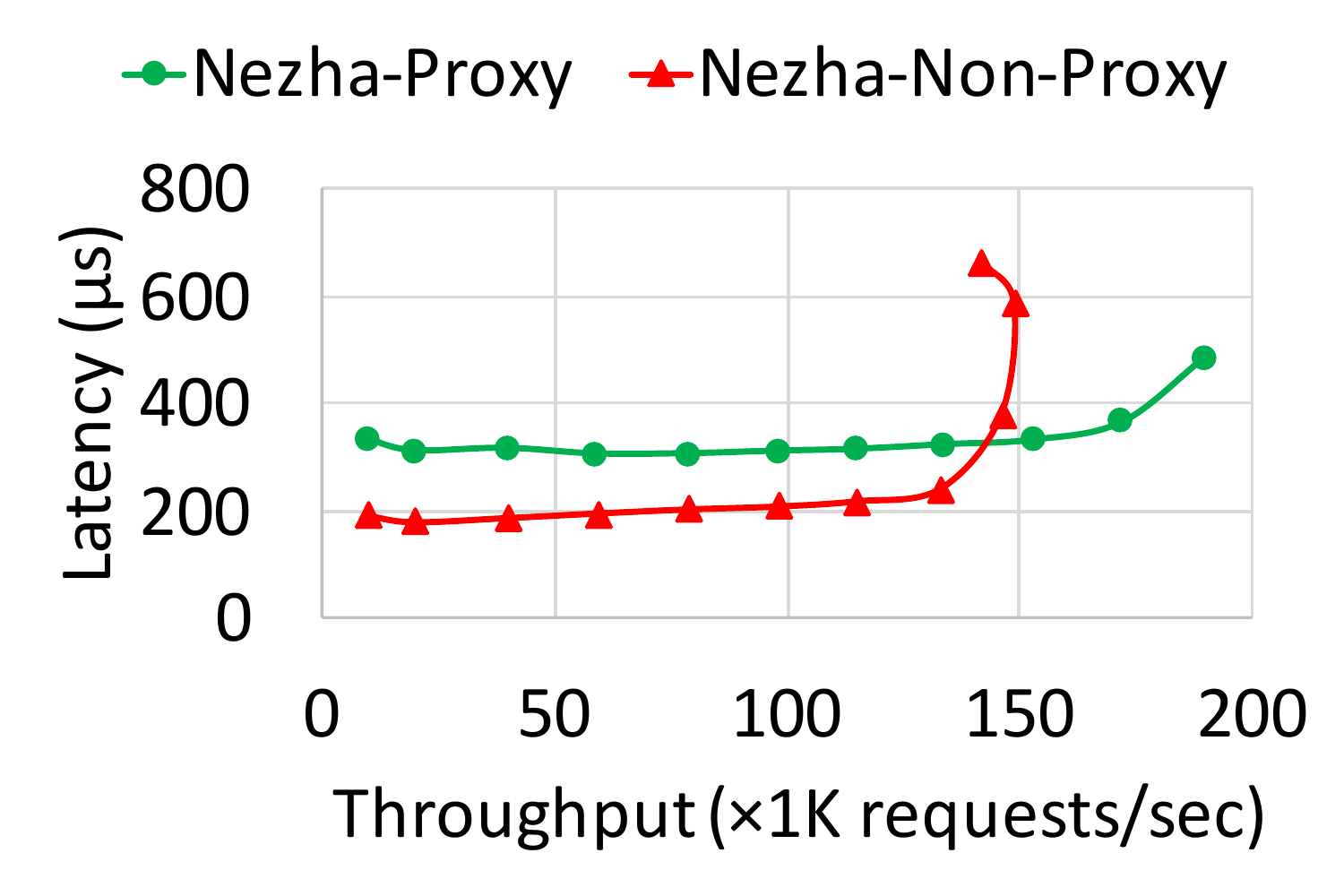}
      }
    \subcaptionbox{CPU cost vs. throughput    \label{proxy-cpu-tp}}
      {
      \includegraphics[width=0.32\linewidth]{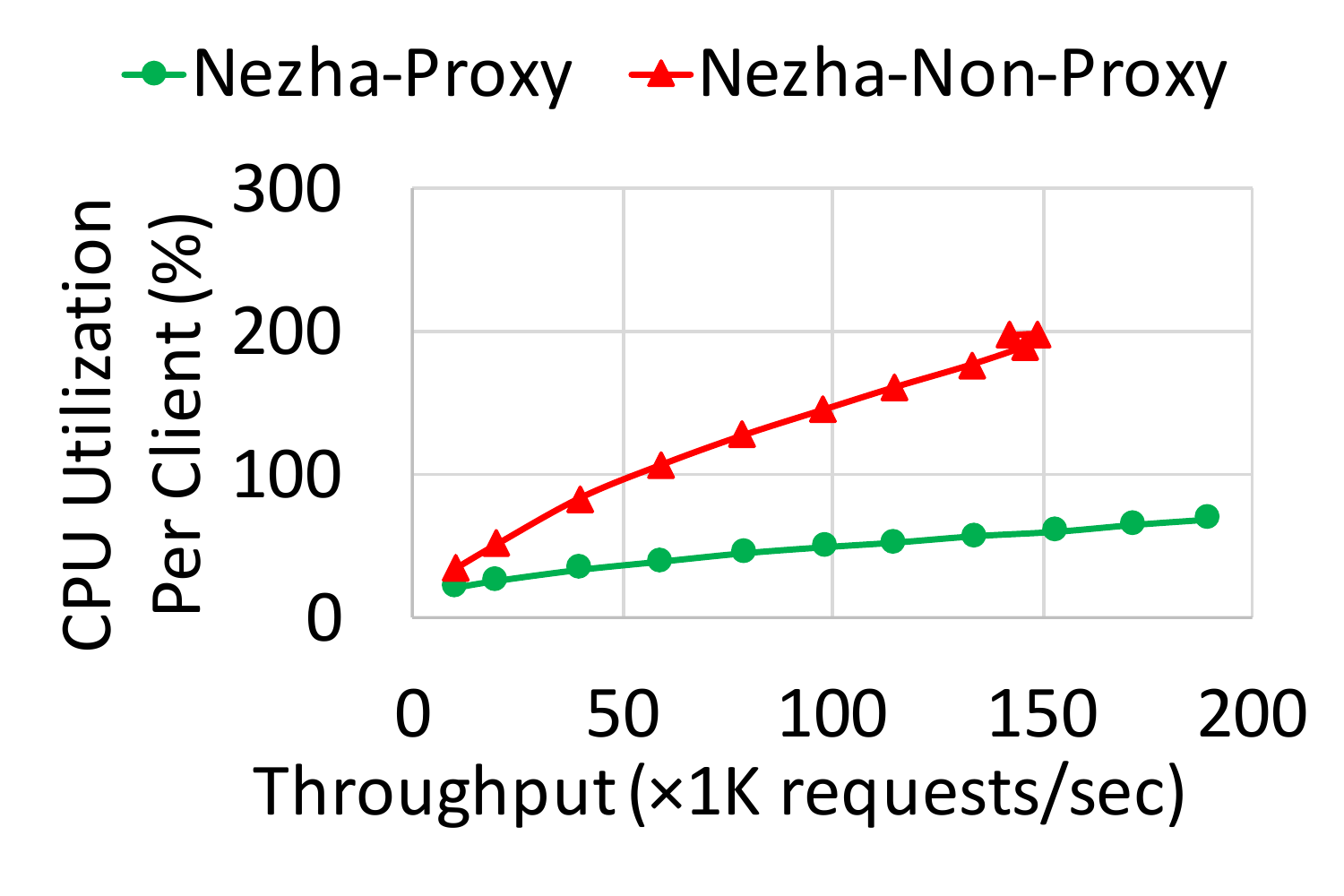}
     }
    \subcaptionbox{Max. client throughput    \label{proxy-individual-client}}
      {
      \includegraphics[width=0.32\linewidth]{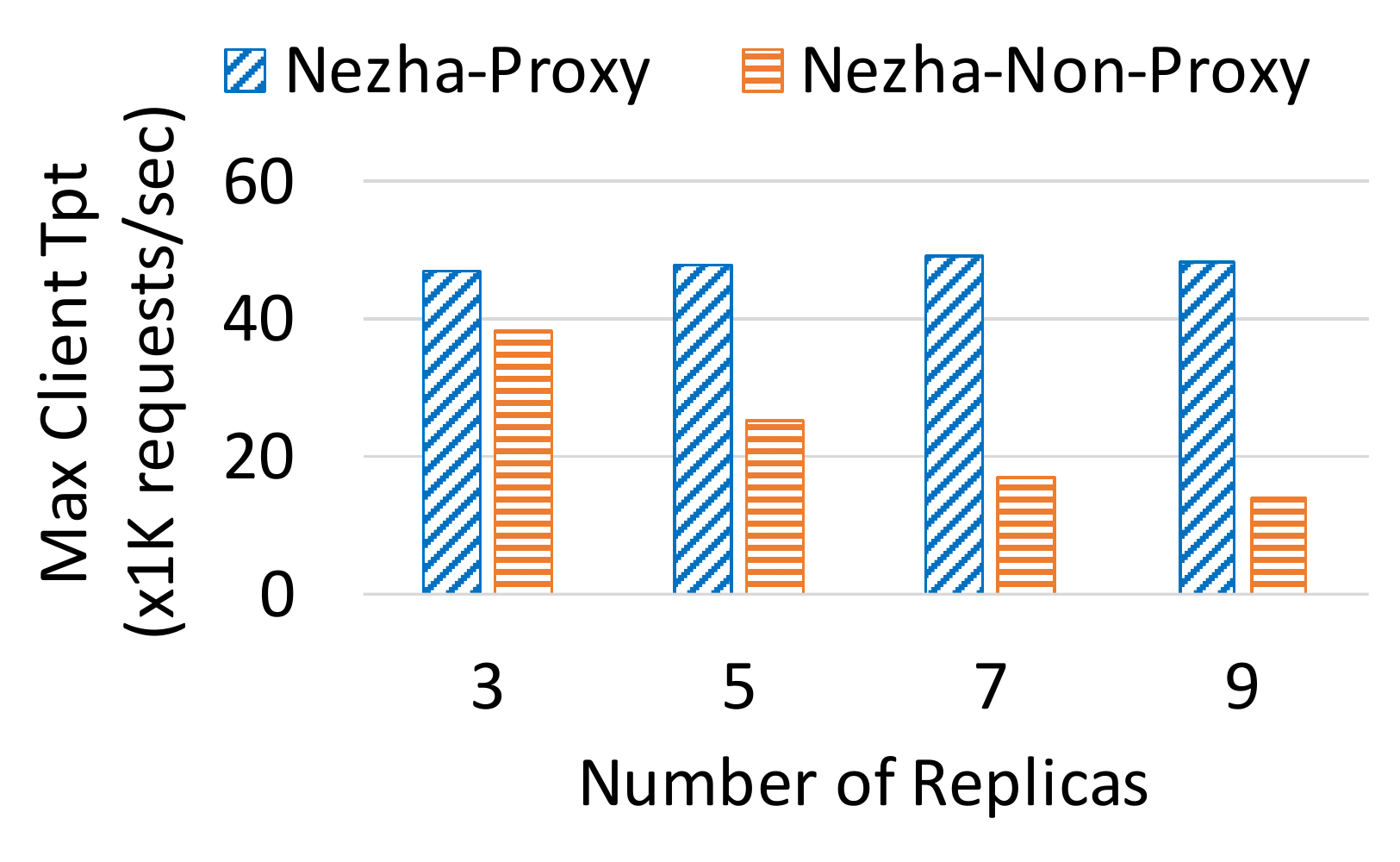}
     }
  \caption{Proxy Evaluation}
  \label{proxy-eval-part}
  \end{minipage}
\end{figure*}

Figure~\ref{tp-scalability-part} shows that, \sysname achieves much higher throughput than the baselines with different number of replicas. However, in open-loop tests with only 10 clients (Figure~\ref{open-loop-tp-scalability-part}), the throughput of \sysname-Non-Proxy distinctly degrades from 187.8K requests/sec to 148.7K requests/sec, as the number of replicas grows. This indicates that the clients become the new bottleneck when submitting at high rates. By contrast, when equipped with proxies, \sysname-Proxy maintains a high throughput regardless of the number of replicas. 



\subsection{Proxy Evaluation}
\label{sec:proxy}


Figure~\ref{proxy-latency-tp} and Figure~\ref{proxy-cpu-tp} compare the two versions of \sysname with 10 open-loop clients and 9 replicas. \sysname-Proxy employs 5 proxies. As clients increase their submission rates, we measure the latency and the average CPU utilization per client. Compared with \sysname-Non-Proxy, which sends 9 messages and receives 17 messages (i.e., 9 \emph{fast-replies} and 8 \emph{slow-replies}~\footnote{Each of the 9 replicas send a \emph{fast-reply} to the client; each of the 8 followers also send a \emph{slow-reply} to the client.}) for each request, \sysname-Proxy incurs 2 extra message delays, but reduces significant CPU usage at the client side. It even achieves lower latency as the throughput grows, because \sysname-Non-Proxy makes the clients CPU-intensive.

Figure~\ref{proxy-individual-client} compares the maximum throughput achieved by one client with/without proxies. Given the same CPU resource,~\footnote{Every client uses one thread for request submission and another for reply handling.} the throughput of the client without proxies declines distinctly as the number of replicas increases. Such bottlenecks can also occur in the other protocols with similar offloading design (e.g., Speculative Paxos, NOPaxos, Domino, CURP). By contrast, when equipped with proxies, the client retains a high throughput regardless of the number of replicas.

\subsection{Comparison in WAN}
\label{sec:wan-comp}

\begin{figure}[!htbp]
    \centering
    \includegraphics[width=9cm]{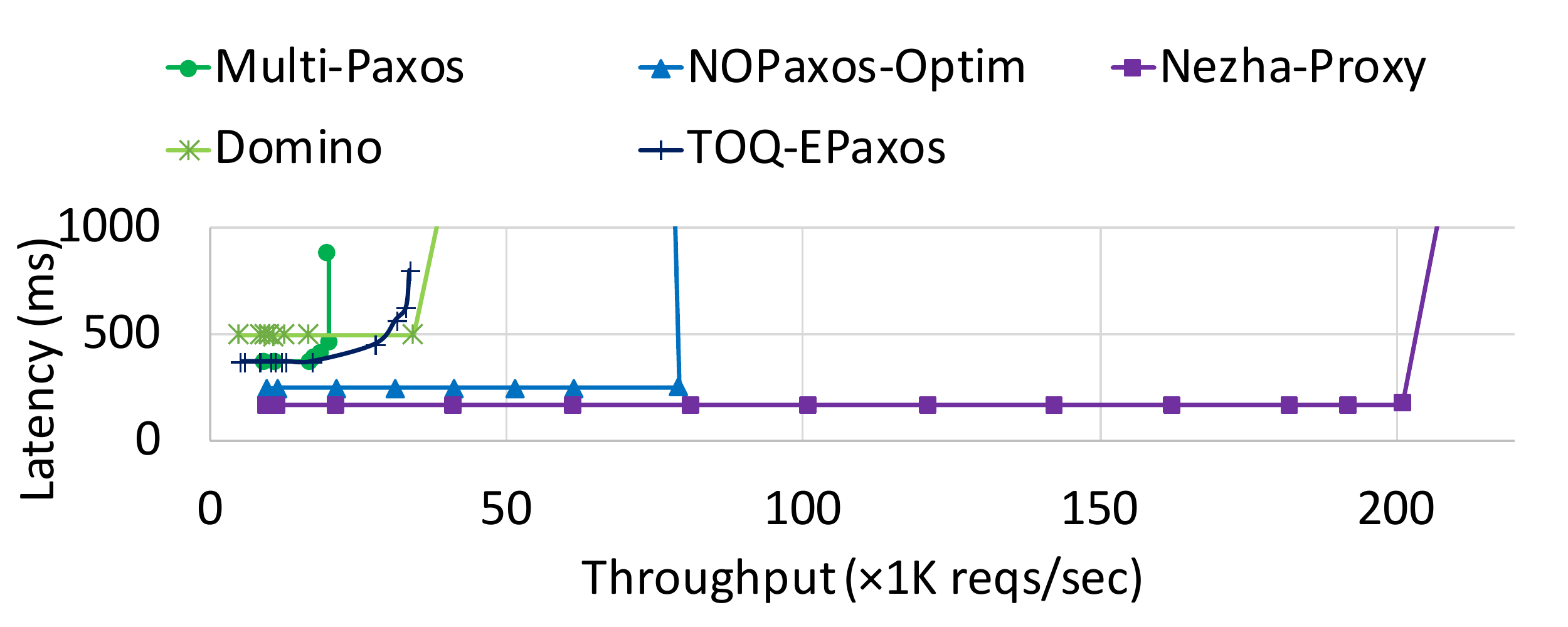}
    \caption{Latency vs. throughput in WAN}
    \label{fig:wan-comp}
\end{figure}

We continue to compare \sysname with the baselines in a wide-area network (WAN). We run the open-loop tests across 5 zones: the 3 replicas are located in \texttt{europe-north1-a}, \texttt{asia-northeast1-a} and \texttt{southamerica-east1-a}, respectively; the 10 open-loop clients are divided into two groups, and distributed in \texttt{us-east1-b} and \texttt{us-west1-a}; correspondingly, the 2 proxies are also distributed in \texttt{us-east1-b} and \texttt{us-west1-a} to serve the clients in their zones.

Compared to the LAN evaluation (Figure~\ref{fig-latency-tp}), \sysname outperforms  all the baselines even more in WAN. As shown in Figure~\ref{fig:wan-comp}, NOPaxos-Optim is the best among the four baselines, but \sysname still outperforms NOPaxos-Optim by 1.51$\times$ in latency and 2.55$\times$ in throughput.
For TOQ-EPaxos, TOQ provides little help to reduce the latency for EPaxos when clients and replicas are located in different zones, because TOQ only leverages clock synchronization to reduce conflicts among replicas, and its fast path still costs 2 WAN RTTs when replicas and clients are separated. By contrast, since \sysname's proxies are stateless and generally deployable, they can be deployed in the same zone as clients, making client-proxy latency as LAN message delay. Therefore, \sysname can achieve 1 WAN RTT in the fast path. We include more discussion in Appendix~\ref{comp-epaxos}.

\subsection{Failure Recovery}

\label{eval-failure-recovery}


\begin{figure*}[!t]
    \begin{minipage}{0.23\linewidth}
        \includegraphics[width=\linewidth]{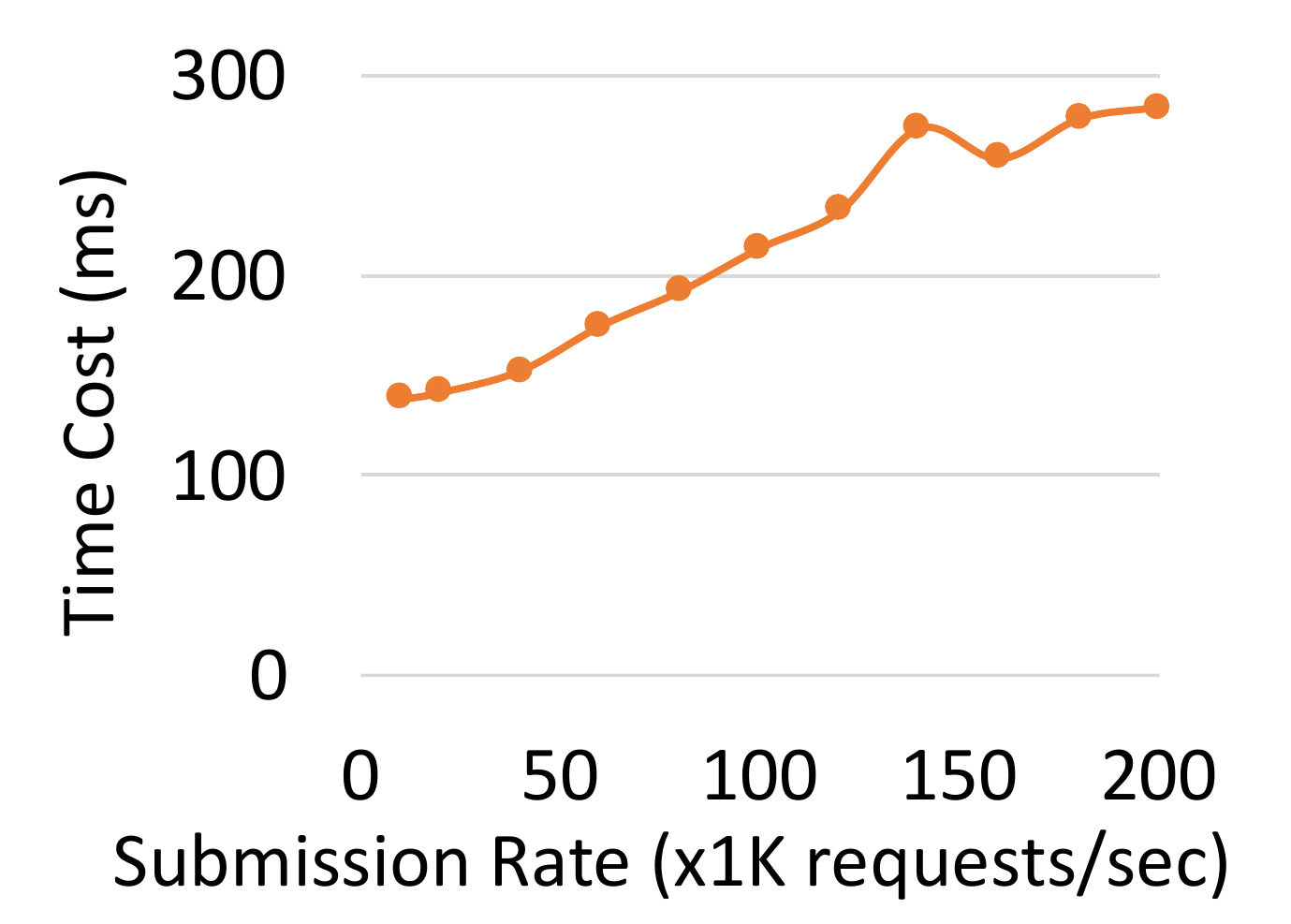}
        \caption{Time cost of view change}
        \label{fig-vc-cost-1}
    \end{minipage}
    \begin{minipage}{0.75\linewidth}
    \centering
    \subcaptionbox{20K requests/sec}
      {
      \includegraphics[width=0.35\linewidth]{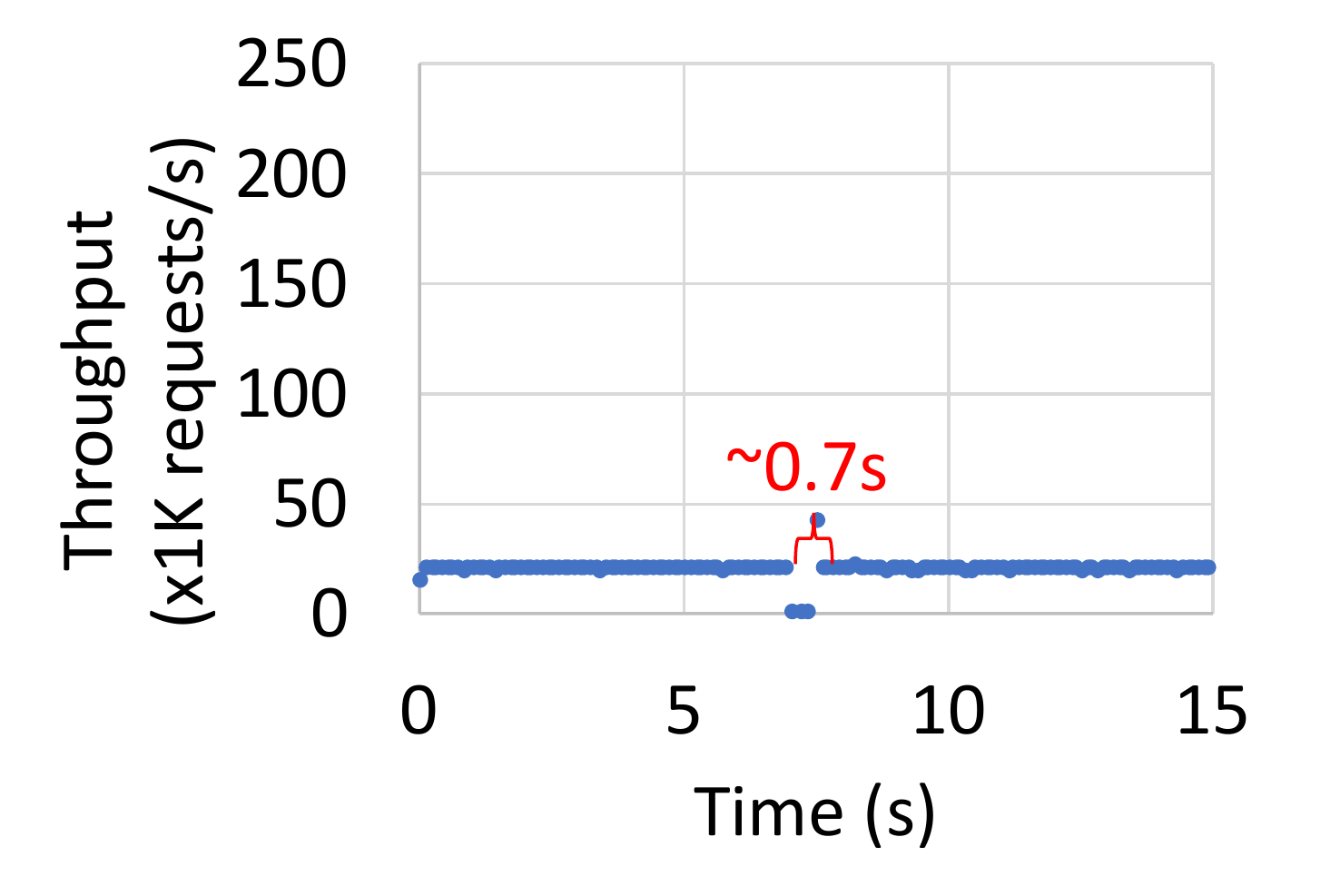}
    \label{recovery-20K}
      }
    \subcaptionbox{100K requests/sec}
      {\includegraphics[width=0.28\linewidth]{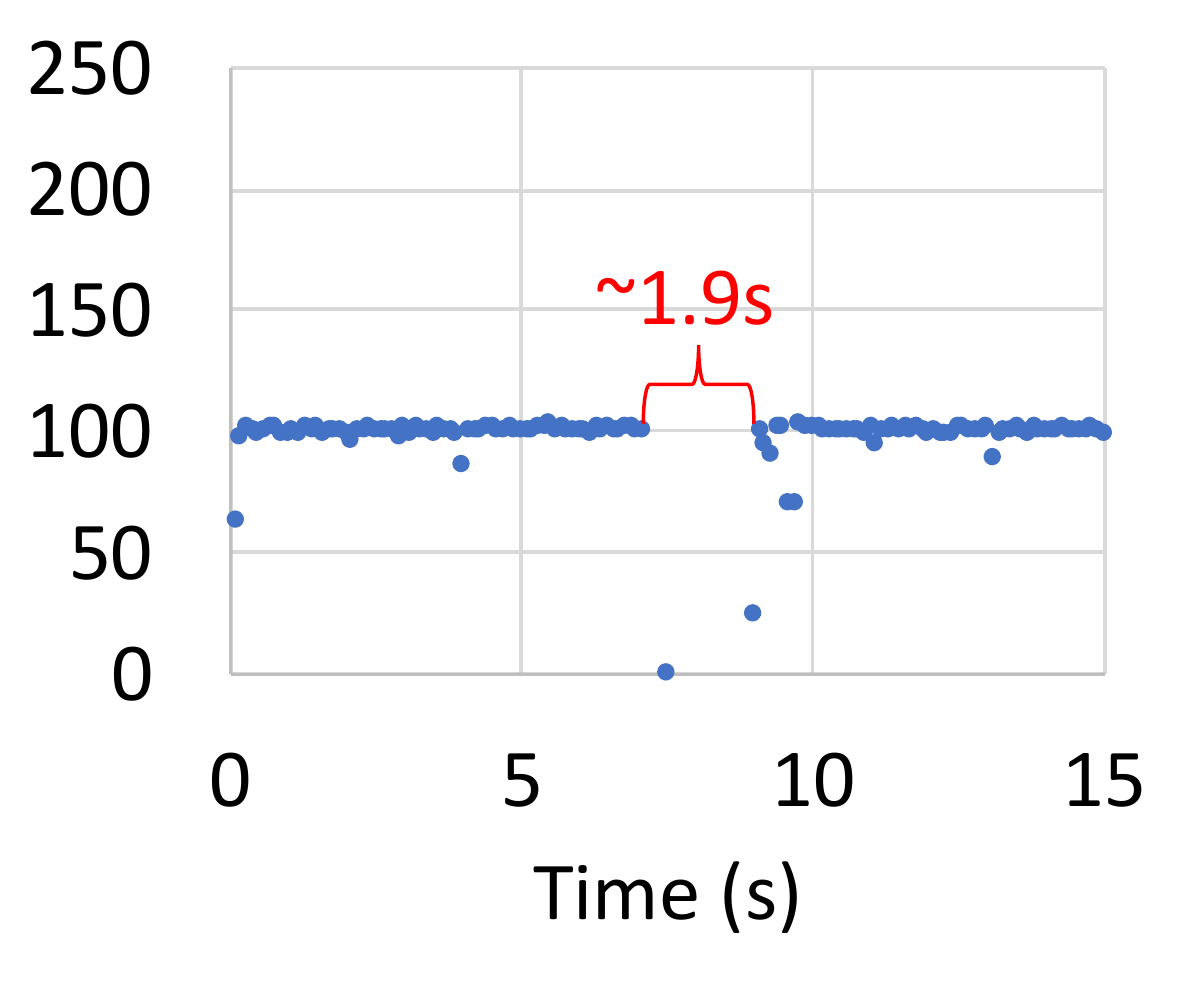}
        \label{recovery-100K}
     }
    \subcaptionbox{200K requests/sec}
      {\includegraphics[width=0.28\linewidth]{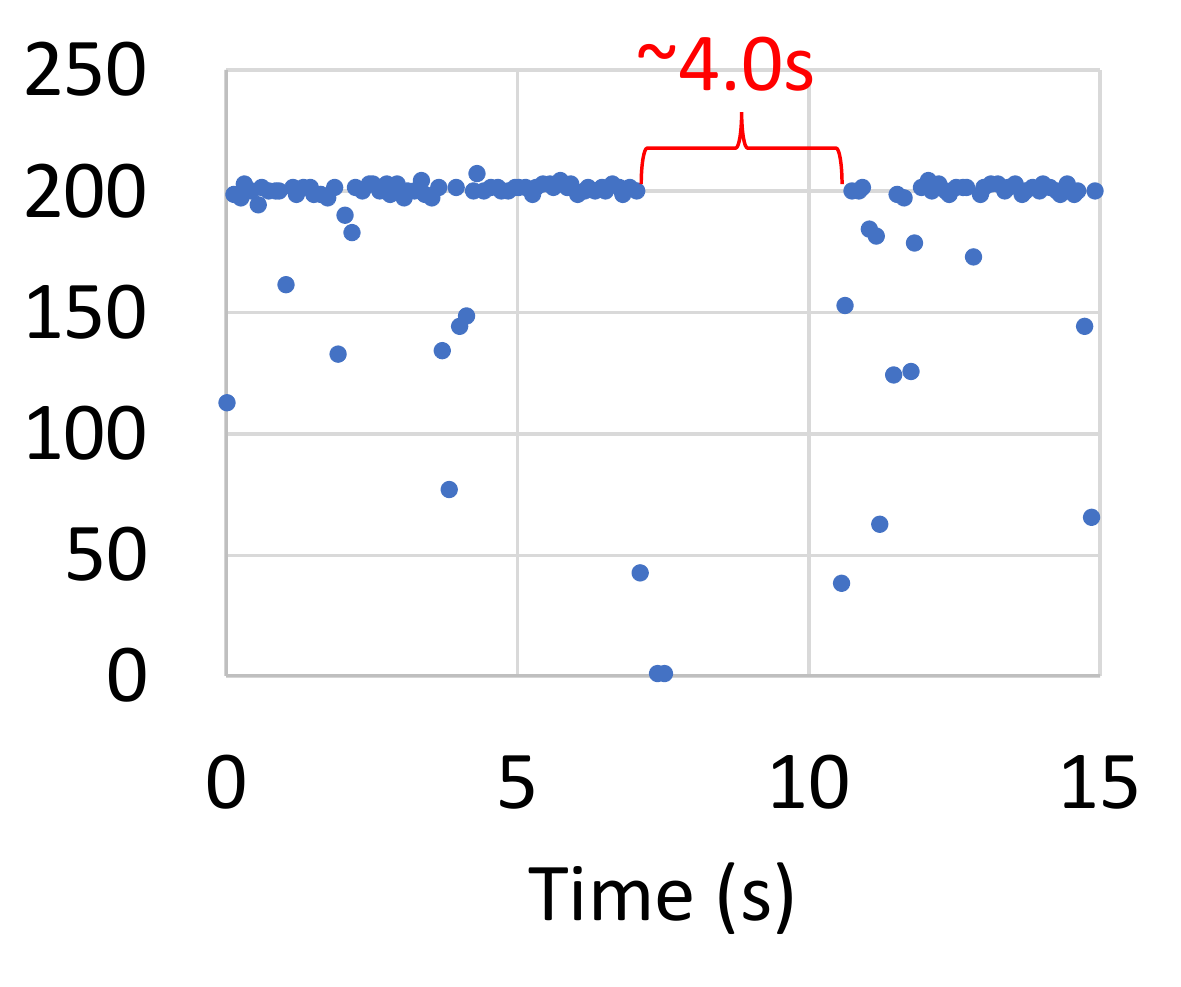}
        \label{recovery-200K}
     }
   \caption{Time cost to recover to the same throughput level}
   \label{fig-recovery-time-1}
  \end{minipage}
  \vspace{-0.4cm}
\end{figure*}

We evaluate failure recovery as shown in Figure~\ref{fig-vc-cost-1} and Figure~\ref{fig-recovery-time-1}. Since follower's crash and recovery do not affect the availability of \sysname (until more than $f$ failures), we mainly focus on the evaluation of the leader's crash and recovery. We study two aspects: (1) How long does it take for the remaining replicas to complete a view change with the new leader elected? (2) How long does it take to recover the throughput to the same level as before crash?

We use 3 replicas and 10 open-loop clients, and vary per-client submission rate from 1K requests/second to 20K requests/second, so the total submission rate varies from 10K requests/second to 200K requests/second. We are using a \emph{null} application during our measurement. We leave for future work the evaluation of \sysname's failure recovery performance for more complex replicated applications.

Under different submission rates, we kill the leader and measure the time taken for a view change. As shown in Figure~\ref{fig-vc-cost-1}, the time taken for a view change grows as the submission rate increases, because there is an increasing amount of state (log) transfer to complete the view change. But the time cost of view change is generally low ( \SI{150}{\milli\second}-\SI{300}{\milli\second}) because of the acceleration idea (\S\ref{sec-checkpoint}) integrated in \sysname.

The time cost to recover the same throughout level (Figure~\ref{fig-recovery-time-1}) is larger than the time cost of view change, because there are other tasks to complete after the replicas enter the new view. For example, replicas need to relaunch the working threads and reinitialize the contexts; replicas need to handle clients' retried requests, which were not responded before crash; followers may need additional state transfer due to lagging too far behind, etc. 

Based on the measured trace, we calculate the throughput every \SI{10}{\milli\second}, and plot the data points in Figure~\ref{fig-recovery-time-1}. Figure~\ref{fig-recovery-time-1} shows that \sysname takes longer to recover to a higher throughput. It takes approximately \SI{0.7}{\second}, \SI{1.9}{\second}, \SI{4.0}{\second}, to recover to the same throughput level under the submission rate of 20K requests/sec, 100K requests/sec, 200K requests/sec, respectively. As a reference to compare, Figure 3.20 in \cite{feiran-craft} evaluates the recovery time for an industrial Raft implementation~\cite{hashicorp-raft}, which takes about 6 seconds to recover to 18K requests/sec.

\subsection{Disk-based Comparison: \sysname vs. Raft}
\label{sec-nezha-vs-raft}

\begin{figure*}[t]
    \begin{minipage}{0.48\linewidth}
    \centering
      \includegraphics[width=1\linewidth]{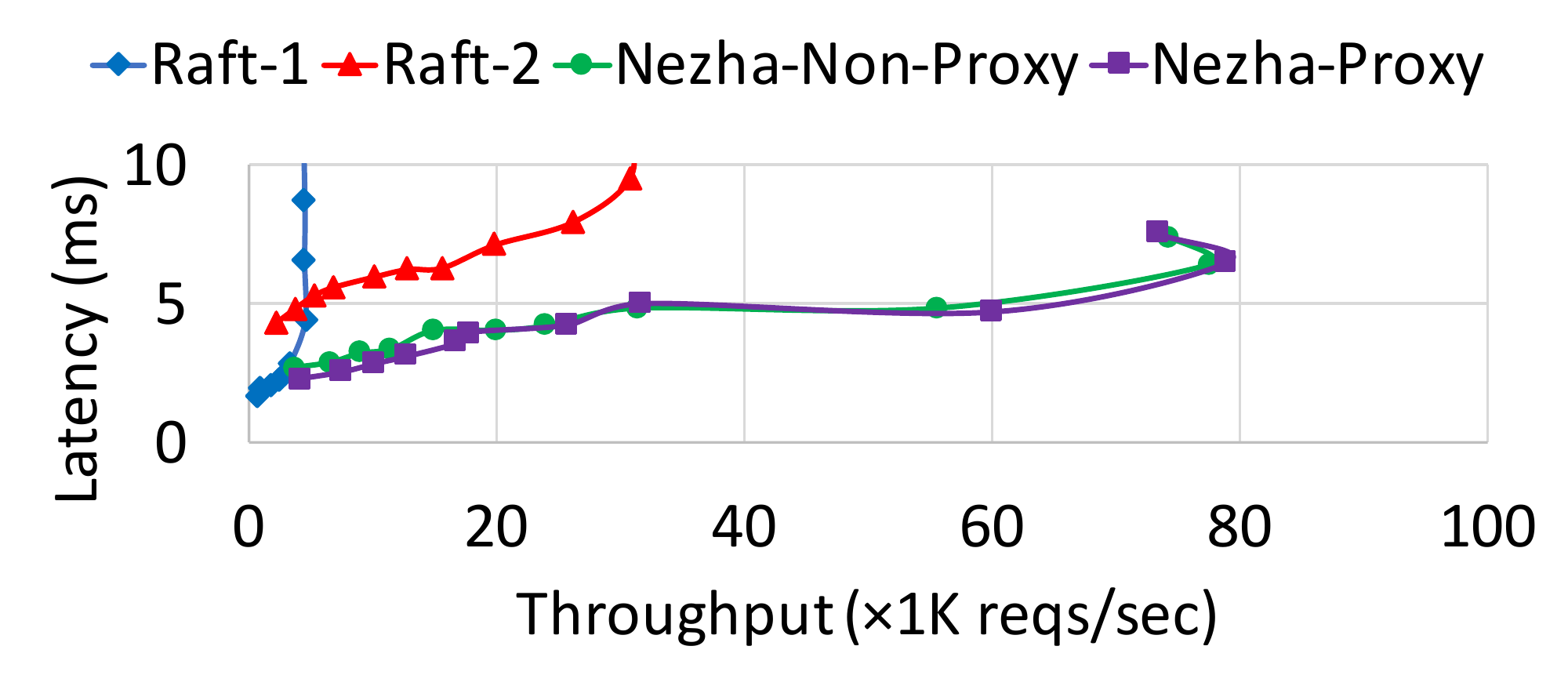}
     \caption{\sysname vs. Raft (closed-loop)}
     \label{figure-nezha-raft-closed}
    \end{minipage}
    \begin{minipage}{0.48\linewidth}
    \centering
      \includegraphics[width=1\linewidth]{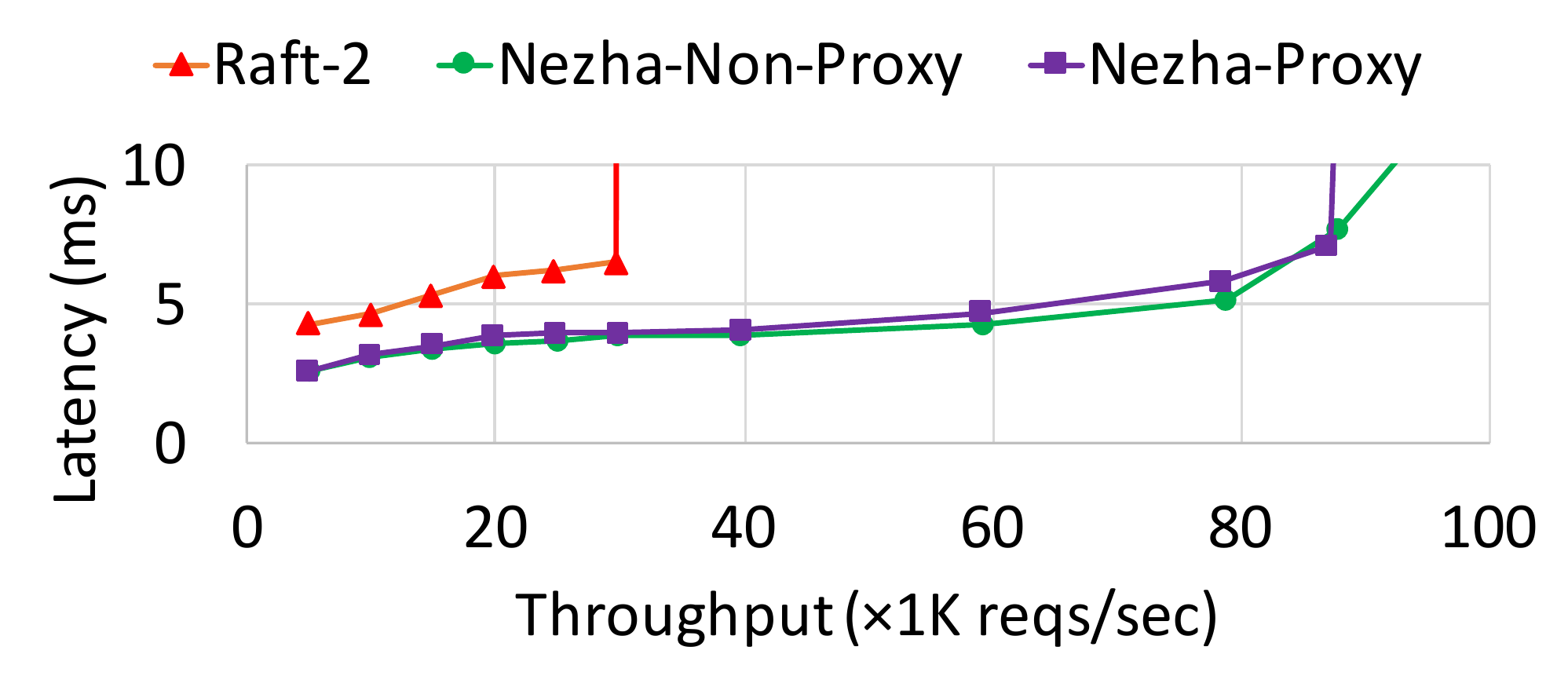}
     \caption{\sysname vs. Raft (open-loop)}
     \label{figure-nezha-raft-open}
    \end{minipage}
\end{figure*}

Raft establishes its correctness on log persistence and relies on stable storage for stronger fault tolerance (e.g. power failure). For a fair comparison to Raft, we convert \sysname from its diskless operation to a disk-based version, making it achieve the same targets as Raft. Before \sysname replicas send replies, they first persist the corresponding log entry (including \emph{view-id} and \emph{crash-vector}) to stable storage. Then, if a replica is relaunched, it can recover its state and replay the \emph{fast-replies}/\emph{slow-replies}. We want to study whether \sysname is fundamentally more I/O intensive than Raft.  

We initially use the original Raft implementation~\cite{atc14-raft} (Raft-1 in Figure~\ref{figure-nezha-raft-closed}), which is written in C++, but uses a slower communication library based on TCP, and involves additional mechanisms (e.g. snapshotting). Raft-1 can only work in closed-loop tests because of its blocking API.  For Raft-1, we use its default batching and pipeline mechanism, and noticed that Raft-1 achieves very low throughput of 4.5K requests/sec on Google Cloud VMs equipped with zonal standard persistent disk~\cite{gcp-disk}. Hence, we implement and optimize Raft (Raft-2), by using the Multi-Paxos code from~\cite{nopaxos-code} as a starting point. For both Raft-2 and \sysname, we tune their batch sizes to reach the best throughput. Our evaluation shows that \sysname outperforms Raft in both closed-loop test (Figure~\ref{figure-nezha-raft-closed}) and open-loop test (Figure~\ref{figure-nezha-raft-open}). We also see that there is little difference in latency with or without a proxy in \sysname because latencies are now dominated by disk writes, not message delays.

\section{Application Performance}
\label{sec:app}

In order to measure \sysname in the context of a replicated application, we port two applications to \sysname and the baseline protocols. Each replicated application uses 3 replicas.

\begin{figure*}[!t]
    \centering
    \begin{minipage}{0.32\textwidth}
        \includegraphics[width=1\linewidth]{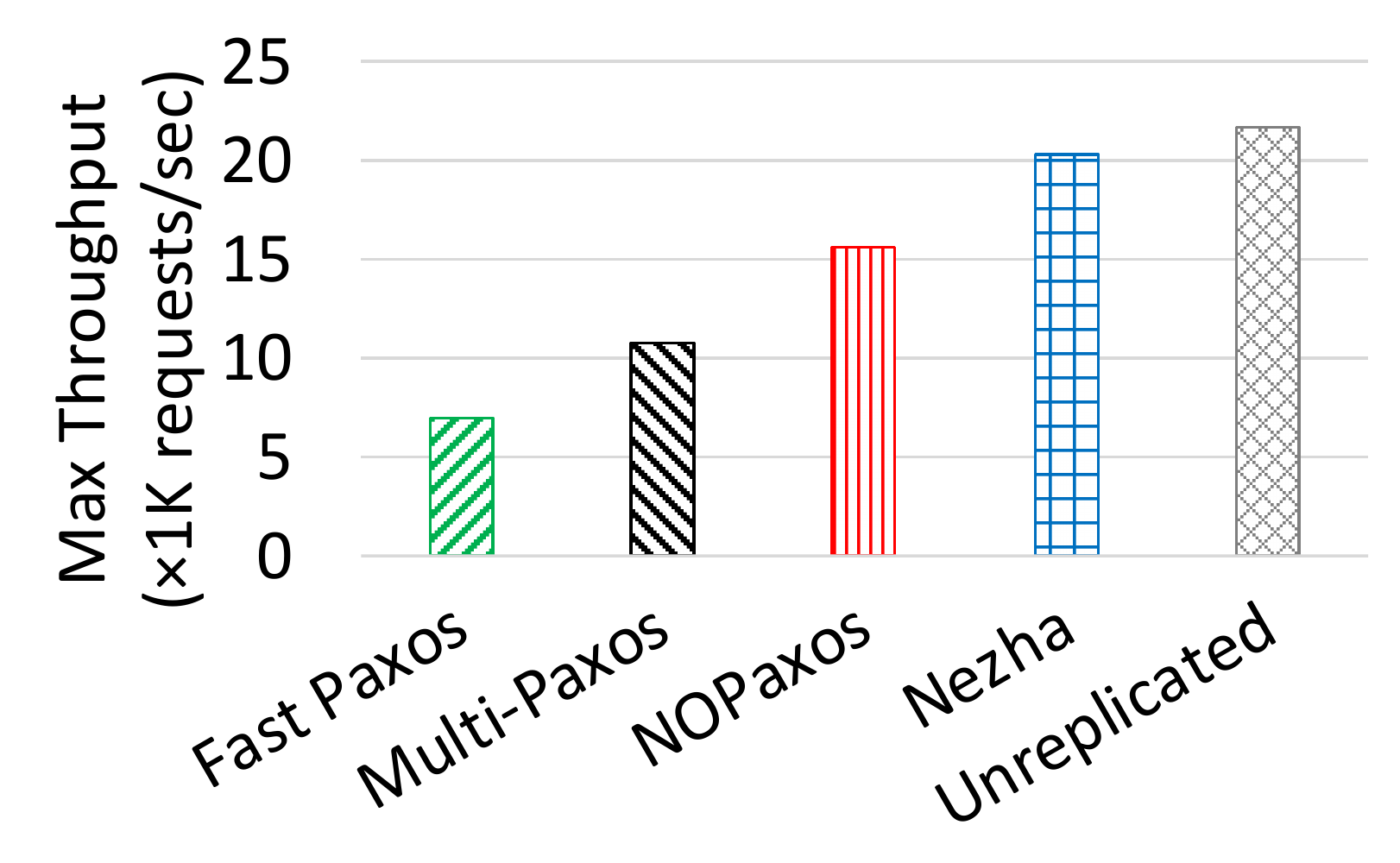}
        \caption{Redis throughput with a 10 ms latency SLO}
        \label{redis-perf}
    \end{minipage}
    \begin{minipage}{0.32\textwidth}
        \includegraphics[width=1\linewidth]{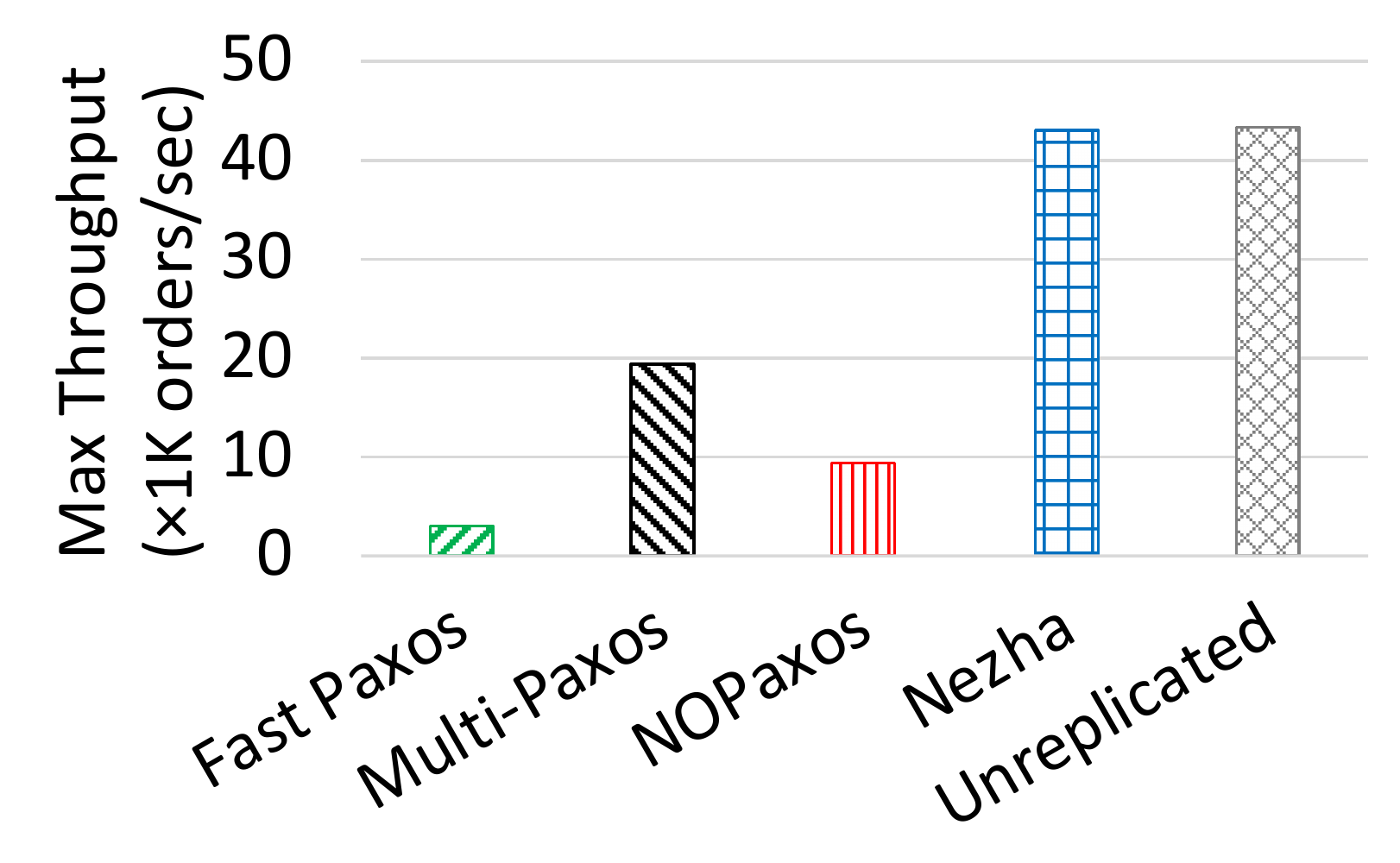}
        \caption{CloudEx throughput}
        \label{cloudex-throughput}
    \end{minipage}
    \begin{minipage}{0.32\textwidth}
        \includegraphics[width=\textwidth]{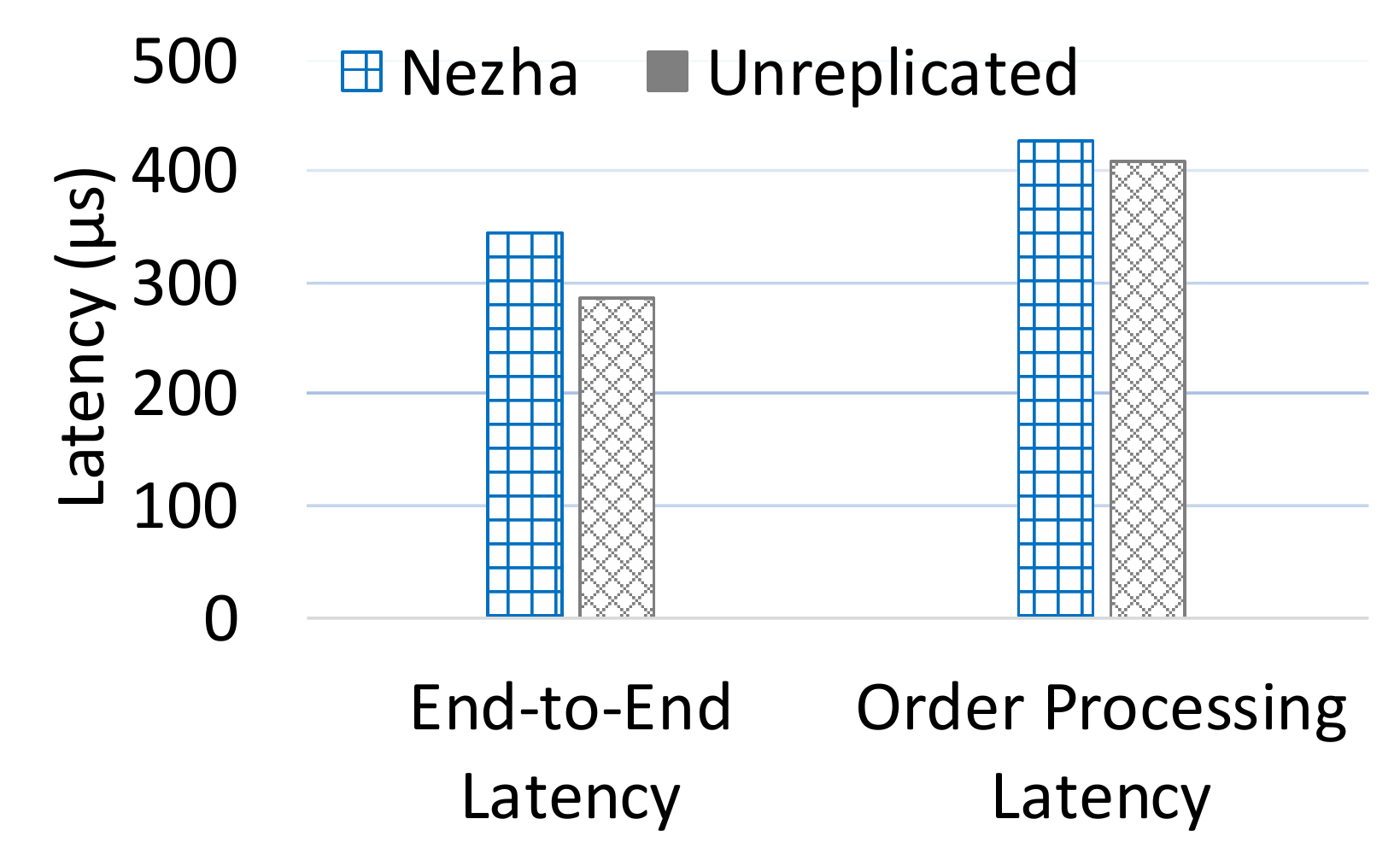}
        \caption{CloudEx latency}
        \label{cloudex-latency}
    \end{minipage}
\end{figure*}

\textbf{Redis}. Redis~\cite{redis} is a typical in-memory key-value store. We choose YCSB-A~\cite{ycsb-workload} as the workload, which operates on 1000 keys with \texttt{HMSET} and \texttt{HGETALL}. We use 20 closed-loop clients to submit requests, which can saturate the processing capacity of the unreplicated Redis. Figure~\ref{redis-perf} illustrates the maximum throughput of each protocol under \SI{10}{\milli\second} service level objective (SLO). \sysname outperforms all the baselines on this metric: it outperforms Fast Paxos by 2.9$\times$,  Multi-Paxos by 1.9$\times$, and NOPaxos by 1.3$\times$. Its throughput is within 5.9\% that of the unreplicated system. 


\textbf{CloudEx.} CloudEx~\cite{hotos21-cloudex} is a research fair-access financial exchange system for the public cloud. There are three roles involved in CloudEx: matching engine, gateways and market participants. To provide fault tolerance, we replicate the matching engine and co-locate one gateway with one proxy. Market participants are unmodified. Before porting it to \sysname, we improved the performance of CloudEx, compared with the version in~\cite{hotos21-cloudex}, by multithreading and replacing ZMQ~\cite{zeromq} with raw UDP transmission. We first run the unreplicated CloudEx with its dynamic delay bounds (DDP) strategy disabled~\cite{hotos21-cloudex}. We configure a fixed sequencer delay parameter ($d_s$) of 200$\mu s$. Similar to~\cite{hotos21-cloudex}, we launch a cluster including 48 market participants and 16 gateways, with 3 participants attached to one gateway. The matching engine is configured with 1 shard and 100 symbols. We vary the order submission rate of market participants, and find the matching engine is saturated at 43.10K orders/second, achieving an inbound unfairness ratio of $1.49\%$.

We then run CloudEx atop the four protocols with the same setting. In Figure~\ref{cloudex-throughput}, only \sysname reaches the throughput (42.93K orders/second) to nearly saturate the matching engine, and also yields a close inbound unfairness ratio of $1.97\%$. 
We further compare the end-to-end latency (i.e., from order submission to the order confirmation from the matching engine) and order processing latency (i.e., from order submission to receiving the execution result from the matching engine.) between \sysname and the unreplicated CloudEx. In Figure~\ref{cloudex-latency}, \sysname prolongs the end-to-end latency by \SI{19.7}{\percent} (\SI{344}{\micro\second} vs. \SI{288}{\micro\second}), but achieves very close order processing latency to the unreplicated version (\SI{426}{\micro\second} vs. \SI{407}{\micro\second}).

\section{Related Work}
\label{sec:related-work}
\Para{Consensus protocols.} Classical consensus protocols, e.g., Multi-Paxos, Raft, and Viewstamped Replication make no distinction between a fast and slow path: all client requests incur the same latency. \sysname uses an optimistic approach to improve latency in the common case. Eve~\cite{osdi12-eve} adopts \emph{execute-verify} architecture, which is similar to \sysname's speculative execution design, but Eve requires application-specific rollback when replicas' states diverge, whereas \sysname does not require such rollback mechanism. Mencius~\cite{osdi08-mencius} exploits a multi-leader design to mitigate the single-leader bottleneck in Multi-Paxos. However, it introduces extra coordination cost among multiple leaders and further, the crash of any of the leaders temporarily stops progress. By contrast, \sysname reduces the leader's bottleneck using proxies and followers' crash does not affect progress. EPaxos~\cite{sosp13-epaxos} can achieve optimal WAN latency in the fast path, but when it comes to the LAN scenarios we focus on, it performs worse than Multi-Paxos~\cite{sigmod19-paxi}. CURP~\cite{nsdi19-curp} can complete commutative requests in 1 RTT, but doesn't take advantage of consistent ordering: hence, it costs up to 3 RTTs even if all witnesses process the non-commutative requests in the same order. SPaxos~\cite{srds12-spaxos}, BPaxos~\cite{bpaxos} and Compartmentalized Paxos~\cite{vldb21-compartmentalized-paxos} address the throughput scaling of consensus protocols with modularity, trading more latency for throughput improvement. The proxy design in \sysname is similar to compartmentalization~\cite{vldb21-compartmentalized-paxos}, but \sysname's proxies are stateless. By contrast, \cite{vldb21-compartmentalized-paxos, srds12-spaxos, bpaxos} use stateful proxies, which complicates fault tolerance.


 

\Para{Network primitives to improve consensus.} Recent works consider building network primitives to accelerate consensus protocols. 4 other primitives are closely related to DOM, namely, mostly-ordered multicast (MOM)~\cite{nsdi15-specpaxos}, ordered unreliable multicast (OUM) ~\cite{osdi16-nopaxos}, timestamp-ordered queuing (TOQ)~\cite{nsdi21-epaxos} and sequenced broadcast (SB)~\cite{eurosys22-sb}. From the perspective of deployability, DOM and TOQ are both based on software clock synchronization whereas MOM and OUM rely on highly-engineered network. This gives DOM and TOQ an advantage over MOM/OUM in environments like the cloud. On the other hand, requests output from MOM and TOQ can still result in inconsistent ordering. By contrast, DOM and OUM guarantee consistent ordering of released requests. DOM's guarantees are stronger than MOM because MOM can occasionally reorder requests, but are weaker than OUM because OUM also provides gap detection. We include a  formal comparison in Appendix~\S\ref{appendix-formulation}. SB is a new primitive for Byzantine fault tolerance. It works in an epoch-based manner and achieves high throughput through load balancing. However, its latency can reach seconds.

 \Para{Clock synchronization applied to consensus protocols.} CRaft~\cite{feiran-craft} and CockRoachDB~\cite{sigmod20-cockroachdb} use clock synchronization to improve the throughput of Raft. However, they base their correctness on the assumption of a known worst-case clock error bound, which is not practical for high-accuracy clock synchronization~\cite{podc84-clock-sync,nancy-clock-bound,podc91-clock-sync-remark}. Domino~\cite{conext20-domino} and TOQ~\cite{nsdi21-epaxos} use clock synchronization to accelerate Fast Paxos and EPaxos respectively. We compare them with \sysname in \S\ref{sec-eval-domino-toq}, and include more details in Appendix~\S\ref{error-trace-domino} and \S\ref{comp-epaxos}.

\section{Conclusion and Future Work}
\label{sec:conclusion}
Recent development of accurate software clock synchronization techniques brings us new opportunities to develop novel consensus protocols to achieve high performance in the public cloud. Leveraging this, we present \sysname in the paper, which can be easily deployed in the public cloud, and achieves both higher throughput and lower latency than baselines. 

We are considering two lines of future work. First, we intend to replace the Multi-Paxos/Raft backend used by some industrial systems (e.g., Kubernetes, Apache Pulsar, etc) so as to boost their performance. Second, we believe DOM can also be applied to other domains. We plan to integrate DOM with existing concurrency control algorithms  (e.g., Two-Phase Locking, Optimistic Concurrency Control, etc) to improve their performance, or invent new concurrency control protocols based on DOM.

\section{Acknowledgements} 
We thank Cisco Systems, Facebook (now Meta), Google, Nasdaq and Wells Fargo for sponsoring our research at Stanford’s Platform Lab.  This work was also supported by NSF-2008048. We thank Shiyu Liu, Feiran Wang, Yilong Geng, Deepak Merugu, Dan Ports, Jialin Li, Ellis Michael, Jinyang Li, Aurojit Panda, Seo Jin Park, Zhaoguo
Wang, Ken Birman, Weijia Song and Eugene Wu for their feedback on our work and help with benchmarking.

\bibliographystyle{ACM-Reference-Format}
\bibliography{nezha}

\clearpage
\appendix
\addcontentsline{toc}{section}{Appendices}
\section*{Appendices} 

In this appendix, we include the following:
\begin{itemize}[leftmargin=*]
    \item Explanation of \sysname's recovery (\S\ref{recovery-algo}). 
    
    \item Correctness proof of \sysname (\S\ref{append-correctness}).
    
    
    \item Further evaluation of \sysname under different workloads (\S\ref{sec:extensive-eval}).
    
    \item Detailed discussion and evaluation about the effect of clock variance on \sysname's performance (\S\ref{clock-eval}).

    
    \item Comparison with Derecho~\cite{tocs19-derecho} (\S\ref{sec-derecho-issue}).
    
        
    \item Analysis of the incorrectness of Domino due to clock skews  (\S\ref{error-trace-domino}).
    \item Formal comparison between DOM, MOM, and OUM (\S\ref{appendix-formulation}).
    \item Discussion about deploying \sysname in WAN and comparing \sysname to EPaxos in WAN environments (\S\ref{comp-epaxos}). 
    \item TLA+ specification of \sysname (\S\ref{sec-nezha-tla}).
    \item Illustration of message delays regarding the protocols listed in Table~\ref{tab:cmp} (\S\ref{sec-protocol-message-delay}).
\end{itemize}

\section{Recovery Protocol and Algorithms}
\label{recovery-algo}

We explain how \sysname leverages the diskless crash recovery algorithm~\cite{viewstamp} from Viewstamped Replication in 3 steps. First, we explain how we adopt the recent concept of crash-vectors~\cite{disc17-dcr, disc17-dcr-techreport} to fix the incorrectness in the crash recovery algorithm. Second, we explain how a replica rejoins \sysname following a crash. Third, we describe how the leader election works if the leader crashes.

\subsection{Crash Vector}
\label{sec-crash-vector-stray-message}
Like Viewstamped Replication, Speculative Paxos and NOPaxos, \sysname also adopts diskless recovery to improve performance. However, in contrast to them, \sysname avoids the effect of \emph{stray messages}~\cite{tdsn21-recovery} (i.e., messages that are sent out but not delivered before replica crash, so that the relaunched replicas forget them) using the \emph{crash-vector}~\cite{disc17-dcr, disc17-dcr-techreport}. \emph{crash-vector} is a vector containing $2f+1$ integer counters corresponding to the $2f+1$ replicas. Each replica maintains such a vector, with all counters initialized as 0s. 

\emph{crash-vectors} can be aggregated by taking the max operation element-wise to produce a new \emph{crash-vector}. During the replica rejoin (\S\ref{replica-rejoin}) and leader change(\S\ref{leader-change}) process, replicas send their \emph{crash-vectors} to each other. Receivers can make their \emph{crash-vectors} more up-to-date by aggregating their \emph{crash-vector} with the \emph{crash-vector} from the sender. Meanwhile, by comparing its local \emph{crash-vector} and the sender's \emph{crash-vector}, the receiver can recognize whether or not the sender's message is a potential \emph{stray message} (refer to~\cite{disc17-dcr-techreport} for detailed description of \emph{crash-vector}).

\sysname uses \emph{crash-vector}s to avoid two types of \emph{stray messages}, i.e. (1) the \emph{stray messages} during the view change process and (2) the \emph{stray messages} (\emph{fast-replies}) during quorum check. (1) has been clearly explained in \cite{disc17-dcr-techreport}, so here we only sketch how \emph{crash-vector} works  preserve the protocol correctness during the view change process. (2) has not been disclosed in prior works, so we will explain more details in \S\ref{general-error-pattern} after we complete the explanation of the replica rejoin (\S\ref{replica-rejoin}) and leader change (\S\ref{leader-change}).

\subsubsection{Stray Message during View Change}
There can be \emph{stray messages} during the view change process: replicas mistakenly elect a leader, whose state falls behind the others,  finally causing permanent loss of committed requests. The \emph{crash-vector} prevents the \emph{stray messages} effect  because it enables the replicas to recognize potential \emph{stray messages} by comparing a \emph{crash-vector} received from a replica with the local \emph{crash-vector}. During recovery, the \textsc{recovering} replica first recovers its \emph{crash-vector} by collecting and aggregating the \emph{crash-vectors} from a majority of \textsc{normal} replicas. Then, the replica increments its own counter (i.e. replica $i$ increments the $ith$ counter in the vector) and tags the new \emph{crash-vector} to the messages sent afterwards. Once the update of \emph{crash-vector} is exposed to the other replicas, they can recognize the \emph{stray messages} sent by the replica before crash (i.e., those messages have a smaller value at the $ith$ counter), and avoid processing those messages. Thus, the recovery will not be affected by \emph{stray messages}.

\subsubsection{Stray Message during Quorum Check}    
Stray messages can also occur during the quorum check in the fast path: some replicas send \emph{fast-replies} and crash after that. 
The reply messages in the fast path (i.e., \emph{fast-reply}) may become \emph{stray messages} and participate into the quorum check, which makes the proxies/clients prematurely believe the request has been persisted to a super-majority of replicas, but actually not yet (i.e. the recovered replicas may not hold the requests after their recovery).  

In brief, the \emph{crash-vector} prevents the effect of such \emph{stray fast-replies}, because we include the information of \emph{crash-vector}s in the \emph{fast-replies} (\S\ref{sec-fast-quorum-check}). When a failed replica rejoins the system (Algorithm~\ref{algo-replica-rejoin}), it leads to the update of \emph{crash-vector}s for the leader and other remaining followers, so these replicas will send \emph{fast-replies} with different \emph{hashes} from the \emph{stray fast-replies} sent by the rejoined replica. Therefore, the stray \emph{fast-replies} from the rejoined replica and the normal \emph{fast-replies} from the other replicas cannot form the super quorum together. After we describe the replica rejoin (\S\ref{replica-rejoin}) and leader change (\S\ref{leader-change}), we come back to explain the details in \S\ref{general-error-pattern}.

\algdef{SE}[EVENT]{Event}{EndEvent}[1]{\textbf{upon}\ \textsc{\small #1}\ \algorithmicdo}{\algorithmicend\ \textbf{event}}%
\algtext*{EndEvent}
\algdef{SE}[DOWHILE]{Do}{doWhile}{\algorithmicdo}[1]{\algorithmicwhile\ #1}%
\makeatletter
\makeatother

\begin{algorithm*}[!htbp]

\small 

\begin{multicols}{2}
\caption{Replica rejoin}
\label{algo-replica-rejoin}
\hspace*{\algorithmicindent} \textbf{Local State}: \Comment{} \\
\hspace*{\algorithmicindent} \emph{nonce},\Comment{A locally unique string on this replica} \\
\hspace*{\algorithmicindent} \emph{C},\Comment{Reply set of \textsc{crash-vector-rep}} \\
\hspace*{\algorithmicindent} \emph{R},\Comment{Reply set of \textsc{recovery-rep}} \\
\hspace*{\algorithmicindent} \emph{r}, \Comment{short for \emph{replica-id} (the message sender)} \\ 
\hspace*{\algorithmicindent} \emph{cv}, \Comment{short for \emph{crash-vector}} \\
\hspace*{\algorithmicindent} \emph{status}, \emph{view-id}, \emph{last-normal-view}, \emph{log}, \emph{sync-point} \Comment{Other attributes} \\

\begin{algorithmic}[1]
  \Event{recover}
    \State $status=$\textsc{recovering} \label{line-reset-status}
    \State $C=\emptyset$
    \State $nonce=\textsc{generate-uuid}$  \label{line-uuid}
    \State \textsc{read-crash-vector}
    \State \emph{cv-set}$=\{ m.cv | m \in C \}$
    \State $cv=$\textsc{aggregate}(\emph{cv-set}$\ \cup\ \{cv\}$)
    \State $cv[r]=cv[r]+1$ \Comment{Increment its own counter}
    \Do
        \State $R=\emptyset$
        \State \textsc{read-recovery-info}
        \State $\emph{highest-view}=\max \{m.v | m \in R \}$
        \State $\emph{leader-id}=\emph{highest-view}\ \%\  (2f+1)$
    \doWhile{($\emph{leader-id}=r$)} 
    \State Pick $m \in R$: $m.v=\emph{highest-view}$
    \State \textsc{state-transfer}(\emph{leader-id})
  \EndEvent

    \Function{read-crash-vector}{}
        \State $m.type$ = \textsc{crash-vector-req}
        \State $m.r=r$
        \State $m.nonce=nonce$
        \LeftComment{Broadcast \textsc{crash-vector-req} to all replicas}
        \For{$i$ $\leftarrow$0 to $2f$}
            \State \textsc{send-message}($m$, $i$) \Comment{Send message $m$ to the replica $i$}
        \EndFor
        
        \State Wait until $|C|\geq f+1$  \Comment{$C$ is initialized as $\emptyset$ by the caller}
        \State \Return
    \EndFunction

    \Function{read-recovery-info}{}
        \State $m.type=$\textsc{recovery-req}
        \State $m.r=r$
        \State $m.cv=cv$
        \LeftComment{Broadcast \textsc{recovery-req} to all replicas}
        \For{$i$ $\leftarrow$0 to $2f$}
            \State \textsc{send-message}($m$, $i$)
        \EndFor
        
        \State Wait until $|R|\geq f+1$
        \Comment{$R$ is initialized as $\emptyset$ by the caller}
        \State \Return
    \EndFunction

    \Function{state-transfer}{$i$}
        \State $m.type=$\textsc{state-transfer-req}
        \State $m.r=r$
        \State $m.cv=cv$
        \State \textsc{send-message}($m$, $i$)
        \State Wait until $status=\textsc{normal}$
        \State \Return
    \EndFunction
    
  \Event{\emph{receiving} crash-vector-req, $m$}
    \If{$status\neq$ \textsc{normal}}
    \State \Return
    \EndIf
    \State $m'.type=$\textsc{crash-vector-rep}
    \State $m'.r=r$
    \State $m'.nonce=m.nonce$
    \State $m'.cv=cv$
    \State \textsc{send-message}($m'$, $m.r$)
  \EndEvent
  
  \Event{\emph{receiving} crash-vector-rep, $m$}
    \If{$status\neq\textsc{recovering}$}
    \State \Return
    \EndIf
    \If{$nonce\neq m.nonce$}
    \State \Return 
    \EndIf
    \State $C=C\ \cup\ \{m\}$
  \EndEvent
  
   \Event{\emph{receiving} recovery-req, $m$}
    \If{$status\neq\textsc{normal}$}
    \State \Return
    \EndIf
    \State $cv=$\textsc{aggregate}($cv$, $m.cv$)
    \State $m'.type=\textsc{recovery-rep}$
    \State $m'.r=r$
    \State $m'.v=\emph{view-id}$
    \State $m'.cv=cv$
    \State \textsc{send-message}($m',m.r$)
  \EndEvent 
     
   \Event{\emph{receiving} recovery-rep, $m$}
    \If{$status\neq\textsc{recovering}$}
    \State \Return
    \EndIf
    \If{\textsc{check-crash-vector}($m$, $cv$)=\textbf{false}}
    \State Resend \textsc{recovery-req} to $m.r$
    \Else \Comment{Remove stray messages and add the fresh one}
    \State $R'=\{m'\in R\ |\  m'.cv[m'.r]<cv[m'.r]\}$
    \State $R=R\cup \{m\}-R'$
    \State $\forall m' \in R'$, resend \textsc{recovery-req} to $m'.r$
    
    \EndIf
  \EndEvent

   \Event{\emph{receiving} state-transfer-req, $m$}
    \If{$status\neq\textsc{normal}$}
    \State \Return
    \EndIf
    \If{\textsc{check-crash-vector}($m$, $cv$)=\textbf{false}}
    \State \Return
    \EndIf
    \State $m'.type=\textsc{state-transfer-rep}$
    \State $m'.log=log$
    \State $m'.v = \emph{view-id}$
    \State $m'.sp= \emph{sync-point}$
    \State $m'.cv= \emph{cv}$
    \State \textsc{send-message}($m',m.r$)
  \EndEvent
 
    \Event{\emph{receiving} state-transfer-rep, $m$}
    \If{$status\neq\textsc{recovering}$}
    \State \Return
    \EndIf
    \If{\textsc{check-crash-vector}($m$, $cv$)=\textbf{false}}
    \State \Return
    \EndIf
    \State $log=m.log$
    \State $\emph{last-normal-view}=\emph{view-id}=m.v$
    \State $log=m.log$
    \State $\emph{sync-point}=m.sp$
    \State $status=\textsc{normal}$
    \Comment{Rejoin as a \textsc{normal} follower}
  \EndEvent
  
  \Function{check-crash-vector}{$m$, $cv$}
        \If{$m.cv[m.r]<cv[m.r]$}
            \Comment{A potential \emph{stray message}}
            \State \Return \textbf{false}
        \Else 
            
            \State $cv=\textsc{aggregate}(\{cv, m.cv\})$ \Comment{Update local $cv$}
            \State \Return \textbf{true}
        \EndIf
        
    \EndFunction

    \Function{aggregate}{\emph{cv-set}}
        \State $ret= [{\underbrace{0 \dots 0}_{2f+1}}]$ 
        \For{$c \in \emph{cv-set}$}
            \For{$i \gets 0$ to $2f$}
            \State $ret[i]=\max(ret[i], c[i])$
            \EndFor
        \EndFor
        \State \Return $ret$ 
    \EndFunction
  
    
\end{algorithmic}
\end{multicols}
\end{algorithm*}

\subsection{Replica Rejoin}
\label{replica-rejoin}
Crashed replicas can rejoin the system as followers. After the replica crashes and is relaunched, it sets its \emph{status} as \textsc{recovering}. Before it can resume request processing, the replica needs to recover its replica state, including \emph{crash-vector}, \emph{view-id}, \emph{log} and \emph{sync-point}. With reference to Algorithm~\ref{algo-replica-rejoin}, we explain how the replica rejoin process works.

\begin{enumerate}[wide, labelwidth=!,nosep,label=\textbf{Step \arabic*:}]
\item The replica sets its \emph{status} as \textsc{recovering} (line 2), and broadcasts the same \textsc{crash-vector-req} to all replicas to request their \emph{crash-vectors}. A \emph{nonce} (line 4) is included in the message, which is a random string locally unique on this replica, i.e., this replica has never used this \emph{nonce}~\footnote{There are many options available to generate the locally unique \emph{nonce} string~\cite{viewstamp,disc17-dcr-techreport}. \sysname uses the universally unique identifier (UUID) (\textsc{generate-uuid} in line 4), which have been widely supported by modern software systems.}.

\item After receiving the \textsc{crash-vector-req}, replicas with \textsc{normal} status reply to the recovering replica with  <\textsc{crash-vector-rep}, \emph{nonce}, \emph{crash-vector}> (line 40-47).

\item The recovering replica waits until it receives the corresponding replies (containing the same \emph{nonce}) from a majority ($f+1$) of replicas (line 23). Then it aggregates the $f+1$ \emph{crash-vectors} by taking the maximum in each dimension (line 7, line 99-104).
After obtaining the aggregated crash vector $cv$, the replica increment its own dimension, i.e. $cv[\emph{replica-id}] = cv[\emph{replica-id}]+1$ (line 8).
\item The recovering replica broadcasts a recovery request to all replicas, which includes its \emph{crash-vector}, i.e. <\textsc{recovery-req}, \emph{cv}> (line 11, line 26-30).
\item After receiving the \textsc{recovery-req}, replicas with \textsc{normal} status update their own \emph{crash-vectors} by aggregating with $cv$, obtained from the request in step 4. Then, these replicas send back a reply including their own \emph{view-id} and \emph{crash-vector}, i.e.  <\textsc{recovery-rep}, \emph{view-id}, \emph{crash-vector}> (line 54-63).
\item The recovering replica waits until it receives the recovery replies from $f+1$ replicas (line 31). If the \textsc{recovery-rep} is not a \emph{stray message}, it updates its own \emph{crash-vector} by aggregating it with the \emph{crash-vectors} included in these replies (line 66); otherwise, it resends \textsc{recovery-req} to that replica, asking for a fresh message (line 67). Because the \emph{crash-vectors} may have been updated (line 66), those \textsc{recovery-rep} which have been received can also become \emph{stray messages} because their \emph{crash-vectors} are no longer fresh enough. Therefore, we also remove them ($R'$ in line 69) from the reply set $R$ (line 70), and resend requests to the related replicas for fresher replies (line 71).

\item The \textsc{recovering} replica picks the highest \emph{view-id} among the $f+1$ replies (line 12). From the highest \emph{view-id}, it knows the corresponding leader of this view (line 13). If the \textsc{recovering} replica happens to be the leader of this view, it keeps broadcasting the recovery request (line 9-14), until the majority elects a new leader among themselves. Otherwise, the \textsc{recovering} replica fetches the \emph{log}, \emph{sync-point}, \emph{view-id} from the leader via a state transfer (line 16, line 33-39). After that, the replica set its \emph{status} to \textsc{normal} and can continue to process the incoming requests. 

\end{enumerate}

Specially, the \textsc{recovering} replica(s) do not participate in the view change process(\S\ref{leader-change}). When the majority of replicas are conducting a view change (possibly due to leader failure), the \textsc{recovering} replica(s) just wait until the majority completes the view change and elects the new leader.

\label{leader-change}
\begin{algorithm*}
\small
\begin{multicols}{2}
\caption{Leader change}
\label{algo-leader-change}
\hspace*{\algorithmicindent} \textbf{Local State}: \Comment{} \\
\hspace*{\algorithmicindent} \emph{V},\Comment{Reply set of \textsc{view-change}} \\
\hspace*{\algorithmicindent} \emph{r}, \Comment{short for \emph{replica-id} (the message sender)} \\ 
\hspace*{\algorithmicindent} \emph{cv}, \Comment{short for \emph{crash-vector}} \\
\hspace*{\algorithmicindent} \emph{last-normal-view}, \Comment{The most recent view in which }\\ \hspace*{\algorithmicindent} \Comment{the replica's \emph{status} is \textsc{normal} } \\
\hspace*{\algorithmicindent} \emph{status}, \emph{view-id}, \emph{log}, \emph{sync-point} \Comment{Other attributes}

\begin{algorithmic}[1]
    \Event{\textsc{suspect leader failure}}
    \Do
        \State \textsc{initiate-view-change}($\emph{view-id}+1$)
    \doWhile{($\emph{status}\neq\textsc{normal}$)} 
    
    \EndEvent

    \Function{initiate-view-change}{$v$}
        \State $status=\textsc{viewchange}$
        \State $\emph{view-id}=v$
        \State $V=\emptyset$
        \LeftComment{Broadcast \textsc{view-change-req} to all replicas}
        \For{$i$ $\leftarrow$0 to $2f$}
            \State \textsc{send-view-change-req}($i$)
        \EndFor
        \LeftComment{Send \textsc{view-change} to the new leader}
        \State \textsc{send-view-change}($v \% (2f+1)$)
        \State Wait until $status=\textsc{normal}$ or \textsc{Timeout}
        \State \Return
    \EndFunction

    \Function{send-view-change-req}{$i$}
        \State $m.type=\textsc{view-change-req}$
        \State $m.v=\emph{view-id}$
        \State $m.cv=cv$
        \State \textsc{send-message}($m$, $i$)
        \State \Return
    \EndFunction

    \Function{send-view-change}{$i$}
        \State $m.type=\textsc{view-change}$
        \State $m.v=\emph{view-id}$
        \State $m.cv=cv$
        \State $m.log=log$
        \State $m.sp=\emph{sync-point}$
        \State $m.lnv=\emph{last-normal-view}$
        \State \textsc{send-message}($m$, $i$)
        \State \Return
    \EndFunction

   \Event{\emph{receiving} view-change-req, $m$}
    \If{$status=\textsc{recovering}$}
    \State \Return
    \EndIf 
    \If{\textsc{check-crash-vector}($m$, $cv$)=\textbf{false}}
        \State \Return
    \EndIf
    \If{$m.v > \emph{view-id}$}
        \State \textsc{initiate-view-change}($m.v$)
    \Else 
         \If{$status=\textsc{normal}$}
            \State \textsc{send-start-view}($m.r$)
        \Else \Comment{The leader is asking for fresher \textsc{view-change}}
            \State \textsc{send-view-change}($m.r$)
        \EndIf 
    \EndIf
  \EndEvent
  
    \Function{send-start-view}{$i$}
        \State $m.type=\textsc{start-view}$
        \State $m.v=\emph{view-id}$
        \State $m.cv=cv$
        \State $m.log=log$
        \State \textsc{send-message}($m$, $i$)
        \State \Return
    \EndFunction

   \Event{\emph{receiving} view-change, $m$}
    \If{$status=\textsc{recovering}$}
    \State \Return
    \EndIf 
    \If{\textsc{check-crash-vector}($m$, $cv$)=\textbf{false}}
    \State \Return
    \EndIf
    \If{$status=\textsc{normal}$}
        \If{$m.v > \emph{view-id}$}
            \State \textsc{initiate-view-change}($m.v$)
        \Else \Comment{The sender lags behind}
            \State \textsc{send-start-view}($m.r$)
        \EndIf
    \ElsIf{$status=\textsc{viewchange}$}
        \If{$m.v > \emph{view-id}$}
            \State \textsc{initiate-view-change}($m.v$)
        \ElsIf{$m.v <\emph{view-id}$}
            \Comment{The sender lags behind}
            \State \textsc{send-view-change-req}($m.r$)
        \Else \Comment{Remove stray messages and add the fresh one}
            \State $V'=\{m'\in V\ | \  m'.cv[m'.r]<cv[m'.r]\}$
            \State $V=V\ \cup\ \{m\} -V'$
            \State $\forall m' \in V'$, resend \textsc{view-change-req} to $m'.r$
            \If{$|V|\geq f+1$}
               \State $log=\textsc{merge-log}(V)$
               \For{$i$ $\leftarrow$0 to $2f$}
                    \State \textsc{send-start-view}($i$)
               \EndFor
               \State $\emph{last-normal-view}=\emph{view-id}$
               \State $status=\textsc{normal}$ \Comment{Leader becomes \textsc{normal}}
                
            \EndIf
        \EndIf
    \EndIf
  \EndEvent
  
 \Function{merge-log}{$V$}
    \State $\emph{new-log}=\emptyset$
    \State $\emph{largest-normal-view}=\max\{m.lnv | m \in V\}$ 
    \State $\emph{largest-sync-point}=\max\{m.sp | m \in V \newline\hspace*{1.5cm} \text{and  } m.lnv= \emph{largest-normal-view}\}$
    \State Pick $m\in V$: $m.lnv=\emph{largest-normal-view}$ and
    \newline\hspace*{1.8cm} $m.sp=\emph{largest-sync-point}$
    \LeftComment{Directly copy entries up to \emph{sync-point}}
    \For{$e\in m.log$}
    \Comment{$m.log$ is already sorted by \emph{deadline}s}
        \If{$e.deadline\leq \emph{largest-sync-point}.deadline$}
            \State \emph{new-log}.append($e$)
        \Else
            \State \textbf{break}
        \EndIf
    \EndFor
    \LeftComment{Add other committed entries beyond \emph{sync-point}}
    \State $\emph{entries}=\{e| e\in m.log \newline\hspace*{2cm} \text{and } e.deadline> \emph{largest-sync-point}.deadline \newline\hspace*{2cm} \text{and }m.lnv=\emph{largest-normal-view}
    \}$
    \For{$e \in entries$}
        \LeftComment{Check how many replicas contain $e$}
        \State $S=\{m|m\in V \text{and } e \in m.log\}$
        \If{$|S|\geq \lceil f/2 \rceil + 1$}
        \State $log$.append($e$)
        \EndIf
    \EndFor
    \State Sort \emph{new-log} by $entries$' $deadline$s
    \State \Return \emph{new-log}
\EndFunction
 
  \Event{\emph{receiving} start-view, $m$}
    \If{$status=\textsc{recovering}$}
    \State \Return
    \EndIf 
    \If{\textsc{check-crash-vector}($m$, $cv$)=\textbf{false}}
        \State \Return
    \EndIf
    \If{$m.v<\emph{view-id}$}
        \State \Return
    \EndIf
    \State $\emph{last-normal-view}=\emph{view-id}=m.v$
    \State $log=m.log$
    \State $\emph{sync-point}=\emph{log}$.last()
    \State $status=\textsc{normal}$ 
    \Comment{Followers become \textsc{normal}}
  \EndEvent
  

\end{algorithmic}

\end{multicols}
\end{algorithm*}
\subsection{Leader Change}
When the follower(s) suspect that the leader has failed, they stop processing new client requests. Instead, they perform the view change protocol to elect a new leader and resume request processing. With reference to Algorithm~\ref{algo-leader-change}, we explain the details of the view change process.

\begin{enumerate}[wide, labelwidth=!,nosep,label=\textbf{Step \arabic*:}]
    \item When a replica fails to receive the heartbeat (i.e., \emph{sync} message) from the leader for a threshold of time, it suspects the leader has failed. Then, it sets its \emph{status} as \textsc{viewchange}, 
    increments its \emph{view-id}, and
    broadcasts a view change request to all replicas including its \emph{crash-vector}, i.e. <\textsc{view-change-req}, \emph{view-id}, \emph{replica-id}, \emph{cv}> (line 6-10) {The view change request will be rebroadcast if the replica times out but is still waiting for the view change process to complete. The same is also true for the view change message described in the next step.}. The replica switches its \emph{status} from \textsc{normal} to \textsc{viewchange}, and enters the view change process. 
    \item After receiving a \textsc{view-change-req} message, the recipient checks the \emph{cv} and \emph{replica-id} with its own \emph{crash-vector} (line 32). If this message is a potential \emph{stray message}, then the recipient ignores it. Otherwise, the recipient updates its \emph{crach-vector} by aggregation. After that, the recipient also participates in the view change (line 35) if its \emph{view-id} is lower than that included in the \textsc{view-change-req} message.
    
    \item All replicas under view change send a message <\textsc{view-change}, \emph{view-id}, \emph{log}, \emph{sync-point}, \emph{last-normal-view}> to the leader of the new view (\emph{replica-id}$=$ \emph{view-id}$\% (2f+1)$) (line 11).     Here \emph{last-normal-view} indicates the last view in which the replica's \emph{status} was \textsc{normal}.
    
    \item After the new leader receives the \textsc{view-change} messages from $f$ followers with matching \emph{view-id}s, it can recover the system state by merging the \emph{log}s from the $f+1$ replicas including itself (line 67). The new leader only merges the \emph{log}s with the highest \emph{last-normal-view}, because a smaller \emph{last-normal-view} indicates the replica has lagged behind for several view changes, thus its \emph{sync-point} cannot be larger than the other replicas with higher \emph{last-normal-view} values. Therefore, it makes no contribution to the recovery and does not need to join. 
    
    
    \item The new leader initializes an empty log list (denoted as \emph{new-log}) (line 74). Among the \textsc{view-change} messages with the highest \emph{last-normal-view}, it picks the one with the largest \emph{sync-point} (line 75-77). Then it directly copies the log entries from that message up to the \emph{sync-point} (line 78-82). 

    \item \label{log-beyond-sync} Afterwards, the new leader checks the remaining entries with larger \emph{deadline}s than \emph{sync-point} (ling 83-88). If the same entry (2 entries are the same iff they have the same <\emph{deadline}, \emph{client-id}, \emph{request-id}>) exists in at least $\lceil f/2 \rceil +1$ out of the $f+1$ \emph{log}s, then leader appends the entry to \emph{new-log}.

    \item After \emph{new-log} is built, the new leader broadcasts <\textsc{start-view}, \emph{cv}, \emph{view-id}, \emph{new-log}> to all replicas (line 68-70).
    
    \item After receiving the \textsc{start-view} message with a \emph{view-id} greater than or equal to its \emph{view-id}, the replica updates its \emph{view-id} and \emph{last-normal-view} (line 97), and replaces its \emph{log} with \emph{new-log} (line 98). Besides, it updates \emph{sync-point} as the last entry in the new \emph{log} (line 98), because all the entries are consistent with the leader. Finally, replicas set their \emph{status}es to \textsc{normal} (line 100), and the system state is fully recovered.
    
    \item After the system is fully recovered, the replicas can continue to process the incoming requests. Recall in \S\ref{sec-commutativity-optimization} that the incoming request is allowed to enter the \emph{early-buffer} if its deadline is larger than \emph{the last released request} which is not commutative. To ensure consistent ordering, the eligibility check is still required for the incoming request even if it is the first one arriving at the replica after recovery. The replica considers the entries (requests) in the recovered \emph{log}, which are not commutative to the incoming request, and chooses the one as \emph{the last released request} with the largest deadline among them. The incoming request can enter the \emph{early-buffer} if its deadline than \emph{the last released request}, otherwise, it is put into the \emph{late-buffer}. 

\end{enumerate}

Note that the view change protocol chooses the leader in a round-robin way ($\emph{view-id} \% (2f+1)$). Specially, a view change process may not succeed because the new leader also fails (as mentioned in~\cite{viewstamp}). In this case (i.e. after followers have spent a threshold of time without completing the view change), followers will continue to increment their \emph{view-id}s to initiate a further view change, with yet another leader.

After the replica rejoin or leader change process, replicas' \emph{crash-vector}s will be updated. Due to packet drop, some replicas may fail to receive the update of \emph{crash-vector}s during the recovery, thus they cannot contribute to the quorum check of the fast path in the following request processing, because their \emph{crash-vector}s are still old and cannot generate the consistent hash with the leader's hash. To enable every replica to obtain the fresh information of \emph{crash-vector}s rapidly, the leader can piggyback the fresh \emph{crash-vector}s in the \emph{sync} messages, so that replicas can check and update their \emph{crash-vector}s as soon as possible.

\subsection{Why Does \emph{crash-vector} Prevent \emph{Stray Message} Effect during Quorum Check?}
\label{general-error-pattern}

The \emph{stray message}s can cause a common bug for most optimistic consensus protocols (e.g., \cite{nsdi15-specpaxos,osdi16-nopaxos,conext20-domino}) when they conduct the quorum check in the fast path. Below we summarize the general pattern to cause the loss of committed requests (durability violation). Then, we explain why \sysname can avoid such problems by use of \emph{crash-vector}.

\Para{General Error Pattern.} Consider a request is delayed in the network, whenever it arrives at one replica, that replica sends a reply and immediately crashes afterwards, then the crashed replica recover from the others and gets an empty log list (because the other replicas have not received the request). After each replica completes such behavior, the client gets replies from all the replicas but actually none of them is holding the request. Such a pattern does not violate the failure model, but causes permanent loss of committed requests.

Reviewing the existing opportunistic protocols,  
Speculative Paxos, NOPaxos and Domino all suffer from such cases. CURP~\cite{nsdi19-curp} can avoid the \emph{stray message} effect by assuming the existence of a configuration manager, which never produces stray messages (e.g., by using stable storage). Whenever the witnesses crash and are relaunched, the configuration manager need to refresh the information for the master replica as well as the clients, so that clients can detect the stray messages during quorum check and avoid incorrectness. 

\sysname avoids such error cases by including the information of \emph{crash-vector} in the hash of \emph{fast-replies} (\S\ref{sec-incremental-hash}), which prevent stray reply messages from forming the super-quorum in the fast path and creating an \emph{illusion} to the proxies/clients. We analyze in more details below.

Regarding the general pattern above, 

\begin{enumerate}[wide,label=(\arabic*)]
    \item When the follower(s) fail, they need to contact the leader and complete the state transfer before their recovery (Algorithm~\ref{algo-replica-rejoin}). 
\begin{itemize}
    \item If the leader has already received the request before the state transfer, then after the follower's recovery, it can remember the \emph{fast-reply} that it has sent before crash, and can replay it. In this case, the \emph{fast-reply} is not a stray message.
    \item If the leader has not received the request before the state transfer, then the leader's \emph{crash-vector} will be updated after receiving the follower's \textsc{state-transfer-req} (line 75-76 in Algorithm~\ref{algo-replica-rejoin}), which includes a different \emph{crash-vector} (the follower has incremented its own counter). Therefore, the \emph{hash} of the leader's \emph{fast-reply} is computed with the aggregated \emph{crash-vector}, and will be different from that included in the \emph{fast-reply} (stray message) sent by the follower before crash, i.e. the leader's \emph{fast-reply} and the followers' stray \emph{fast-replies} cannot form a super quorum. 
\end{itemize}
    \item When the leader fails, based on Algorithm~\ref{algo-leader-change}, the view change will elect a new leader. \emph{crash-vectors} ensure the view change process is not affected by \emph{stray messages}. After the view change is completed, the \emph{view-id} is incremented. At least $f+1$  replicas after the view change will send \emph{fast-replies} with higher \emph{view-id}s. Because the quorum check requires reply messages have matching \emph{view-id}s, the \emph{stray fast-replies} (sent by the old leader) can not form a super quorum together with the \emph{fast-replies} sent by the replicas after the view change.
 
\end{enumerate}

\sysname's slow path does not suffer from \emph{stray message} effect, because there is causal relation between the leader's state update (advancing its \emph{sync-point}) and followers' sending \emph{slow-replies}. 
\begin{enumerate}[wide,label=(\arabic*)]
    \item When followers crash and recover, they copy the state from the leader. The followers' state before crash is no fresher than their recovered state, so the followers have no \emph{stray slow-replies}, i.e. the followers can remember the \emph{slow-replies} they have sent before crash and can replay them.
    \item When the leader crashes and recovers, it can only rejoin as a follower replica after the new leader has been elected (\S\ref{replica-rejoin}), so the old leader's reply messages before crash have smaller \emph{view-id}s, compared with the \emph{slow-replies} of replicas after the view change. With matching \emph{view-id}s, these reply messages cannot form a quorum together in the slow path. 
\end{enumerate}

\subsection{Reconfiguration}
While it has not been implemented, \sysname can also use the standard reconfiguration protocol from Viewstamped Replication~\cite{viewstamp} (with its incorrectness fixed by \emph{crash-vector}~\cite{disc17-dcr,disc17-dcr-techreport}) to change the membership of the replica group, such as replacing the failed replicas with the new ones that have a new disk, increasing/decreasing the number of replicas in the system, etc. 


\section{Correctness Proof of \sysname}
\label{append-correctness}
With the normal behavior described in \S\ref{sec-fast-quorum-check}$\sim$\S\ref{sec-slow-quorum-check},
we can prove that the recovery protocol of \sysname guarantees the following correctness properties.

\begin{itemize}
    \item \textbf{Durability:} if a client considers a request as committed, the request survives replica crashes.
    \item \textbf{Consistency:} if a client considers a request as committed, the execution result of this request remains unchanged after the replica's crash and recovery.
    \item \textbf{Linearizability:} A request appears to be executed exactly once between start and completion. The definition of linearizability can also be reworded as: if the execution of a request is observed by the issuing client or other clients, no contrary observation can occur afterwards (i.e., it should not appear to revert or be reordered). 
\end{itemize}

\subsection{Proof of Durability}
The client/proxy 
considers $req$ as committed after receiving the corresponding quorum or super quorum of replies. Since the quorum checks on both the fast path and slow path require the leader's reply, a committed request indicates that the request must have been accepted by the leader. If a follower crashes, it does not affect durability because the recovered followers directly copy \emph{log} from the leader via state transfer (Step 7 in \S\ref{replica-rejoin}) before serving new requests. Hence, we consider the durability property during leader crashes. 
\begin{enumerate}[wide,label=(\arabic*)]
    \item If the client/proxy commits $req$ in the fast path, it means the request has been replicated to the leader and at least $f+\lceil f/2 \rceil$ followers. 
    When the leader crashes, among any group of $f+1$ replicas, $req$ exists in at least $\lceil f/2 \rceil+1$ of them because of quorum intersection. Hence, $req$ will be added to the \emph{new-log} in Step 6 in \S\ref{leader-change}, and eventually recovered.
    \item If the client/proxy commits $req$ in the slow path, it means $req$ has been synced with the leader by at least $f+1$ replicas, i.e., there are at least $f+1$ replicas containing a \emph{sync-point} whose \emph{deadline} is greater than or equal to $req$'s \emph{deadline}. Due to quorum intersection, there will at least one replica which has the \emph{sync-point} in Step 4 of \S\ref{leader-change}. Therefore, $req$ will be directly copied to \emph{new-log} in Step 5 of \S\ref{leader-change}, and eventually recovered. 
\end{enumerate}

\subsection{Proof of Consistency}
Without considering the acceleration of recovery mentioned in \S\ref{sec:recovery}, we prove consistency. It is also easy to check that the recovery acceleration is a performance optimization that does not affect the consistency property. So, ignoring acceleration of recovery for simplicity, followers do not execute requests. Thus, we only need to consider the leader's crash and recovery. We assume the client/proxy has committed $req$ before the leader crash. 
\begin{enumerate}[wide,label=(\arabic*)]
    \item If the client/proxy commits $req$ in the fast path, it means at least $f+\lceil f/2 \rceil$ followers have consistent log entries with the leader up to this request $req$. Therefore, on the old leader, all the log entries \emph{before} $req$ are \emph{also} committed, because they also form a super quorum with consistent hashes. So, they can survive crashes and be recovered in Steps 5 and 6 of \S\ref{leader-change}. Additionally, consider an uncommitted request $ureq$, which is not commutative to $req$ and has a smaller \emph{deadline} than $req$, it cannot be appended by any of the $f+\lceil f/2 \rceil + 1$ replicas which have appended $req$, because the \emph{early-buffer} of \primname only accepts and releases requests in the ascending order of deadlines (\S\ref{sec-dom}). Even if all the other $\lfloor f/2 \rfloor$ have appended $ureq$, they fail to satisfy the condition in Step 5/Step 6, so $ureq$ cannot appear in the recovered \emph{log}s to affect the execution result of $req$.
    
    \item If the client/proxy commits $req$ in the slow path, it means at least $f$ followers have consistent log entries with the leader up to $req$, i.e., the \emph{deadline}s of their \emph{sync-points} are greater than or equal to the \emph{deadline} of $req$. Therefore, on the old leader, all the log entries before $req$ are committed, and they can survive crashes and be recovered in Step 5 of \S\ref{leader-change}. Additionally, if the follower's \emph{log} contains the request $ureq$, which is not commutative to $req$ and has a smaller deadline than $req$, but does not exist on the leader, then $ureq$ cannot appear in the recovery \emph{log} of the new leader. This is because,   
    based on the workflow of the slow path (\S\ref{sec-slow-quorum-check}), the follower advances its \emph{sync-point} strictly following \emph{sync} messages from the leader. Since the \emph{sync} message does not include the $ureq$'s 3-tuple \emph{<client-id, request-id, deadline>}, the follower will delete $ureq$ before updating its \emph{sync-point}. Therefore, it is impossible for $ureq$ to appear in the recovered \emph{log}s and affect the execution result of $req$.
\end{enumerate}

After recovery, the survived log entries will be executed by the new leader according to the ascending order of their \emph{deadline}s, thus the same execution order is guaranteed and provides the consistent execution result for $req$.

\subsection{Proof of Linearizability}
We assume there are two committed requests, denoted as \emph{req-1} and \emph{req-2}. The submission of \emph{req-2} is invoked after the completion of \emph{req-1}, i.e. the client has observed the execution of \emph{req-1} before submitting \emph{req-2}. We want to prove that no contrary observation can occur after crash and recovery. Here we assume \emph{req-1} and \emph{req-2} are not commutative with each other, because the execution of commutative requests cause no effect on each other, regardless of their execution order.

 Since \emph{req-2} is invoked after the completion of \emph{req-1}, \emph{req-2} must have a larger \emph{deadline} than \emph{req-1}, otherwise, it cannot be appended to the \emph{log}. Based on the durability property, \emph{req-1} and \emph{req-2} will be both recovered after a crash. According to the recovery algorithm, the new leader still executes the two requests based on their \emph{deadline}s. Therefore, the execution of \emph{req-1} on the new leader cannot observe the effect of \emph{req-2}. By contrast, while executing \emph{req-2}, the effect of \emph{req-1}'s execution has already been reflected in the leader's replica state. Therefore, no contrary observation (i.e., revert or recorder) can occur after the crash and recovery.
 

\section{Evaluation of \sysname under Different Workloads}
\label{sec:extensive-eval}

\begin{figure*}[!t]
  \centering
  \begin{minipage}{.98\linewidth}
    \centering
    \subcaptionbox{Read ratio=10\%, skew factor=0}
      {\includegraphics[width=0.32\linewidth]{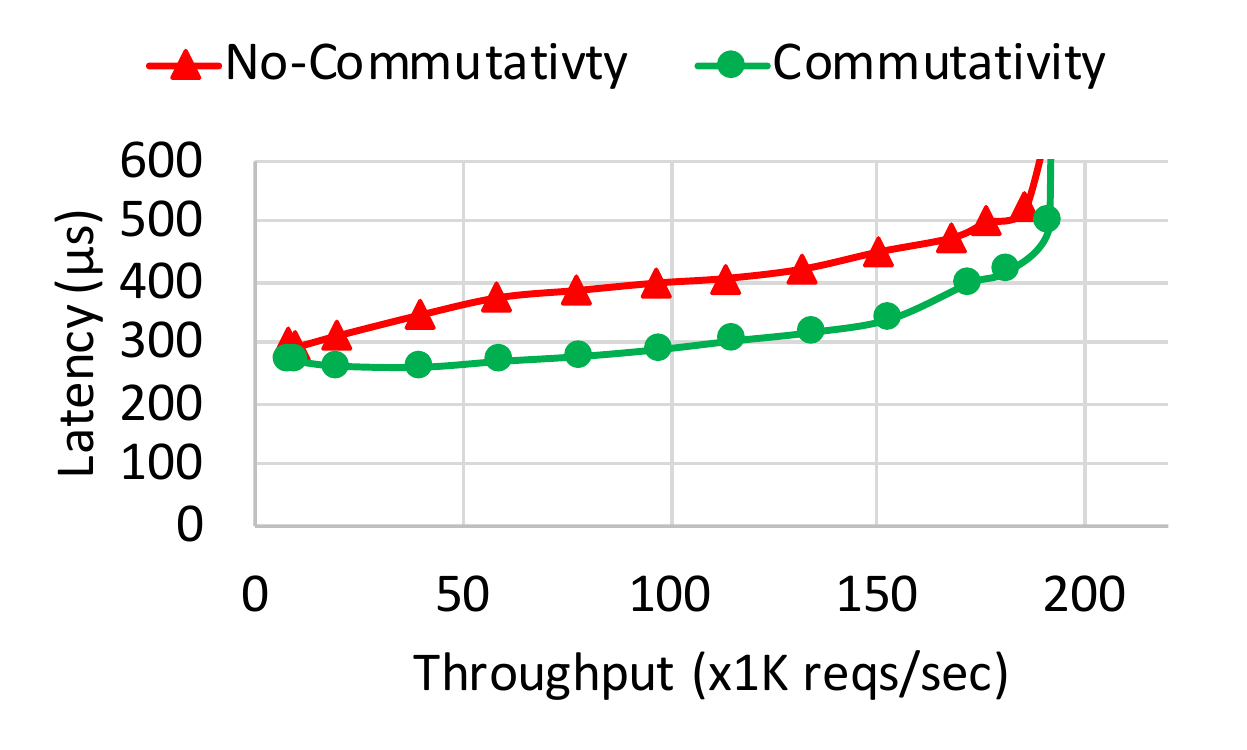}}
    \subcaptionbox{Read ratio=10\%, skew factor=0.5}
      {\includegraphics[width=0.32\linewidth]{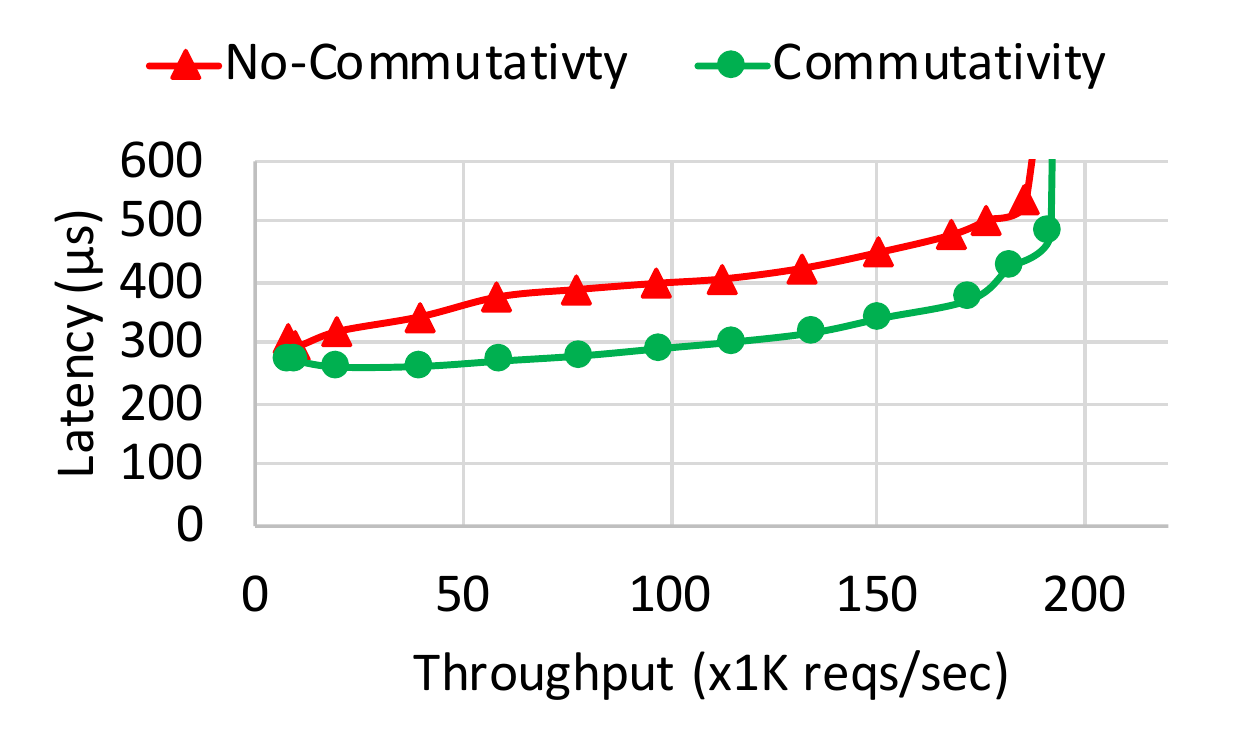}}
    \subcaptionbox{Read ratio=10\%, skew factor=0.99}
      {\includegraphics[width=0.32\linewidth]{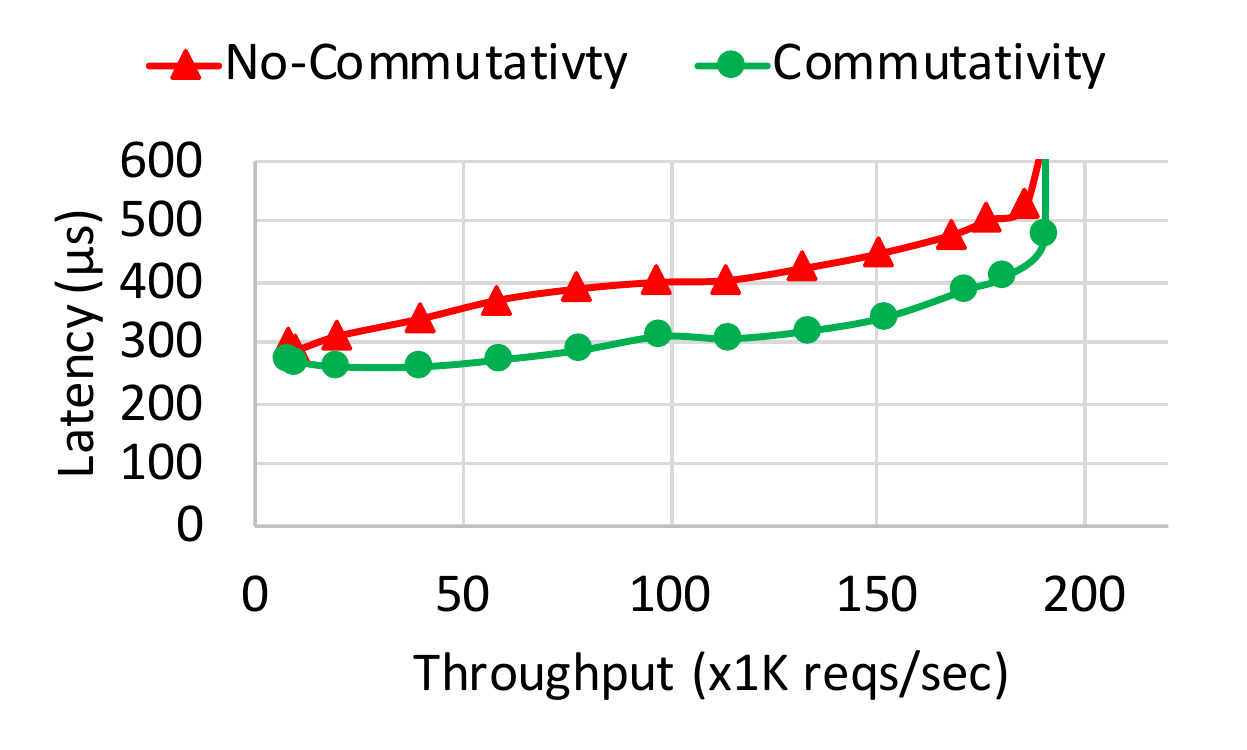}}
  \end{minipage}

  \begin{minipage}{.98\linewidth}
    \centering
    \subcaptionbox{Read ratio=50\%, skew factor=0}
      {\includegraphics[width=0.32\linewidth]{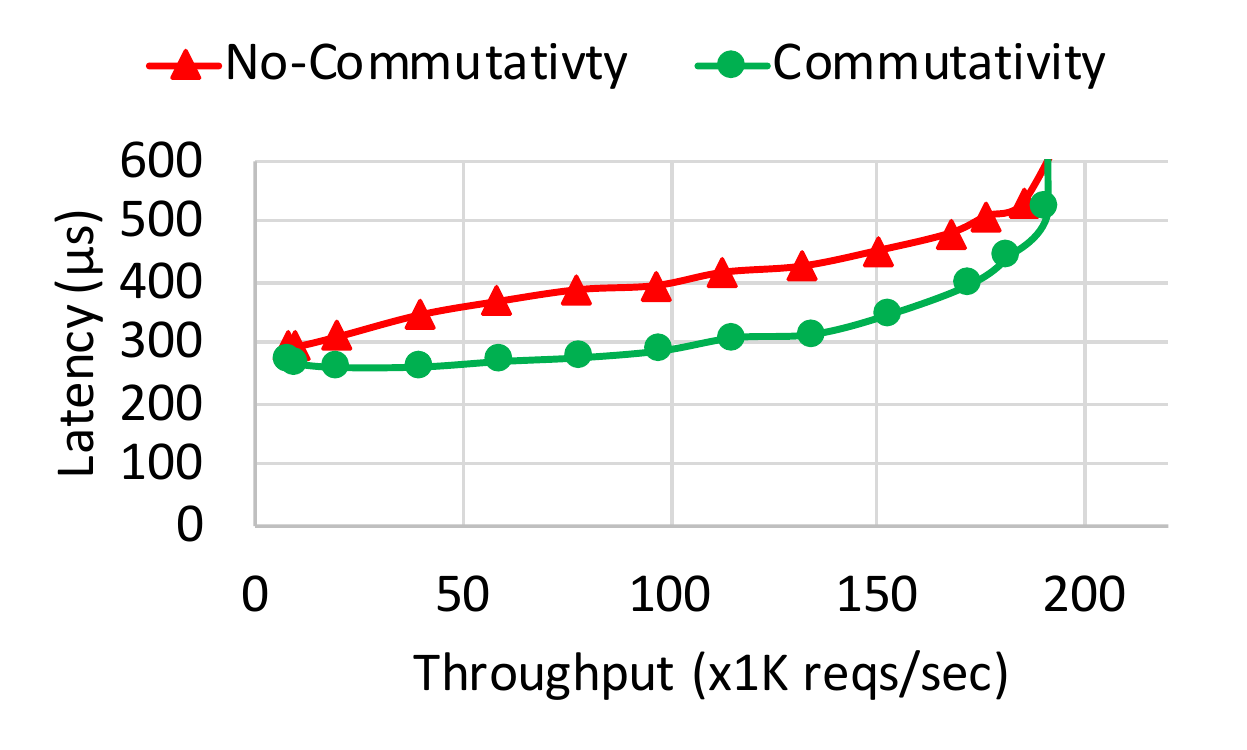}}
    \subcaptionbox{Read ratio=50\%, skew factor=0.5}
      {\includegraphics[width=0.32\linewidth]{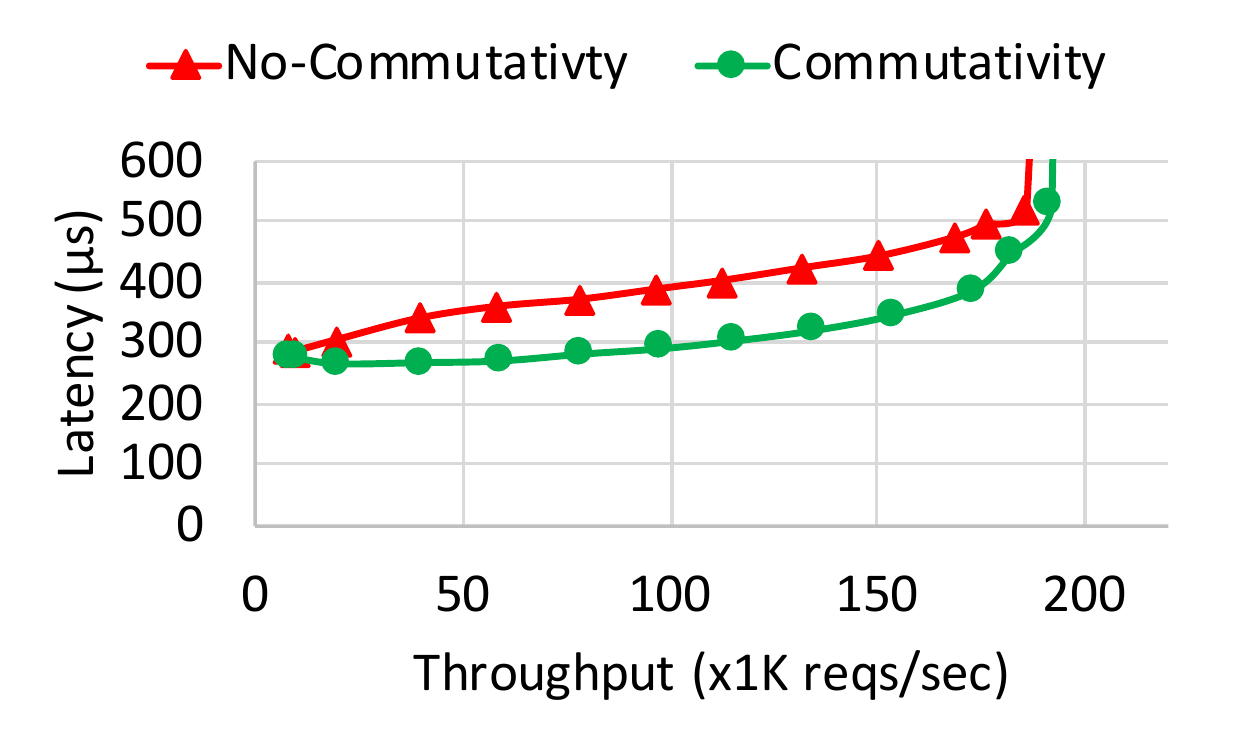}}
    \subcaptionbox{Read ratio=50\%, skew factor=0.99}
      {\includegraphics[width=0.32\linewidth]{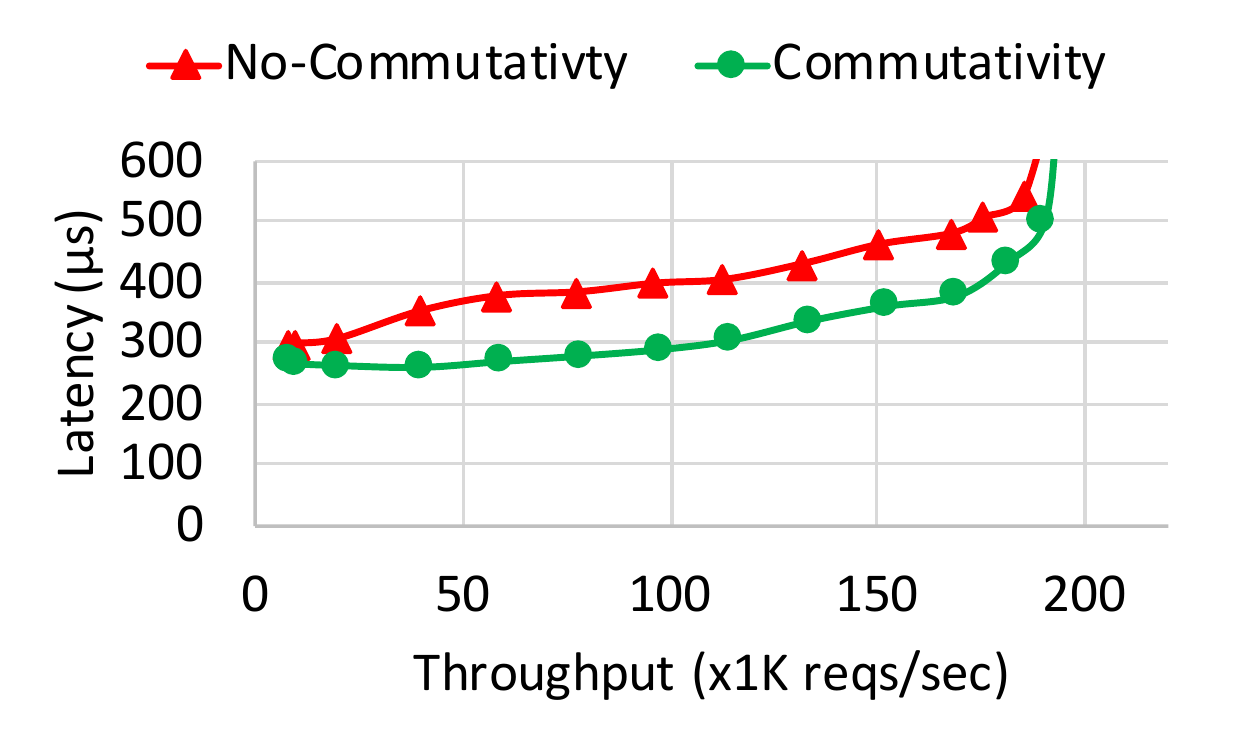}}
  \end{minipage}
  
   \begin{minipage}{.98\linewidth}
    \centering
    \subcaptionbox{Read ratio=90\%, skew factor=0}
      {\includegraphics[width=0.32\linewidth]{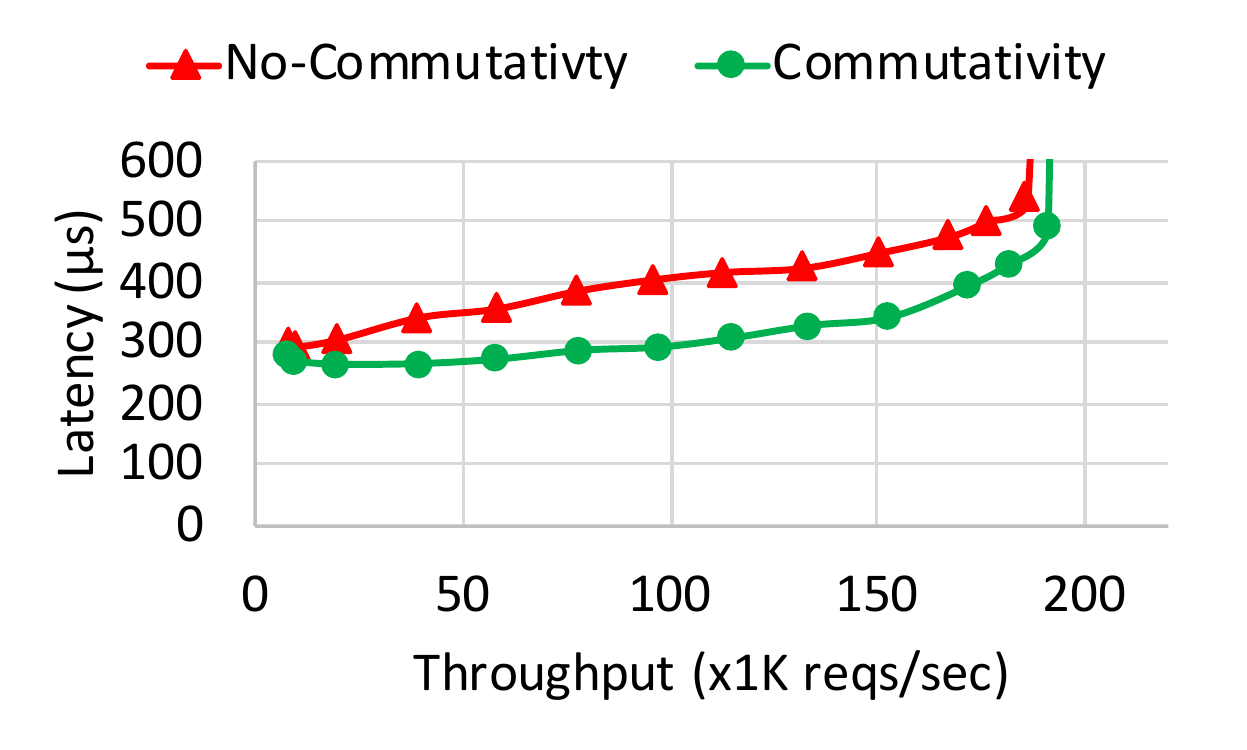}}
    \subcaptionbox{Read ratio=90\%, skew factor=0.5}
      {\includegraphics[width=0.32\linewidth]{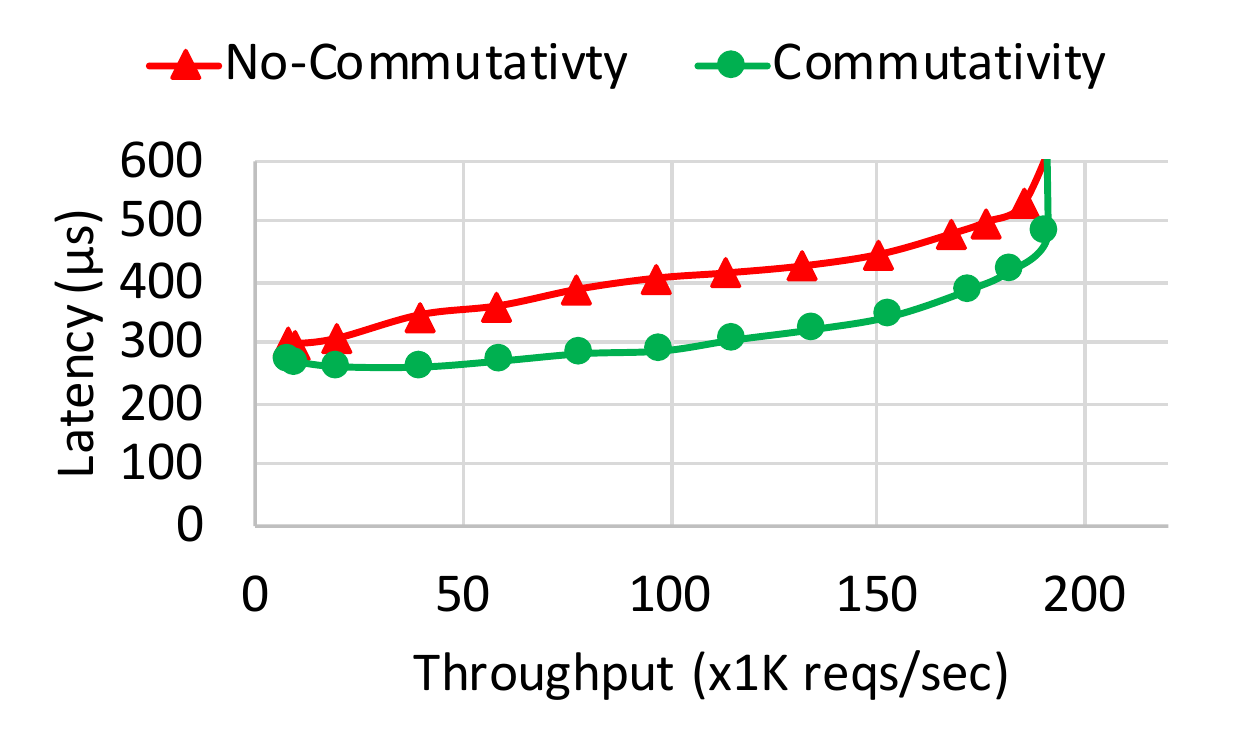}}
    \subcaptionbox{Read ratio=90\%, skew factor=0.99}
      {\includegraphics[width=0.32\linewidth]{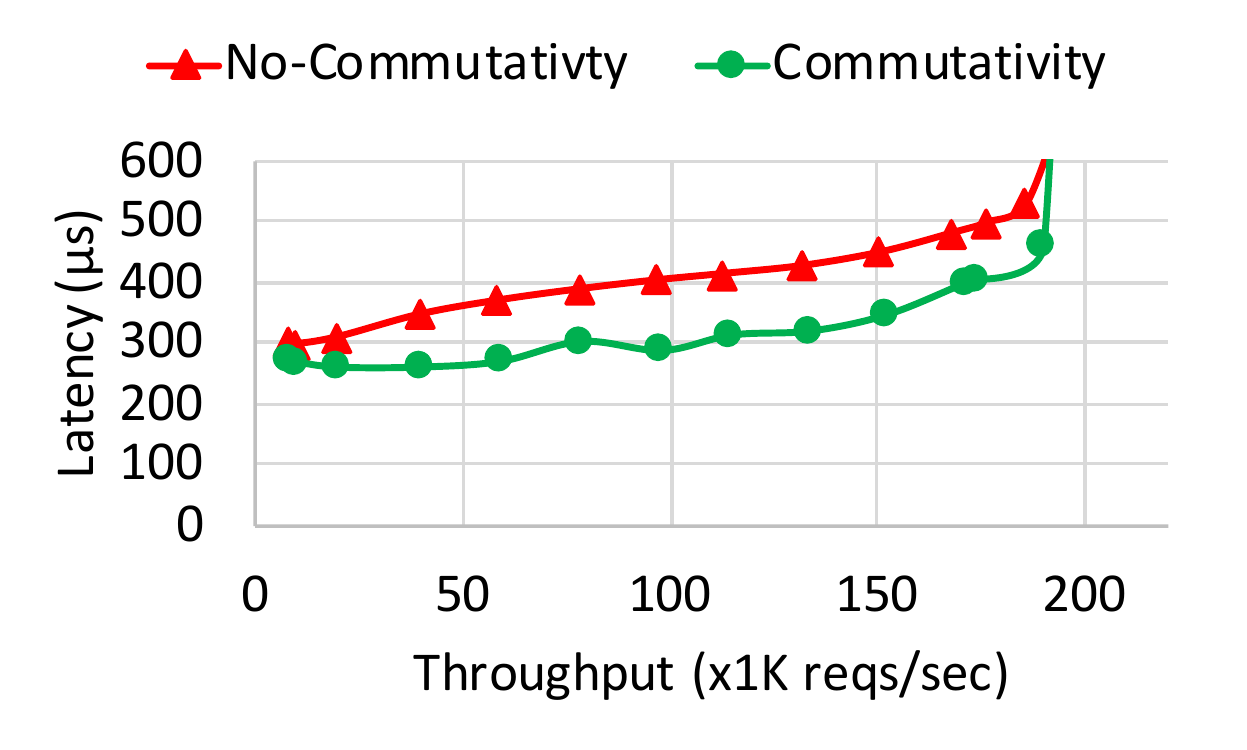}}
  \end{minipage}
  \caption{Latency vs. throughput (open-loop)}
  \label{fig-extensive-eval-closed}
\end{figure*}

We adopt the similar approach as~\cite{nsdi21-epaxos} to conduct extensive evaluation under different workloads: we maintain 1 million unique keys and choose different values of read ratio and skew factors to generate different workloads.
As for the read ratio, we choose three different values, i.e. read-10\% (write-intensive), read-50\% (medium) and read-90\% (read-intensive). As for the skew factor, we also choose three different values, i.e. skew-0.0 (evenly distributed), skew-0.5 (medium) and skew-0.99 (highly skewed). The combination of the two dimensions create 9 different workloads. We measure the median latency and throughput under each workload, as shown in Figure~\ref{fig-extensive-eval-closed}. Considering the variance in cloud environment, we run each test case for 5 times and plot the average values.

Although the latency in the cloud can vary over time~\cite{nsdi13-bobtail,hotos15-tail-latency,silo} and introduces some noise to performance results, in general, the commutativity optimization remains effective across all workloads and helps reduces the latency by \SI{7.7}{\percent}-\SI{28.9}{\percent}: under low throughput, the effectiveness of the commutativity optimization is not distinct because the No-Commutativity variant can also keep a high fast commit ratio ($\sim$\SI{75}{\percent}). However, as the throughput increases, the fast commit ratio of No-Commutativity variant drops distinctly but the commutativity variant can still maintain a high fast commit ratio (\SI{80}{\percent}-\SI{97}{\percent}), so the commutativity optimization becomes more effective. Then, as the throughput continues to grow, it reaches closer to the capacity of the replicas and even overloads the replicas, so  the reduction of latency becomes less distinct again and eventually negligible.

\section{Evaluation under Different Clock Variance}
\label{clock-eval}

\subsection{Explanation of Clock Assumption}
\label{explanation-clock}
\sysname depends on clock synchronization for performance but not for correctness. An accurate clock synchronization provides favorable conditions for \sysname to achieve high performance. Here ``accurate'' means the clocks among replicas and proxies are synchronized with a small error bound \emph{in most cases}, but note that \sysname does not assume a deterministic worst-case error bound, which is impractical because Huygens is built atop a probabilistic model (SVM), and the Huygens agents (or other clock synchronization algorithms) can also fail while the consensus protocol is still running without awareness of that.

Besides, \sysname's correctness does not require the assumption of monotonously increasing clock time either.
In other words, even if the local clock time goes back and forth (this can happen because Huygens or other clock synchronization algorithms may correct the clocks with some negative offset), \sysname's correctness is still preserved thanks to the entrance condition of the \emph{early-buffer}. Recall in \S\ref{sec-dom}, the eligibility check to enter the \emph{early-buffer} is to compare the incoming request's deadline with the deadline of the last released one (rather than the replica's local clock time). Requests in the \emph{early-buffer} are organized with a priority queue and released according to their deadline order. The clock skew can only cause requests to be released prematurely, but the released requests all follow the ascending order of deadlines, therefore, the invariant of uniform ordering is preserved by DOM. Establishing protocol correctness independent of clock skew is desirable and we will show in \S\ref{error-trace-domino} that the other protocol, Domino, loses its correctness due to clock skew (i.e. it can lose committed requests permanently if replica clocks go from large values to small values).

\subsection{Quantifying the Effect of Bad Clock Synchronization on \sysname Performance}
\label{quantify-clock}
Although we have not experienced significant clock skew in our evaluation, it is worthwhile to quantify the effect on \sysname performance imposed by different clock synchronization quality. To simplify the discussion below, we consider most VM/server's clocks are synchronized to the reference clock time within a tight bound, whereas the other ones suffer from distinct skew and are not well synchronized with the reference clock time. Thus, we mainly focus on three categories.

\begin{enumerate}
    \item The leader replica's clock is badly synchronized with the other VMs.
    \item The follower replica's clock is badly synchronized with the other VMs.
    \item The proxy's clock is badly synchronized with the other VMs.  
\end{enumerate}

\Para{Method.} To create the effect of bad clock synchronization, we choose one or multiple target VMs (i.e. the leader replica, or the follower replica, or the proxies) and inject artificial offsets when the clock APIs are called on the VM. To be more specific, we generate random offsets based on normal distribution $N(\mu, \sigma)$. For each test case, we choose different mean values ($\mu$) and standard deviation ($\sigma$) for the distribution to mimic bad clock synchronization of different degrees. When the clock API is called, instead of returning the clock value, we take an offset sample from the distribution and add it to the clock value, and then return this summed value, to make its clocks faster/slower than the others. 

\begin{figure*}[!t]
    \begin{minipage}{1\linewidth}
    \centering
    \subcaptionbox{Impact of the leader's clock synchronization quality on \sysname performance  \label{fig-leader-clock}}
      {
      \includegraphics[width=0.32\linewidth]{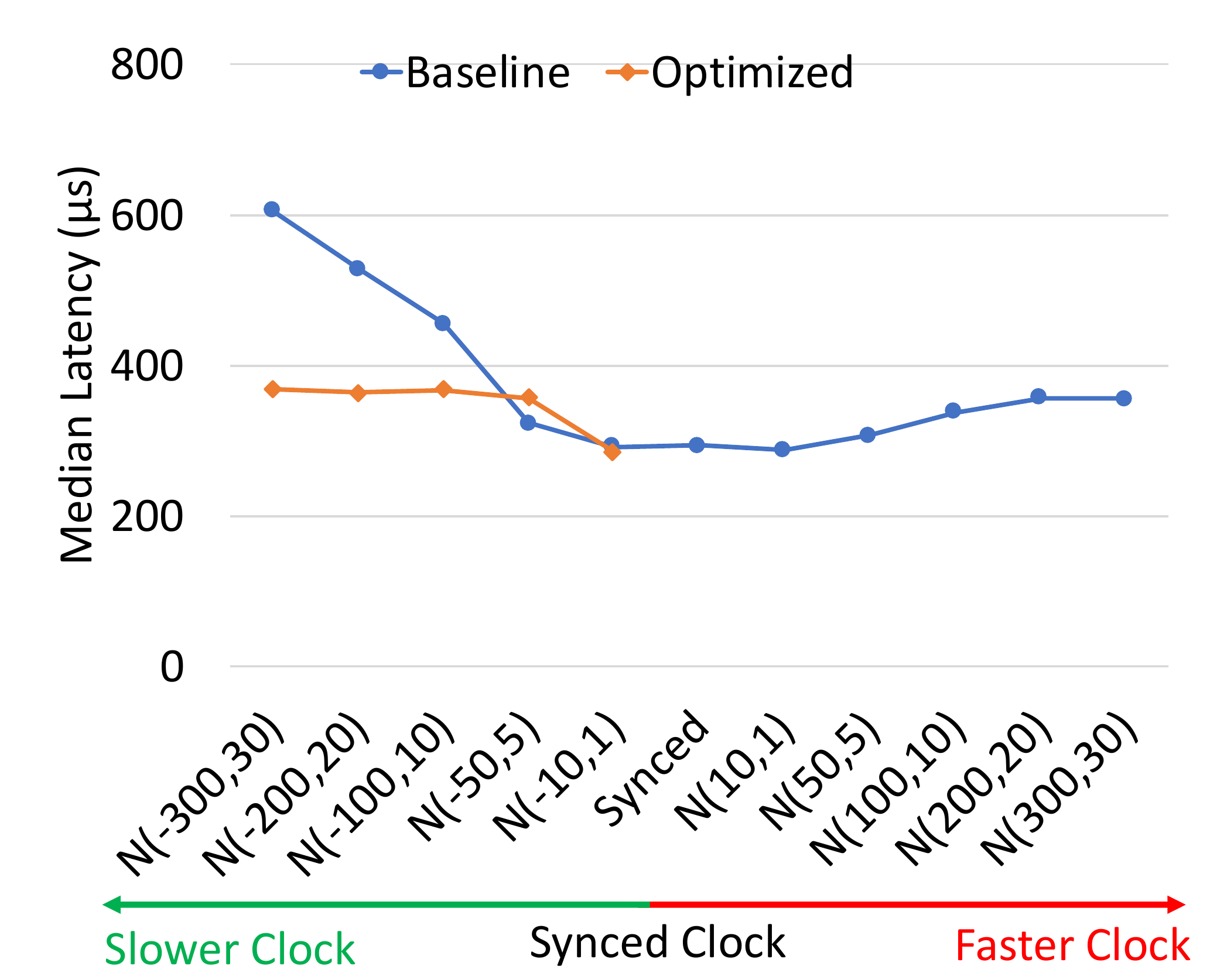}

      }
    \subcaptionbox{Impact of the follower's clock synchronization quality on \sysname performance  \label{fig-follower-clock}}
      {\includegraphics[width=0.32\linewidth]{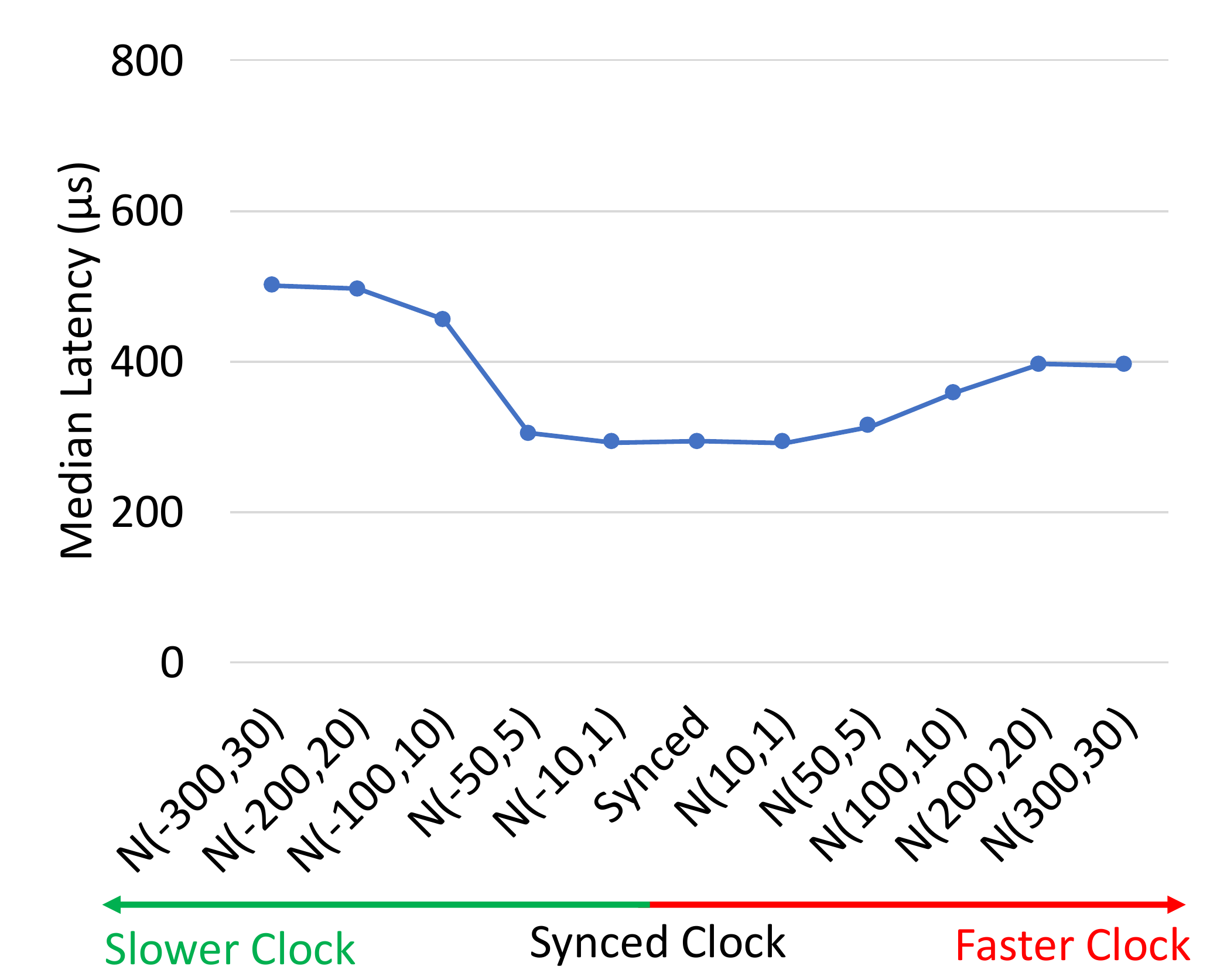}
     }
    \subcaptionbox{Impact of the proxies' clock synchronization quality on \sysname performance  \label{fig-proxy-clock}}
      {\includegraphics[width=0.32\linewidth]{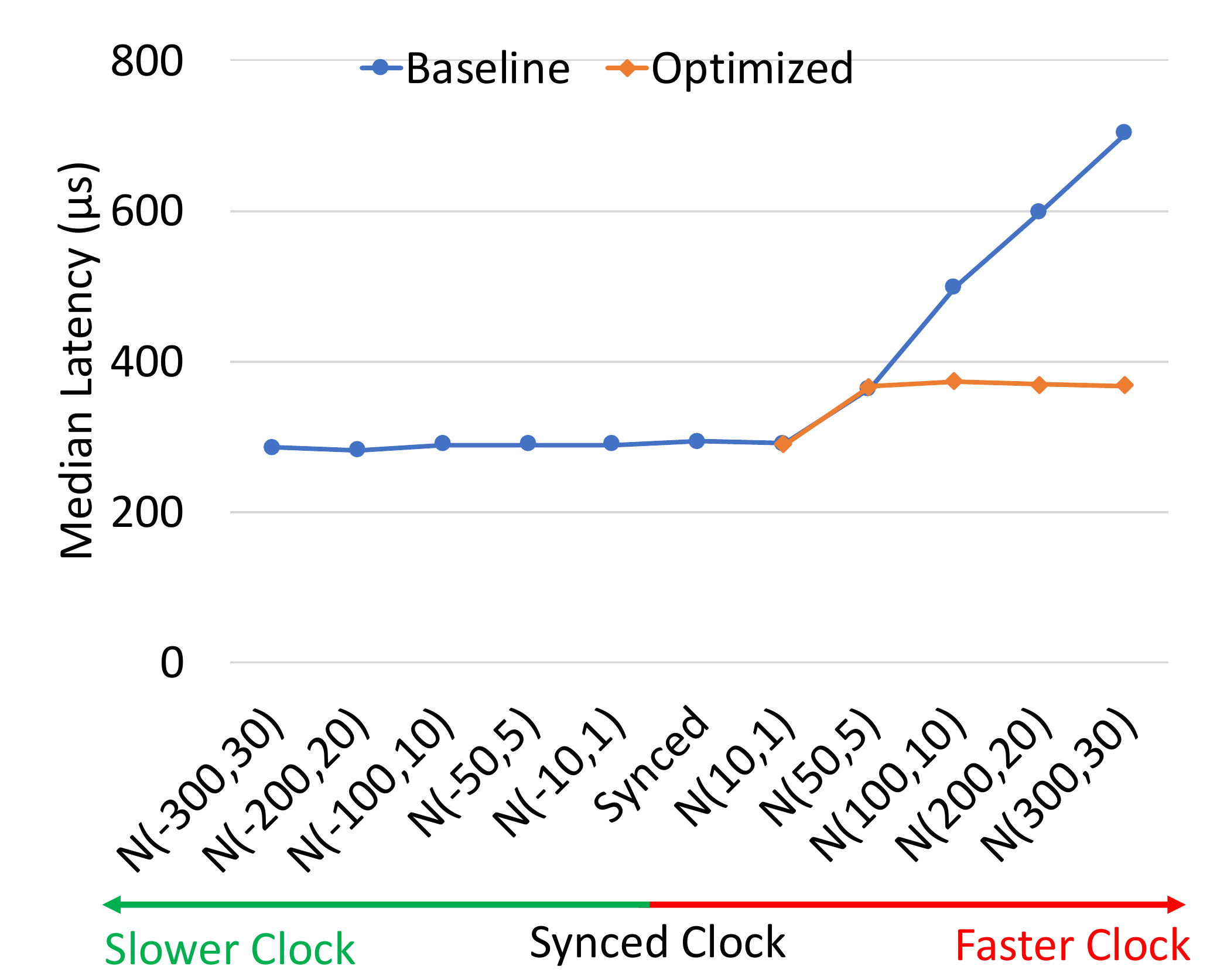}
     }
   \caption{\sysname latency vs. clock synchronization quality}
   \label{fig-clock-eval}
  \end{minipage}
\end{figure*}

\Para{Test Setting}. Similar to the setting in \S\ref{sec-latency-throughput}, we set up 3 replicas and 2 proxies, and use 10 open-loop clients to submit at 10K request/sec each, thus yielding a throughput $\sim$100K request/sec. We measure the latency for each test case and study how the latency evolves as the clock synchronization quality varies. We maintain the same parameters for the adaptive latency bound formula (refer to \S\ref{sec-dom}). Specifically, the sliding window size is 1000 to calculate the moving median $M$; $beta=3$; $D=200\mu s$. During our tests, we observe the $\sigma_S$ and $\sigma_R$ returned by Huygens are both very small, typically $1-2\mu s$. We choose 10 different normal distributions (as shown in Figure~\ref{fig-clock-eval}) to mimic bad clock synchronization of different degrees, from the slowest clock to the fastest clock.

For example, $N(-300,30)$ indicates that the mean value of the normal distribution is \SI{-300}{\micro\second} with a standard deviation of  \SI{30}{\micro\second}. When we choose an offset (typically a large negative value) from this distribution and add it to the clock value, it will make the clock value smaller than the synchronized clock value by hundreds of microseconds, i.e., the clock becomes slower than the other clocks due to the offset we have added.

\subsubsection{Bad Clock Synchronization of Leader Replica}
\label{sec-leader-bad-clock}

As shown in Figure~\ref{fig-leader-clock}, when the leader's clock fails to be synchronized with the other VMs and goes faster or slower, it will inflate the latency performance of \sysname. Comparing the faster-clock cases and the slower-clock cases, we can see that a slower clock on the leader replica causes more degradation than a faster clock. 

\textbf{When the leader replica has a slower clock}, it will accept most requests into its \emph{early-buffer} but keep them for a much longer time. The requests cannot be committed until the leader releases it. Therefore, the slower the leader's clock is, the long latency \sysname will suffer from. 

\textbf{When the leader replica has a faster clock}, it will cause two main effects. First, the leader replica will prematurely release requests with large deadlines, causing the subsequent requests unable to be accepted by its \emph{early-buffer}, so the subsequent requests can only be committed in the slow path. Second, the leader will provide overestimated one-way delay (OWD) values and piggyback them to the proxies (recall that the OWD is calculated by using leader's receiving time to subtract the proxies' sending time) and cause the proxies to use large latency bound (i.e. the max of the estimated OWDs from all replicas) for its following requests multicast. However, the second effect is mitigated by DOM-Rs, because we use the clamping function: when the estimated OWD goes beyond the scope of $[0,D]$, it will use $D$ as the estimated value. Therefore, the negative impact due to the leader's slower clock is constrained. The major impact is that more requests can only be committed in the slow path, which can degrade the latency performance, but the degradation is bounded.

\subsubsection{Bad Clock Synchronization of Follower Replica}
\label{sec-follower-bad-clock}

As shown in Figure~\ref{fig-follower-clock}, similar to the cases in Figure~\ref{fig-leader-clock}, the follower's bad clock synchronization also inflates the latency. However, the negative impact of the follower's bad clock synchronization is less distinct than leader's bad clock synchronization: both a faster clock and a slower clock of the follower only cause bounded degradation of latency performance.

\textbf{When the follower has a faster clock}, it may prematurely release requests with large deadlines and cause subsequent requests not accepted by the \emph{early-buffer} (similar to the case where the leader has a faster clock). But eventually the request can be committed in the slow path, so the slow-path latency will bound the degradation. 

\textbf{When the follower has a slower clock}, it will hold the requests in its \emph{early-buffer} for longer time. However, if the leader and the other follower(s) have well synchronized clocks, they can still form a simple majority to commit the request in the slow path. Therefore, this follower's slower clock can not degrade the latency without bounds (whereas the leader's slower clock can). 

The major negative impact caused by the follower's faster/slower clock is that, it will lead to inaccurate estimation of OWDs. If the follower has a faster clock, it will piggyback large OWDs to the proxies, thus causing the proxies to choose large latency bound for the following requests. If the follower has a slower clock, it will piggyback small OWDs (or even negative OWDs) to the proxies. However, thanks to the clamping operation during the latency bound estimation, the latency bound will fall back to $D$ ($D=\SI{200}{\micro\second}$) when the estimated OWD goes too large or negative. In this way, the negative impact of follower's faster/slower clock is constrained.

\subsubsection{Bad Clock Synchronization of Proxy}
\label{sec-proxy-bad-clock}

As shown in Figure~\ref{fig-proxy-clock}, the proxies' having slower clocks do not cause degradation of the latency performance, so long as replicas have well synchronized clocks. However, the proxies' having faster clocks can lead to unbounded degradation of latency performance. 

\textbf{When the proxies have slower clocks}, it does not affect the latency so long as replicas are still well synchronized with each other. This is because, although proxies' slower clocks cause smaller sending time, it also leads to larger OWD, which is calculated by the replicas using its local clock time to subtract the sending time. The OWDs are piggybacked to the proxies and eventually lead to large latency bound. Therefore, although the clocks of proxies  lag behind, the over-estimated latency bound compensate the lag, and summing up the sending time and latency bound still yields a proper deadline. Therefore, the latency performance does not degrade when proxies have slower clocks.

\textbf{When the proxies have faster clocks}, the latency can go up without bound. When there is only a small clock offset occurring (e.g. $N(10,1)$), i.e. the proxies' clocks are not fast enough, it will not degrade the latency performance of \sysname. This is because, although the faster clock leads to a larger sending time, it also leads to a smaller latency bound, summing them up still yields a proper deadline. However, when proxies' clocks are too fast (e.g. $N(300,30)$), the sending time becomes even larger than the receiving time obtained at replicas. In this case, replicas will get negative OWD values, so the estimated OWD will be clamped to $D$ and piggybacked to the proxies. Then proxies will use $D$ as the latency bound. Since the proxies' clocks are already faster than the replicas' clocks, the request deadline will become much larger than the replicas' clock time when the request arrives at replicas, leading to long holding delay in the \emph{early-buffer}, and eventually causing much degradation of \sysname's latency performance. 

\subsubsection{Optimization: Bounding Latency Degradation}

Reviewing the cases described in \S\ref{sec-leader-bad-clock}-\S\ref{sec-proxy-bad-clock}, we note that the leader's slower clock and proxies' faster clocks can cause unbounded latency performance degradation to \sysname. Although such cases do not affect \sysname's correctness and can hardly be long-lasting in practice (because Huygens will keep monitoring its agents and correct the error bounds), we propose an optimization strategy to bound the latency even when such cases of bad clock synchronization become long-lasting. The key idea of the optimization is to let leader force the request to be committed in the slow path.

Recall in the design of DOM (\S\ref{sec-dom}), DOM-R will not accept the request into \emph{early-buffer} only if its deadline is smaller than the last released one, which is not commutative to this request (\S\ref{sec-commutativity-optimization}). We can enforce the entrance condition of \emph{early-buffer}: If the request's deadline is much larger than the current clock time of the leader, which means the request will suffer from a long holding delay if it is put into the \emph{early-buffer}, then the leader also modifies its deadline to be slightly larger than the last released one and then put it into the \emph{early-buffer}. This step is similar to \circled{3} in Figure~\ref{fig-workflow-slow}. The difference is, here we modify a large deadline to a smaller one so as to make it release from the \emph{early-buffer} earlier. By contrast, step \circled{3} in Figure~\ref{fig-workflow-slow} is to modify a small deadline to a larger one, so that it will not violate uniform ordering with previously released requests from the \emph{early-buffer}.

The effectiveness of the optimization is shown in Figure~\ref{fig-leader-clock} and Figure~\ref{fig-proxy-clock}. We configure a threshold for the leader replica: if the request's deadline is larger than the replica's current clock time by \SI{50}{\micro\second}, then the request will not be directly put into the \emph{early-buffer} (as the baseline does). Instead, the leader replica modifies the request's deadline to be slightly larger than the deadline of the last released request (which is not commutative to this request), so that the request can enter the leader's \emph{early-buffer} and be released much earlier without violating uniform ordering. Eventually, the request can be committed in the slow path. After installing the optimization strategy, we can see from Figure~\ref{fig-leader-clock} and Figure~\ref{fig-proxy-clock} that, the degradation of the latency performance becomes bounded, which provides \sysname with stronger resistance to bad clock synchronization. 

In this section, we only discuss the three typical cases. Theoretically, there exists some possibility that these cases can happen simultaneously, which creates even more complicated scenarios. For example, when proxies and the leader replica both have slower clocks, the effect due to the bad clock synchronization can be counteracted to some extent. However, the optimization strategy discussed here is still effective to bound the latency degradation and help \sysname to resist the impact of bad clock synchronization.    



\section{Derecho in the Public Cloud}
\label{sec-derecho-issue}
Derecho~\cite{tocs19-derecho,spindle} is a recent high-performance state machine replication system. It works with both RDMA and TCP, and achieves very high throughput with RDMA. Since Derecho is also deployable in public cloud (with TCP), we intended to compare \sysname with Derecho in the public cloud. 

We follow the guidelines from the Derecho team~\cite{derecho-deployment}: First, we try to tune the configuration parameters for Derecho and reproduce the performance number in~\cite{spindle} by using bare-metal machines. We set up a cluster in Cloudlab~\cite{atc19-cloudlab}. We use 3 \texttt{c6525-100g} instances (equipped with 100GB RDMA NICs) to form a Derecho subgroup size of 3. We use ramfs~\cite{ramfs-wiki} as the storage backend for Derecho to avoid disk writes. Then, we evaluate the throughput of Derecho in all-sender mode and one-sender mode. As for the all-sender mode, Derecho yields the throughput of 626K request/sec with 1KB message size and 634K request/sec with 100B message size. As for the one-sender mode, Derecho yields the throughput of 313K request/sec with 1KB message size and 305K request/sec with 100B message size. These numbers are close to the reported number in~\cite{spindle}, which convinces us that the configuration parameters have been properly set.

Then, we keep using the cluster and the configuration files for Derecho, but switch the backend from RDMA to TCP. After switching to TCP, we find Derecho's performance drops much: with 100B message size, the all-sender mode achieves the throughput of 17.4K request/sec with the median latency of \SI{2.33}{\milli\second}; the one-sender mode achieves the throughput of 5.68K request/sec with the median latency of \SI{2.35}{\milli\second}. The throughput becomes even lower after we move back to Google Cloud: with 100B message size, the all-sender mode achieves the throughput of 16.5K request/sec with the median latency of \SI{2.0}{\milli\second}; the one-sender mode achieves the throughput of 4.93K request/sec with the median latency of \SI{2.54}{\milli\second}. 

We speculate that the low performance of Derecho is due to \texttt{libfabric} it uses for communication. Although \texttt{libfabric} supports both RDMA and TCP communication, it is mainly optimized for RDMA, and the TCP backend is mainly used for test and debug~\cite{libfabric}. We expect Derecho can achieve much higher performance if equipped with a better TCP backend. Therefore, we think the comparison is unfair to Derecho and do not include it.


\section{Error Traces of Domino}
\label{error-trace-domino}
Domino~\cite{conext20-domino} is a recently proposed solution to achieve consensus with clock synchronization. When clock skew happens, clients may consider the request as committed, but eventually the request is lost from the replicas, leading to durability violation. The key reason for the durability violation is because clocks cannot always maintain monotonically increasing values. This lack of monotonicity is true both for NTP, used by Domino and Huygens, which \sysname uses. However, \sysname does not rely on clock monotonicity for correctness.\footnote{NTP is known to be unable to guarantee monotonicity~\cite{clock-monotonically,hotnet15_ntp_behavior, srds17_causalspartan}. Huygens synchronizes clocks by tuning the clock frequency instead of adding offsets to clock time. Thus it can provide monotonically increasing clock time if Huygens does not fail. However, when Huygens fails and is restarted, or when Huygens changes its reference clock, it cannot guarantee monotonicity either.} In this section, we will use an error trace to demonstrate Domino's durability violation due to the non-monotonicity of clock time. 


\textbf{Error Trace 1:} There are 5 replicas in Domino, and we denote them as R0-R4. Suppose R0 is the DFP (Domino’s Fast Paxos) leader and the others are the followers. There are two requests included in the trace, denoted as request-1 and request-2.
\begin{enumerate}
    \item R1-R5's clocks are synchronized. R1-R4 report their current clock time T to the coordinator R0, indicating they have accepted no-ops for all log positions before T (as described in 5.3.2 of Domino paper~\cite{conext20-domino}).
    \item R0 receives request-1 with predefined arrival time T+1. So R0 accepts this request.
    \item R0 intends to execute request-1. Before execution, R0 broadcasts request-1 with the other replicas.
    \item R1 and R2 also accept request-1 and reply to R0, whereas R3 and R4 do not receive request-1 from either the client or the replica because the request is dropped.
    \item R0 considers request-1 is committed because it has received the majority of replies (R1, R2 and itself). R0 considers it safe to execute the request, because R1-R4 have reported T to R0, and R1 and R2 also accept request-1.
    \item R0 executes the request, but has not broadcast the execution to learners (i.e. the other replicas). 
    \item R1 and R2 fail (i.e. so the NTP services of R1 and R2 also fail). When R1 and R2 are relaunched, the NTP services on their nodes are reinitialized, but the reinitialized NTP gives a time T1, which is smaller than T.
    \item R3 and R4's NTP services encounter a skew and get a clock time T2, which is smaller than T. 
    \item The client submits request-2, which has a pre-arrival time smaller than T but larger than both T1 and T2.
    \item R1-R4 all accept request-2 and send replies to the client. The client considers request-2 as committed. 
    \item R1-R4 waits for the notification from the coordinator R0, when the notification arrives, R1-R4 will do either (a) replace request-2 with no-op and only execute request-1 or (b) execute both requests but with request-2 first and request-1 second.
\end{enumerate}

The choice between (a) and (b) in Step 11 depends on how Domino implements the coordination between the leader and the other learners (followers), which is not shown in the Domino paper~\cite{conext20-domino}. We have studied the implementation~\cite{domino-repo} of Domino and found that, followers will choose (a) because DFP leader will also  broadcast the log positions (refer to \texttt{NonAcceptTime} variable in~\cite{domino-repo}) which the leader fills no-ops. When followers choose (a), Domino violates durability because request-2, which have been considered committed, is lost permanently. As an alternative, if followers choose (b), consistency will be violated: After the DFP leader (i.e., coordinator) fails and one replica among R1-R4 becomes the new leader, it will have different system state (which executes both request-2 and request-1) from the old leader (which only executes request-1)

Furthermore, the durability property is a necessary condition for the consistency and linearizability properties (as we defined in \S\ref{append-correctness}): 
\begin{itemize}
    \item Because one committed request can affect the execution result of the subsequent requests, the loss of it will lead to different execution results for the subsequent requests, thus violating consistency. 
    \item Because the committed request can be observed by clients, the loss of it causes contrary observation afterwards, thus violating linearizability.
\end{itemize}
Hence, Error Trace 1 has shown that Domino can violate durability, and consequently violate consistency and linearizability property.

Since the Domino paper~\cite{conext20-domino} does not specify the consistency model it targets, we speculate that Domino is based on eventual consistency~\cite{queue08_eventual_consistency, sosp95_eventual_consistency, sigmod13-bolt-on, jepson_consistency}, because the stronger consistency model, casual consistency,~\footnote{The three consistency models can be sorted from the weakest to the strongest: eventual consistency < causal consistency < linearizability. A more detailed comparison can be found in~\cite{jepson_consistency}.} is not guaranteed by Domino's protocol design: suppose one Domino client has two requests, i.e., request-1 and request-2. If request-1 influences the creation of request-2 but the client assigns a larger arrival time to request-1, then the other clients can see the result of request-2 earlier than the result of request-1. Thus Domino does not provide casual consistency.~\footnote{The example of casual consistency is adapted from the example in~\cite{sigmod13-bolt-on}.} However, eventual consistency cannot be guaranteed either when durability is violated~\cite{blog10_eventual_consistency_data_loss}.

\sysname avoids such error cases because \sysname exploits synchronized clocks to \emph{reduce packet reordering in the network, rather than to directly decide ordering with clock time}.  The design of the \emph{early-buffer} maintains the invariant of consistent ordering regardless of clock skew/failure, because the eligibility check for the request to enter the \emph{early-buffer} is to compare its deadline with the last released one (\S\ref{sec-dom}). Even after the replica fails and recovers, the consistent ordering invariant is still guaranteed: in this case, the \emph{last released request} is the last appended entry in the recovered \emph{log} (Step 9 in \S\ref{leader-change}). Therefore, \sysname's correctness is independent of the clock behavior. However, the clock synchronization indeed affects the performance of \sysname. For example, if the clock time of the replica becomes much faster and goes to a very large value, it can release some requests with very large deadlines. The large deadlines will be used in the eligibility check of the \emph{early-buffer}, making the subsequent requests unable to enter the \emph{early-buffer} and trigger the slow path more frequently. We have discussed in \S\ref{clock-eval} different cases of bad clock synchronization and their impact on \sysname.

By contrast, \emph{Domino directly uses the clock time for ordering}, and does not expect that the clocks can also give a smaller time than before, violating monotonicity (Step 7 and Step 8 in Error Trace 1), which leads to the incorrectness of the protocol.

\section{Formal Comparison of Different Primitives}
\label{appendix-formulation}

Concretely, the mostly ordered multicast (MOM) primitive~\cite{nsdi15-specpaxos} used by Speculative Paxos creates a network environment to make most requests arrive at all replicas in the same order. The ordered unreliable multicast (OUM) primitive~\cite{osdi16-nopaxos} used by NOPaxos ensures ordered delivery of requests without a reliability guarantee using a programmable switch as a request sequencer. By contrast, the deadline-ordered-multicast (DOM) primitive used by \sysname leverages clock synchronization to guarantee consistent ordering, so as to ease the work for replication protocols to achieve state consistency (i.e. to satisfy both consistent ordering and set equality). In this section, we aim to make a formal comparison among the three primitives. 

\subsection{Notation}
\begin{itemize}[leftmargin=*]
    \item Replicas: $R_1, R_2,\dots$ 
    \item Messages: $M_1, M_2, \dots$
    \item $a(M_i, R_k)$:  the arrival time of  $M_i$ at $R_k$. It is  the \textbf{reference time which is not accessible by replicas}, replicas can only get an approximate $\hat{a}(M_i,R_k)$ by calling their local clock API once $M_i$ arrives.
    \item $r(M_i,R_k)$: the reference time at which $M_i$ is released by the primitive to $R_k$'s protocol. It is the reference time. The primitive does not deliver a reference time  $r(M_i,R_k)$ to replication protocols, instead, it delivers an approximate $\hat{r}(M_i,R_k)$ by calling the clock API before releasing $M_i$.
    \item $S(M_i)$: the sequential number of $M_i$ given by Sequencer.(OUM Oracle Information). 
    \item $D(M_i)$: the \textbf{planned} deadline of $M_i$ to arrive at all replicas (DOM Oracle Information). Replicas know the value of $D(M_i)$ but cannot decide when is exactly the time point of $D(M_i)$ by simply checking their local clock. 
    
    
    
    
\end{itemize}



\subsection{Definition}
\label{def-primitive}
\begin{itemize}[leftmargin=*]
    \item Packet drop: $M_i$ is lost to $R_x$ if $r(M_i, R_x)=\infty$
    \item consistent ordering: $R_1$ and $R_2$ are said to be \emph{consistently ordered} (denoted as $UO(R_1,R_2,M_1,M_2)$) with respect to $M_1$ and $M_2$ if:
    \begin{itemize}
        \item $r(M_1, R_1)> r(M_2, R_1)$ and $r(M_1, R_2)> r(M_2, R_2)$
        \item Or $r(M_1, R_1)< r(M_2, R_1)$ and $r(M_1, R_2)< r(M_2, R_2)$
    \end{itemize}

    For simplicity, we omit discussing the edge case $r(M_1,R_1)=r(M_1,R_1)$ and/or $r(M_1,R_2)=r(M_1,R_2)$, which can be categorized into either of the two aforementioned outcomes. Similar edge cases are also omitted in the discussion of \S\ref{primitive-action}.
    
    \item Set equality: $R_1$ and $R_2$ are \emph{set-equal} with respect to $M_1$ (denoted as $SE(R_1,R_2, M_1)$) if
    \begin{itemize}
        \item $r(M_1, R_1)=\infty$ and $r(M_1, R_2)= \infty$
        \item Or $r(M_1, R_1)< \infty$ and $r(M_1, R_2)< \infty$
    \end{itemize}
    
    Set equality is similar to the term \emph{reliable delivery} in NOPaxos~\cite{osdi16-nopaxos}. While NOPaxos describes the property from the network perspective, our description is more straightforward by describing it from the replica perspective. 
    
    \item Consistency: $R_1$ and $R_2$ are consistent if
    \begin{equation*}
\begin{split}
\forall M_i,M_j:\quad 
  & UO(R_1,R_2,M_i,M_j) \\
  & \And SE(R_1,R_2,M_i) 
  \And  SE(R_1,R_2,M_j)
\end{split}
  \vspace{-0.2cm}
    \end{equation*}
  
   Satisfying both \emph{UO} and \emph{SE} property is equivalent to implementing an atomic broadcast primitive~\cite{atomic-broadcast-survey}, which is as hard as the consensus protocol.
\end{itemize}




\subsection{Primitive Actions}
\label{primitive-action}
Given a replica $R_k$, and two messages $M_1$ and $M_2$, we can formally describe the actions of the three primitives as follows.
\subsubsection{MOM}

\begin{equation}
\begin{split}
    &r(M_1, R_k) = a(M_1,R_k) \\
    &r(M_2, R_k) = a(M_2,R_k)  
\end{split}
\end{equation}
$r(*,*)$ is completely determined by $a(*,*)$ without guaranteeing consistent ordering.

\subsubsection{OUM}
Without loss of generality, 
the OUM Oracle gives $S(M_1)<S(M_2)$.


If $a(M_1,R_k)<a(M_2,R_k)$ (\textbf{Branch 1}), then
\begin{equation}
    \begin{split}
    &r(M_1,R_k)=a(M_1,R_k)\\
    &r(M_2,R_k)=a(M_2,R_k)
    \end{split}
    \label{eqn:oum1}
\end{equation}

Otherwise $a(M_1,R_k)>a(M_2,R_k)$ (\textbf{Branch 2}), then
\begin{equation}
    \begin{split}
    &r(M_1,R_k)= \infty\\
    &r(M_2,R_k)=a(M_2,R_k)
    \end{split}
    \label{eqn:oum2}
\end{equation}

Equation~\ref{eqn:oum1} captures the case where $M_{1}$ and $M_{2}$ arrive in an order \textit{consistent} with their sequence numbers. 

Equation~\ref{eqn:oum2} captures the case where $M_{1}$ and $M_{2}$ arrive in an order \textit{inconsistent} with their sequence numbers, in which case $M_{1}$ is immediately declared lost.

consistent ordering is guaranteed by OUM because messages arrive at different replicas either consistent with their sequence numbers (which are unique to a message and not a replica) or messages are declared lost.

\subsubsection{DOM}
\label{orm-form}
To simplify the following comparison analysis, we assume the local clock of each replica is monotonically increasing, which is a common assumption in clock modeling~\cite{podc84-clock-sync,lamport-clock-sync,LUNDELIUS1984190}. It is worth noting that, this assumption does not always hold in practice (we have explained this in \S\ref{error-trace-domino}). Here, we are not focusing on the protocol correctness (\sysname's correctness does not require this assumption). Instead, we only adopt this assumption to simplify the comparison between DOM and MOM/OUM. {Since the non-monotonicity of clock time occurs rarely, such rare cases will not affect the conclusions we derived on the \emph{overall performance} of DOM and \sysname}.


DOM can satisfy the monotonically increasing property as follows: DOM tracks the returned value every time it calls the clock API. If the returned value is smaller than the last one (i.e. violating the monotonically increasing property), DOM disposes of the value and retries the clock API. When the replica fails, DOM can rely on the replication protocol to recover the committed logs, and then it starts using the clock time which is larger than the deadline of the last log entry. In this way, DOM guarantees that each replica clock follows monotonically increasing property.


The \emph{monotonically increasing clock time} leads to the following fact:
\begin{equation}
\begin{split}
    r(M_1,R_k)<r(M_2,R_k) \iff
    \hat{r}(M_1,R_k)<\hat{r}(M_2,R_k) \\
    a(M_1,R_k)<a(M_2,R_k) \iff
    \hat{a}(M_1,R_k)<\hat{a}(M_2,R_k) \\
\end{split}
\label{eqn-fact}
\end{equation}

Without loss of generality, assume DOM Oracle decides two deadlines $D(M_1)$ and $D(M_2)$, satisfying $D(M_1)<D(M_2)$.

If $\hat{a}(M_1,R_k)<\hat{a}(M_2,R_k)$ or $\hat{a}(M_1,R_k)<{D}(M_2)$ (\textbf{Branch 3}), then
\begin{equation}
    \begin{split}
    &\hat{r}(M_1,R_k)=\max\{{D}(M_1), \hat{a}(M_1,R_k)\}\\
    &\hat{r}(M_2,R_k)=\max\{{D}(M_2), \hat{a}(M_2,R_k)\}
    \end{split}
    \label{eqn:orm1}
\end{equation}
Based on the condition and the formula, it is easy to check that $\hat{r}(M_1,R_k)<\hat{r}(M_2,R_k)$, thus ${r}(M_1,R_k)<{r}(M_{2},R_k)$.

Otherwise $\hat{a}(M_1,R_k)>\hat{a}(M_2,R_k)$ and $\hat{a}(M_1,R_k)>D(M_2)$ (\textbf{Branch 4}), then

\begin{equation}
    \begin{split}
    &\hat{r}(M_1,R_k)=\infty \\
    &\hat{r}(M_2,R_k)=\max\{D(M_2), \hat{a}(M_2,R_k)\}
    \end{split}
    \label{eqn:orm2}
\end{equation}

Equation~\ref{eqn:orm1} captures the cases where either $M_{1}$ arrives  before $M_{2}$ arrives, or $M_{1}$ arrives before $M_{2}$'s deadline (based on the local clock of the replica). In both cases, $M_{1}$ can be released to the protocol before $M_{2}$ is released.

Equation~\ref{eqn:orm2} captures what happens when $M_{1}$ arrives after both $M_{2}$'s deadline and $M_{2}$'s arrival. Here, $M_{1}$ has to be declared lost (so that both $\hat{r}(M_1,R_k)$ and $r(M_1,R_k)$ are $\infty$) and $M_{2}$ is released to the protocol.

consistent ordering is guaranteed by DOM because messages are released to replicas according to their deadline's order. Those which have violated the increasing deadline order will not be released by DOM and should be handled by the replication protocol.  



\subsection{Understanding Difference between MOM, OUM and DOM}

\begin{figure}[!htbp]
    \centering
    \includegraphics[width=8cm]{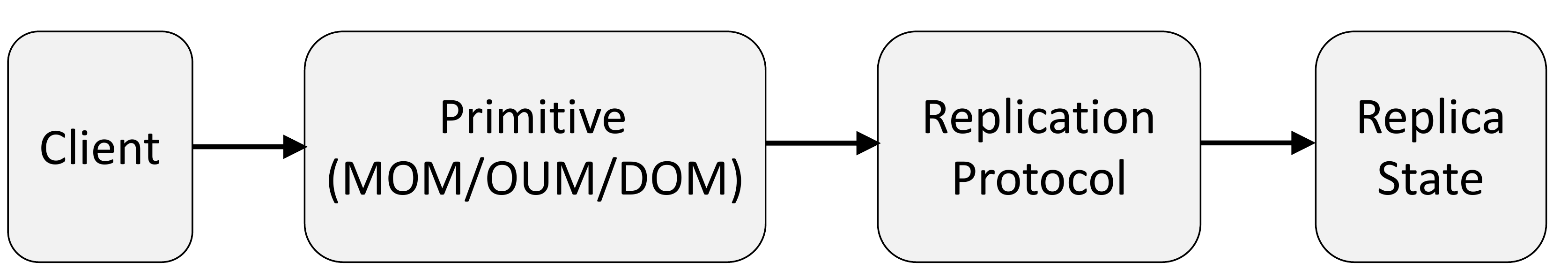}
    \caption{General Model of Speculative Paxos/NOPaxos/\sysname}
    \label{prim-model}
\end{figure}

As shown in Figure~\ref{prim-model}, primitives are decoupled from the replication protocol. None of the primitives guarantees consistency defined in \S\ref{def-primitive}.
The primitives are just used to create favorable message sequences for the replication protocol to achieve consistency more efficiently. 

Considering two replicas $R_1$ and $R_2$, with $M_1$ and $M_2$ arriving at both replicas. We aim to study the question: When equipped with MOM/OUM/DOM, how likely (easily) can $R_1$ and $R_2$ reach consistency without sacrificing liveness (i.e. both $M_1$ and $M_2$ should appear in the consistent replica state) given the message sequences output from the primitive?

\begin{itemize}[wide, labelwidth=!,nosep]
    \item MOM simply relies on the highly-engineered network to remove inconsistency flavor, and its output is exactly the output of the network. There is no guarantee on either consistent ordering or set equality. When it comes to the general network (e.g. public cloud), Rarely both $r(M_1,R_1)<r(M_2,R_1)$ and $r(M_1,R_2)<r(M_2,R_2)$ (or the other direction $r(M_1,R_1)>r(M_2,R_1)$ and $r(M_1,R_2)>r(M_2,R_2)$) are satisfied at the same time, thus most consensus work still needs to be undertaken by the replication protocol.
    
    \item OUM is potentially better than MOM, because it does the serialization between the clients and replicas with a standalone sequencer, so that the reordering occurrence in the path between clients and the sequencer does not matter. However, when reordering happens in the path between the sequencer and replicas, it leads to \textbf{Branch 2}, thus the replication protocol (e.g. NOPaxos) has to handle the loss of $M_1$ for the affected replicas (e.g. fetching from other replicas or starting gap agreement). Although the consistency property is still satisfied if all replicas take \textbf{Branch 2}, that leads to liveness problem: the client which submits $M_1$ has to retry a new submission. If it goes to the extreme case, when clients submit a series of requests and only the one with the largest sequential number arrives first on all replicas, then all the other requests are declared loss by OUM. In this case, the replicas reach consistency, but little progress is made.
    
    \item DOM performs better than OUM in general network because it maintains stronger resistance to reordering. Based on the Equation~\ref{eqn-fact}, we can easily derive that \textbf{Branch 3} of DOM is a superset of \textbf{Branch 1} of OUM. In other words, replicas equipped with DOM are more likely to take DOM's ``good'' branch (i.e. \textbf{Branch 3}), whereas replicas equipped with OUM are less likely to take OUM's ``good'' branch (i.e. \textbf{Branch 1}). However, DOM's strong resistance is obtained at the expense of extra pending delay. According to Equation~\ref{eqn:orm2}, even when $M_1$ and $M_2$ come in order and before their deadlines, they still need to be held until $D(M_1)$ and $D(M_2)$. By contrast, OUM can immediately present $M_1$ and $M_2$ to the replication protocol, according to Equation~\ref{eqn:oum1}.

\end{itemize}

\subsection{Why does Clock Synchronization Matter to DOM?}
Clock synchronization affects the effectiveness of DOM for two reasons. First, clock synchronization affects whether DOM can resist the reordering. Second, clock synchronization is closely related to the measurement of client-replica one-way delay, thus (indirectly) affecting whether the client can decide a proper deadline for its messages (requests). We use two cases to illustrate how bad clock synchronization and bad deadlines can affect DOM's effectiveness, and use one case to illustrate the effective DOM with good clock synchronization and proper deadlines.

\textbf{Bad Case-1:} Bad clock synchronization. $M_1$ and $M_2$ arrive at $R_1$ out of order but $a(M_1,R_1)<D(M_2)$. Meanwhile, the two messages arrive at the other replicas in order. If $R_1$'s clock had been well synchronized with the reference clock, $\hat{a}(M_1,R_1)$ should be very close to $a(M_1,R_1)$, leading to $\hat{a}(M_1,R_1)<D(M_2)$, and then DOM should be able to rectify the reordering on $R_1$, so that it outputs the consistent message sequence as the others. However, $R_1$'s clock fails at that time and gives a very large $\hat{a}(M_2,R_1)$ that leads to $\hat{a}(M_2,R_1)>D(M_3)$. In this case, DOM becomes ineffective and $R_1$ takes \textbf{Branch 4}, leaving more consensus work for the replication protocol to complete. 

\textbf{Bad Case-2:} Improper deadline. Suppose the clock synchronization goes wrong on some replicas (e.g. $R_2$), and  the clocks on the problematic replicas are much faster than the reference clock, so the one-way delay (OWD) measurement gives very large value and elevates the latency bound estimation (\S\ref{sec-dom}). When $M_1$ and $M_2$ are given very large deadlines $D(M_1)$ and $D(M_2)$. The replicas (e.g. $R_1$) will take \textbf{Branch 3} and DOM is able to rectify possible reordering. However, $M_1$ suffers from the pending time of $D(M_1)-\hat{r}(M_1,R_1)$ whereas $M_2$ suffers from the pending time of $D(M_2)-\hat{r}(M_2,R_1)$ on $R_1$ (Assume $R_1$'s clock is well synchronized with the reference clock).

\textbf{Good Case:} Clocks are well synchronized and $D(M_i)$s are properly decided, i.e. $D(M_i)$ is close to (but slightly larger than) the arrival time $a(M_i, R_x)$ regarding most replicas.  In this case, when the network is good, DOM delivers the message to the replication protocol with both consistent ordering and set equality, just like MOM and OUM. More than that, when the network causes message reordering, both MOM and OUM will present the reordering effect to the replication protocol, and triggers the replication protocol to take extra effort. Specifically, MOM presents non-uniformly ordered messages to the replication protocol which causes Speculative Paxos to go to the slow path and costly rollback; OUM presents consistent order messages with gaps (equation~\ref{eqn:oum2}), which also causes NOPaxos to go to the slow path and make the following messages pending before the gap is resolved. By contrast, so long as the out-of-order message ($M_1$) does not break the deadline ($D(M_2)$) of the message ($M_2$), the reordering between $M_1$ and $M_2$ can be rectified by DOM in equation~\ref{eqn:orm1} and is insensible to the replication protocol, so that the workload of replication protocol (\sysname) is much relieved.

\section{\sysname vs. EPaxos in WAN}
\label{comp-epaxos}

EPaxos~\cite{sosp13-epaxos,nsdi21-epaxos} is a consensus protocol that proves to outperform Multi-Paxos in Wide Area Netowrk (WAN) scenario. EPaxos fully exploits the fact that Local Area Network (LAN) message delays are negligible when compared with WAN message delays. Therefore, EPaxos distributes its replicas across multiple zones. Such design enjoys two benefits: First, the long-distance (cross-zone) communication between replicas is fully controlled by the service providers, so the service providers can use private backbone network to provide better quality of service. By contrast, if replicas are co-located together and far away from clients. The long distance from clients to replicas is out of control, which may cause longer latency and more frequent message drop. Second, Although EPaxos also incurs 2 RTTs in the fast path, one of them is LAN RTT (i.e. client$\xrightarrow[]{}$replica and replica$\xrightarrow[]{}$client message delays) that can be ignored. Therefore, EPaxos claims to achieve optimal RTT (1 WAN RTT) in the fast path and 2 RTTs in the slow path, which makes it outperform Multi-Paxos in latency. Besides, by using the multi-leader design and commutativity, EPaxos also enjoys less throughput bottleneck compared with Multi-Paxos.

While in this paper we mainly focus on LAN deployment and have shown that \sysname outperformed TOQ-EPaxos in LAN (Figure~\ref{fig-latency-tp}), we would like to highlight that \sysname is also deployable in WAN environment, and we have demonstrated in Figure~\ref{fig:wan-comp} that, \sysname can also earn more advantages over EPaxos when deployed in WAN. We analyze the advantages below. 

\subsection{Latency} 
When deployed in WAN, \sysname shares the same benefit as EPaxos: \sysname deploys its stateless proxies in every zone, so the client$\xrightarrow[]{}$proxy and  proxy$\xrightarrow[]{}$client message delays are also LAN message delays that can be ignored. Therefore, \sysname also achieves 1 WAN RTT as EPaxos, but \sysname achieves only 1.5 WAN RTTs in the slow path, compared with 2 WAN RTTs achieved by EPaxos. 

Besides, \sysname earns more performance advantages over EPaxos when there are more zones than replicas. For instance, consider a 3-replica consensus protocol with 10 different zones, and clients are evenly distributed in every zone, EPaxos cannot benefit all clients regarding the latency. Since there are only three replicas, at most the clients in three zones can enjoy 1 WAN RTT to commit their requests in the fast path. The majority of clients (70\%) still suffer 2 WAN RTTs to commit in the fast path, and even worse (3 WAN RTTs) to commit in the slow path. The large number of zones makes EPaxos lose most of its latency benefit. To let all clients enjoy 1 WAN RTT fast path, EPaxos has to deploy one replica in each zone (i.e. 10 replicas), but in that case, the quorum check will become much heavier and more interference/conflicts among replicas can occur. In contrast, \sysname \emph{distributes proxies instead of replicas across zones, and proxies are highly scalable}. Regardless of the number of zones, \sysname can still maintain 1 WAN RTT for all clients, so long as sufficient proxies are deployed in every zone. 

Besides, when data center failure is not considered (i.e. the number of zone failures is assumed to be 0), \sysname can even co-locate all replicas in the same zone and connect them with high-end communication (e.g. DPDK, RDMA). In this case, inter-replica communication is also LAN message delays, and \sysname can achieve optimal WAN RTT (1 WAN RTT) for both fast path and slow path, which gives \sysname more latency benefit than EPaxos. 

\subsection{Throughput}
While EPaxos uses multiple leaders to mitigate single-leader bottleneck, \sysname adopts an alternative design: \sysname still maintains single leader but offloads most workload to proxies. The proxy brings two major advantages for \sysname regarding the throughput. First, the inter-replica communication is much more lightweight because the leader only multicast index messages (rather than request messages) to other followers, which have much smaller sizes than requests and can be batched to amortize the communication cost. Second, replicas do not undertake quorum checks, and proxies can conduct the quorum check concurrently. Although EPaxos can share the workload of request multicast and quorum check among replicas, the number of replicas is limited and it is still likely that the quorum check workload can overwhelm the capacity of multiple leaders. However, the number of proxies in \sysname can be considered without constraint (i.e. as many as Huygens can support), and \sysname can deploy as many proxies as needed to tackle the workload of request multicast and quorum check. Therefore, we expect \sysname can also achieve higher throughput than EPaxos.

\subsection{Clock Synchronization in WAN}

As mentioned in our paper, the performance of \sysname is closely related to the synchronization performance of clocks. A reasonable concern about deploying \sysname in WAN is that the clock error can become very large and cause performance degradation. Such concerns prove to be unnecessary. According to the discussion with the developer team of Huygens, when deployed in public cloud across multiple data centers, the clock accuracy provided by Huygens will be in  the order of 10s of microseconds, with occasional spikes if the WAN link is unstable. Such claims have been verified in \cite{nsdi21-epaxos} and our experiments (\S\ref{sec:wan-comp}), which evaluate Huygens in the WAN setting and observes the clock offsets between \SI{20}{\micro\second} and \SI{1}{\milli\second}. Considering the inter-datacenter latency is usually tens of or even hundreds of milliseconds (as shown in Figure 5 of \cite{nsdi21-epaxos}), the synchronization performance of Huygens is sufficient for \sysname to achieve fast consensus in WAN. 

\section{Nezha TLA+ Specification}
\label{sec-nezha-tla}
The TLA+ specification of \sysname is available at the anonymous repository 
\url{https://github.com/Steamgjk/Nezha/blob/main/docs/Nezha.tla}.

\section{Illustration of Message Delays in Different Protocols}
\label{sec-protocol-message-delay}
We illustrate the workflow for each of the protocols we listed in Table~\ref{tab:cmp}.

\begin{figure}[H]
\vspace{-0.5cm}
\centering
    \includegraphics[width=6cm]{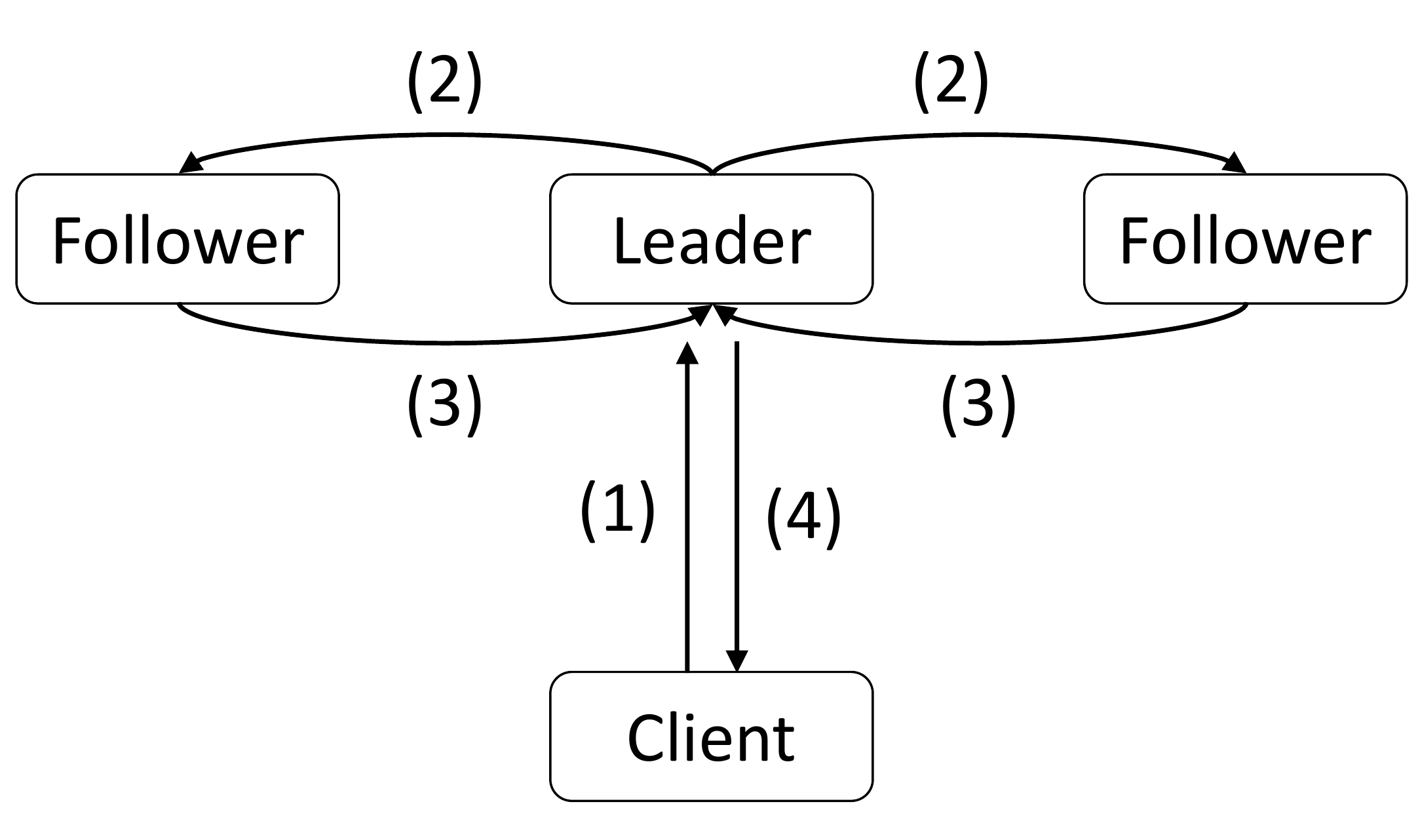}
    \caption{Message Delay of Multi-Paxos}
    \label{fig:mp-md}
\end{figure}

Multi-Paxos/Raft requires 4 message delays to commit a request (Figure~\ref{fig:mp-md}).

\begin{figure}[H]
    \centering
    \includegraphics[width=6cm]{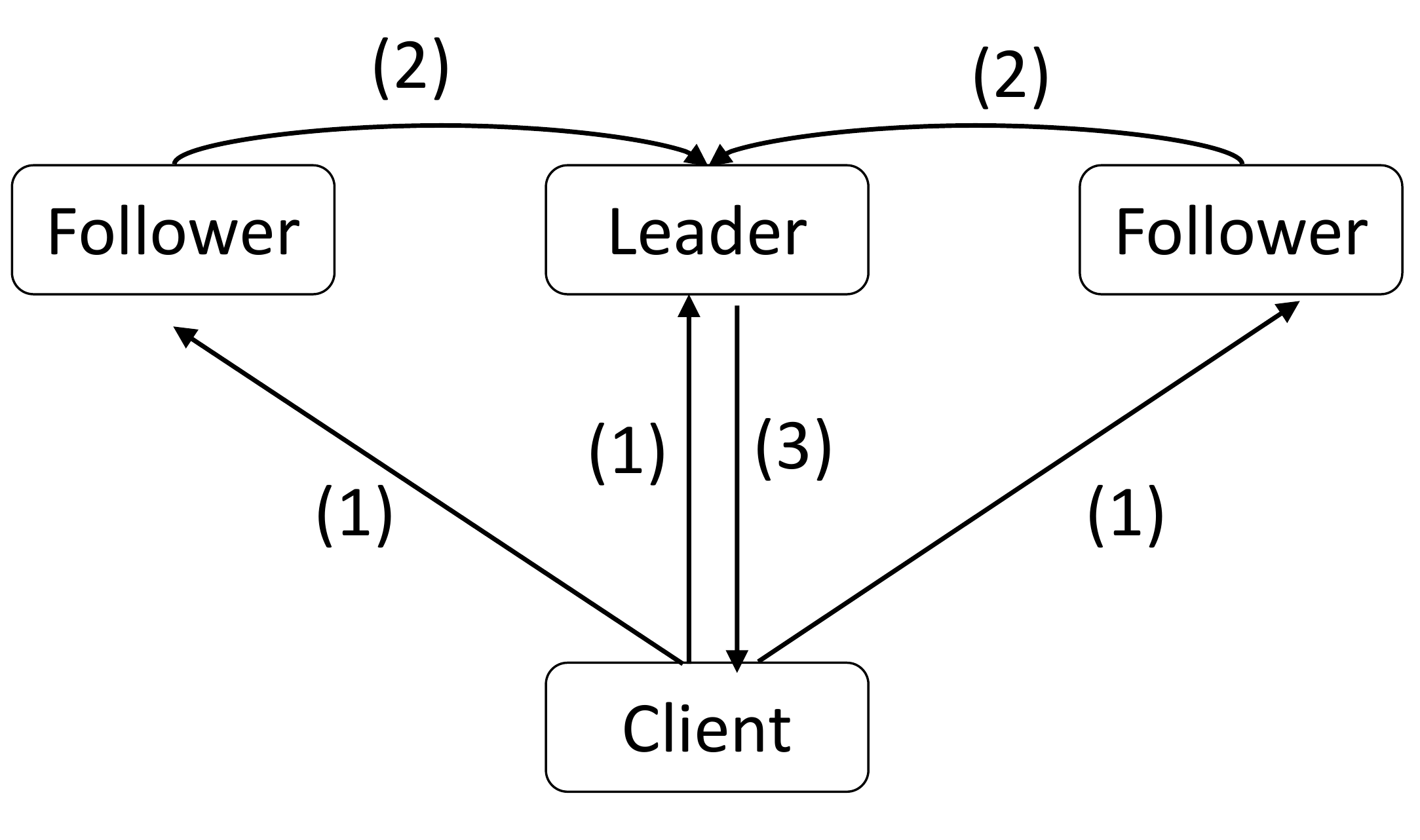}
    \caption{Message Delay of Fast Paxos (Fast Path)}
    \label{fig:fp-fast-md}
\end{figure}

\begin{figure}[H]
    \centering
    \includegraphics[width=6cm]{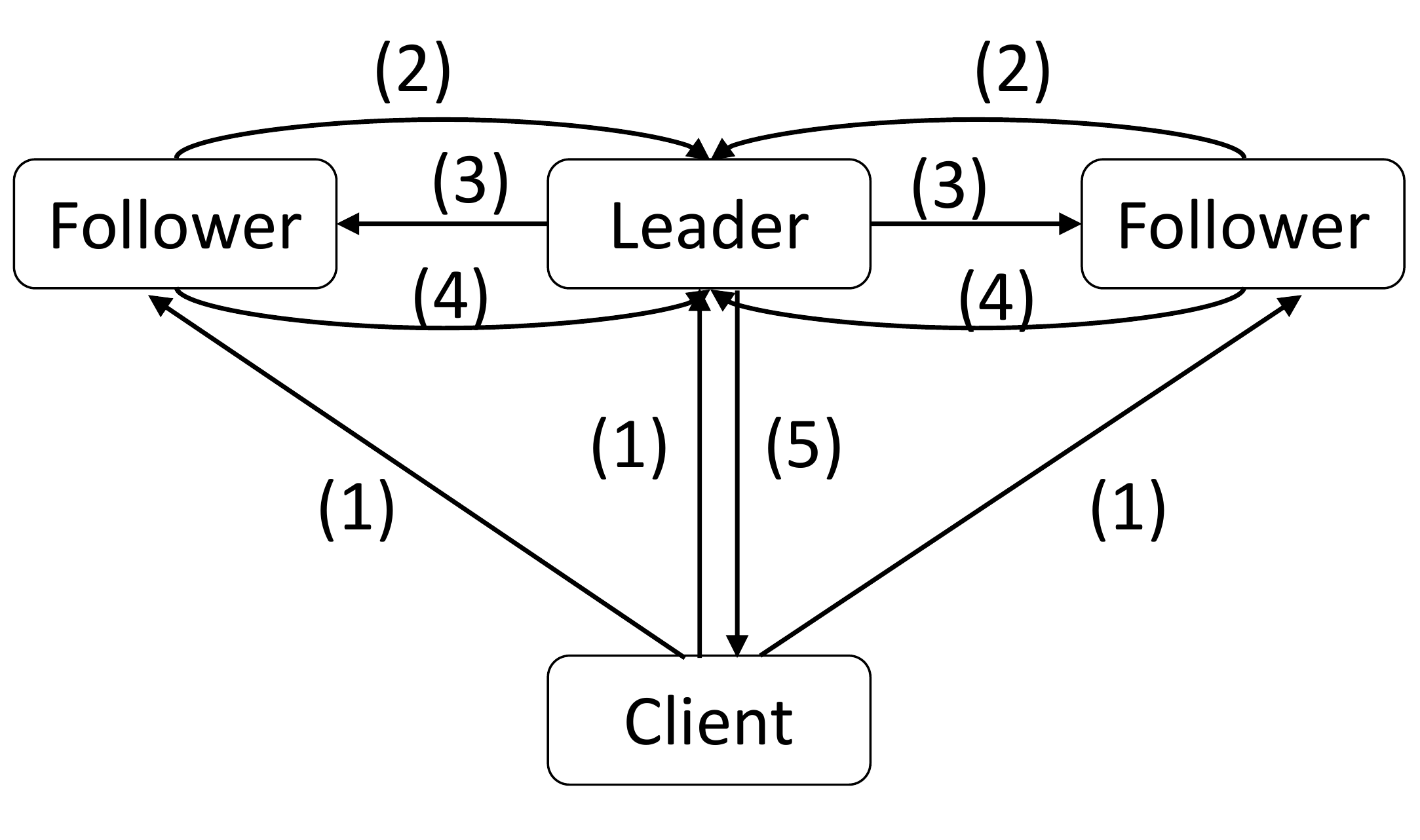}
    \caption{Message Delay of Fast Paxos (Slow Path)}
    \label{fig:fp-slow-md}
\end{figure}

Fast Paxos requires 3 message delays to commit a request in the fast path (Figure~\ref{fig:fp-fast-md}) and 5 message delays in the slow path (Figure~\ref{fig:fp-slow-md}).

\begin{figure}[H]
\vspace{-0.2cm}
    \centering
    \includegraphics[width=6cm]{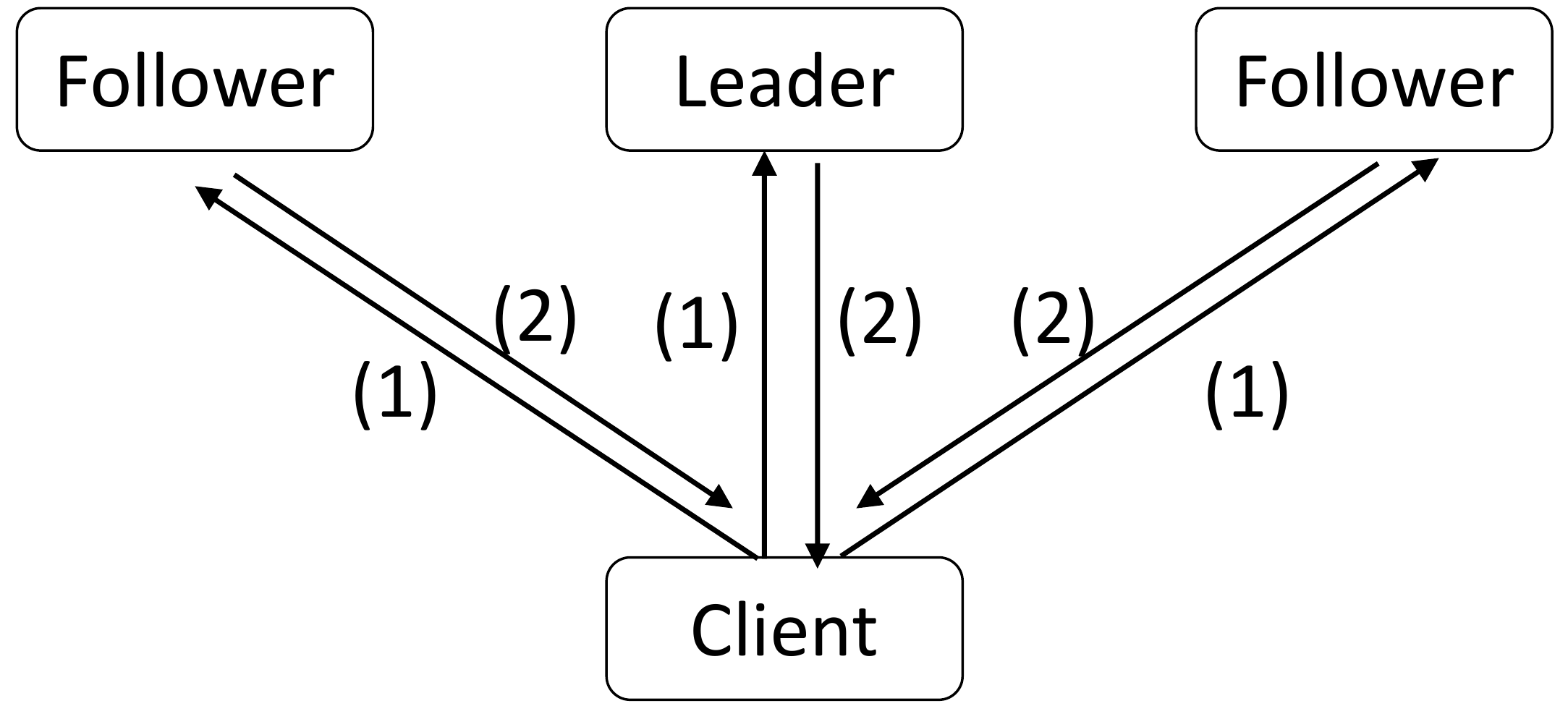}
    \caption{Message Delay of Speculative Paxos (Fast Path)}
    \label{fig:spec-fast-md}
\end{figure}

\begin{figure}[H]
\vspace{-0.5cm}
    \centering
    \includegraphics[width=6cm]{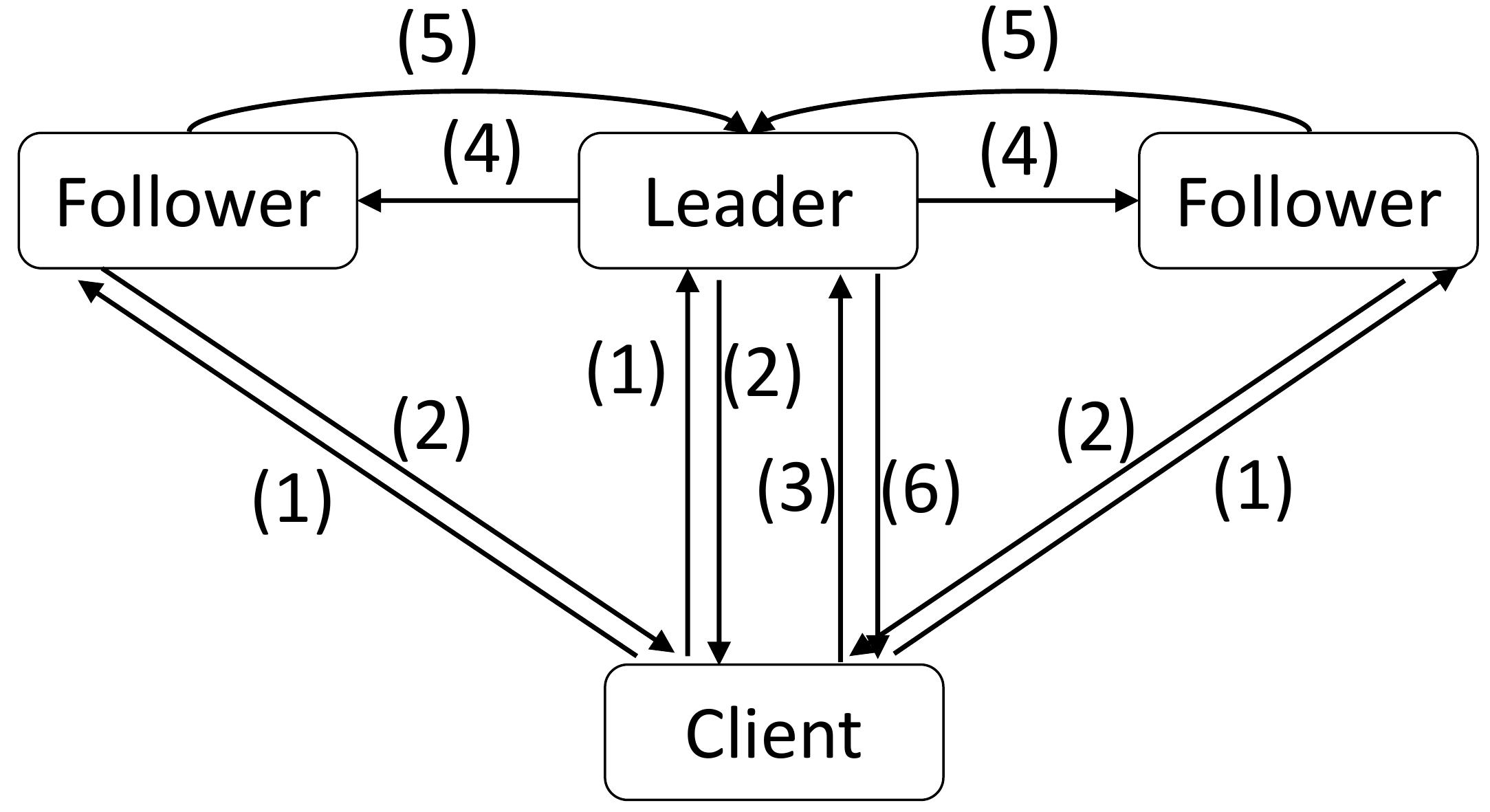}
    \caption{Message Delay of Speculative Paxos (Slow Path)}
    \label{fig:spec-slow-md}
\end{figure}

Speculative Paxos requires 2 message delays to commit a request in the fast path (Figure~\ref{fig:spec-fast-md}) and 6 message delays in the slow path (Figure~\ref{fig:spec-slow-md}).

\begin{figure}[H]
    \centering
    \includegraphics[width=8cm]{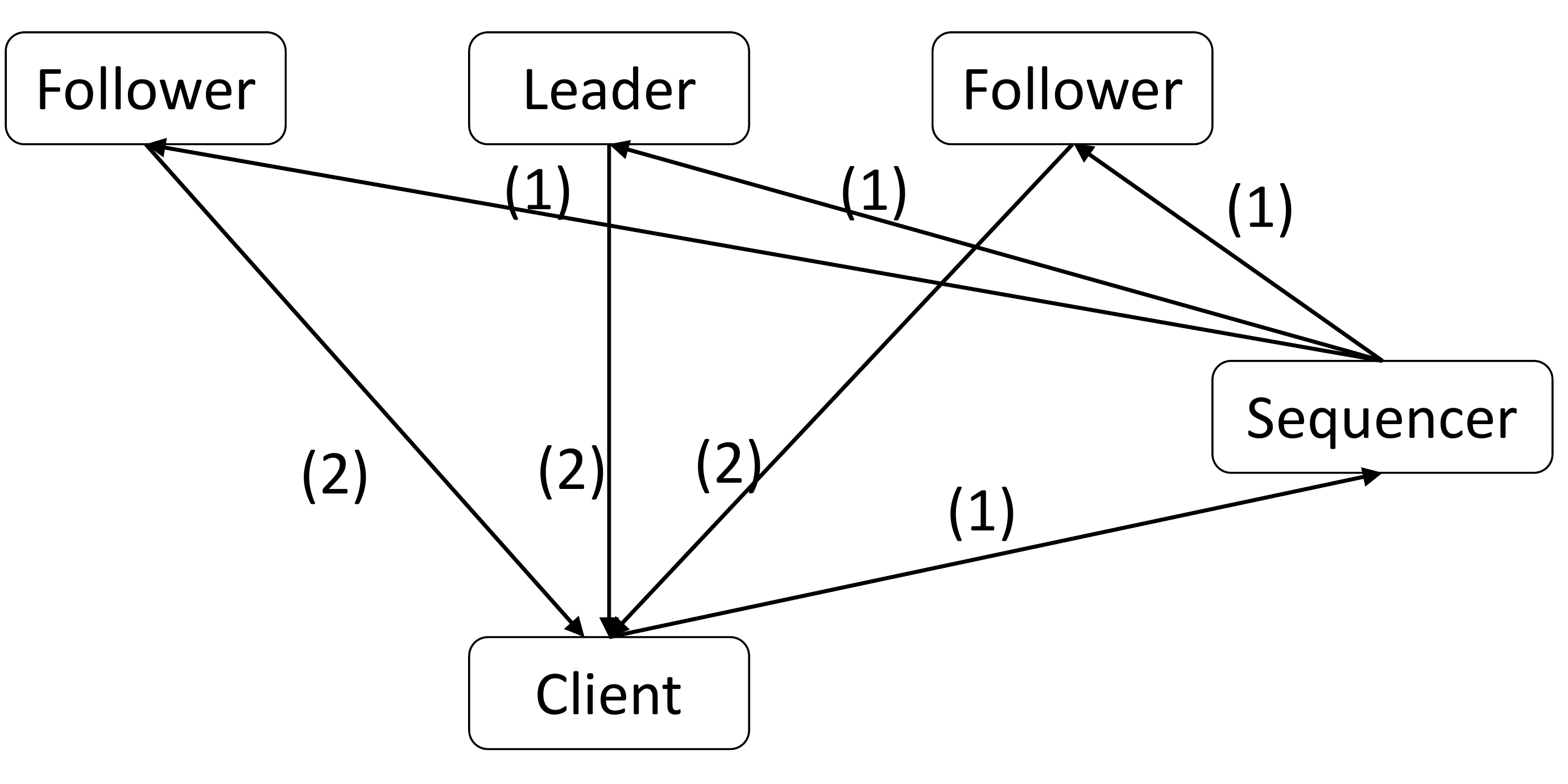}
    \caption{Message Delay of NOPaxos (Fast Path)}
    \label{fig:nopaxos-fast-md}
\end{figure}

\begin{figure}[H]
    \centering
    \includegraphics[width=8cm]{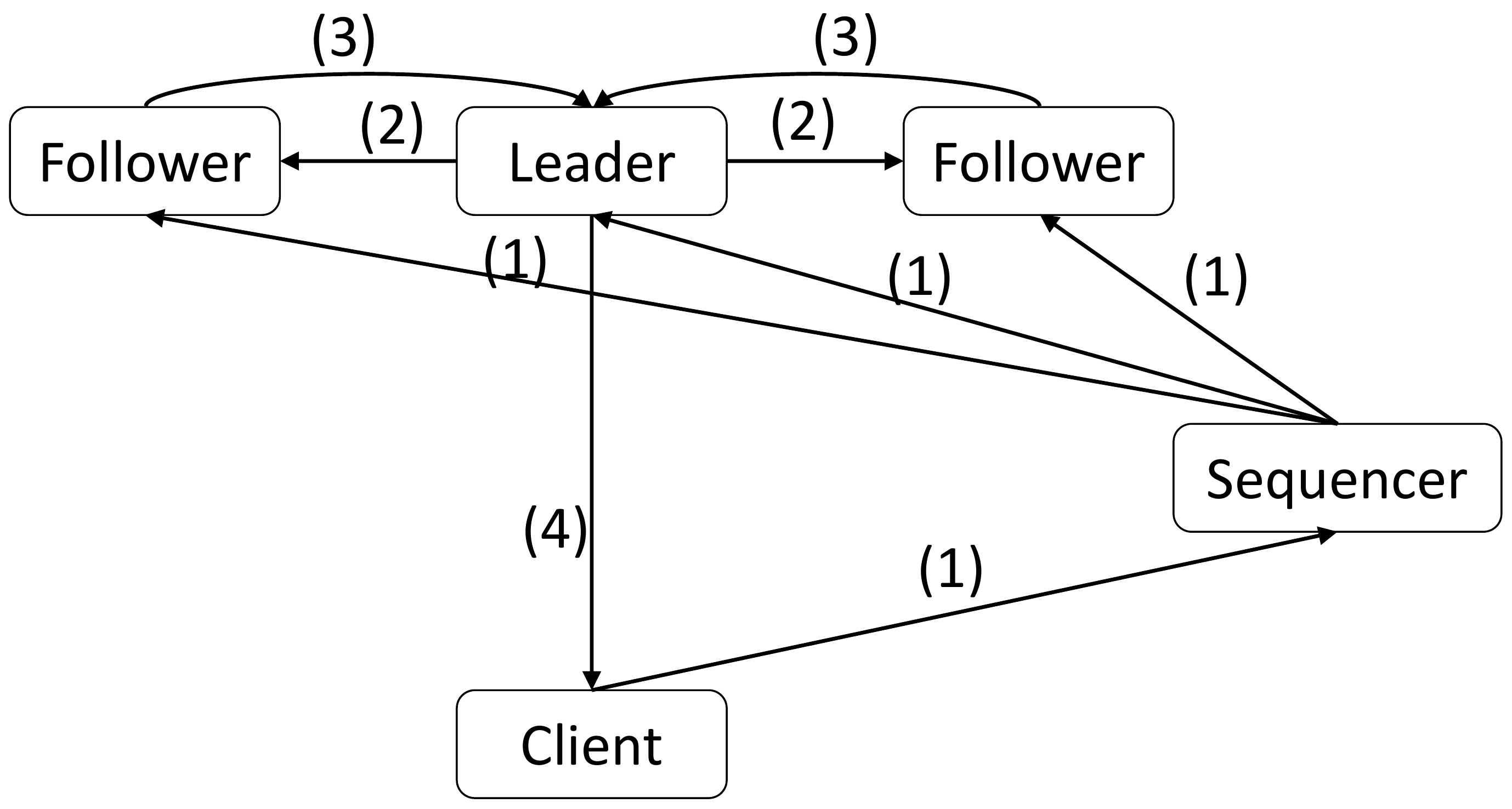}
    \caption{Message Delay of NOPaxos (Slow Path)}
    \label{fig:nopaxos-slow-md}
\end{figure}

NOPaxos requires 2 message delays to commit a request in the fast path (Figure~\ref{fig:nopaxos-fast-md}) and 4 message delays in the slow path (Figure~\ref{fig:nopaxos-slow-md}). Note that when the sequencer is a software sequencer (e.g., a standalone VM) instead of a hardware sequencer (e.g, programmable switch), there will be one extra message delay added to both the fast path and the slow path.

\begin{figure}[H]
    \centering
    \includegraphics[width=6cm]{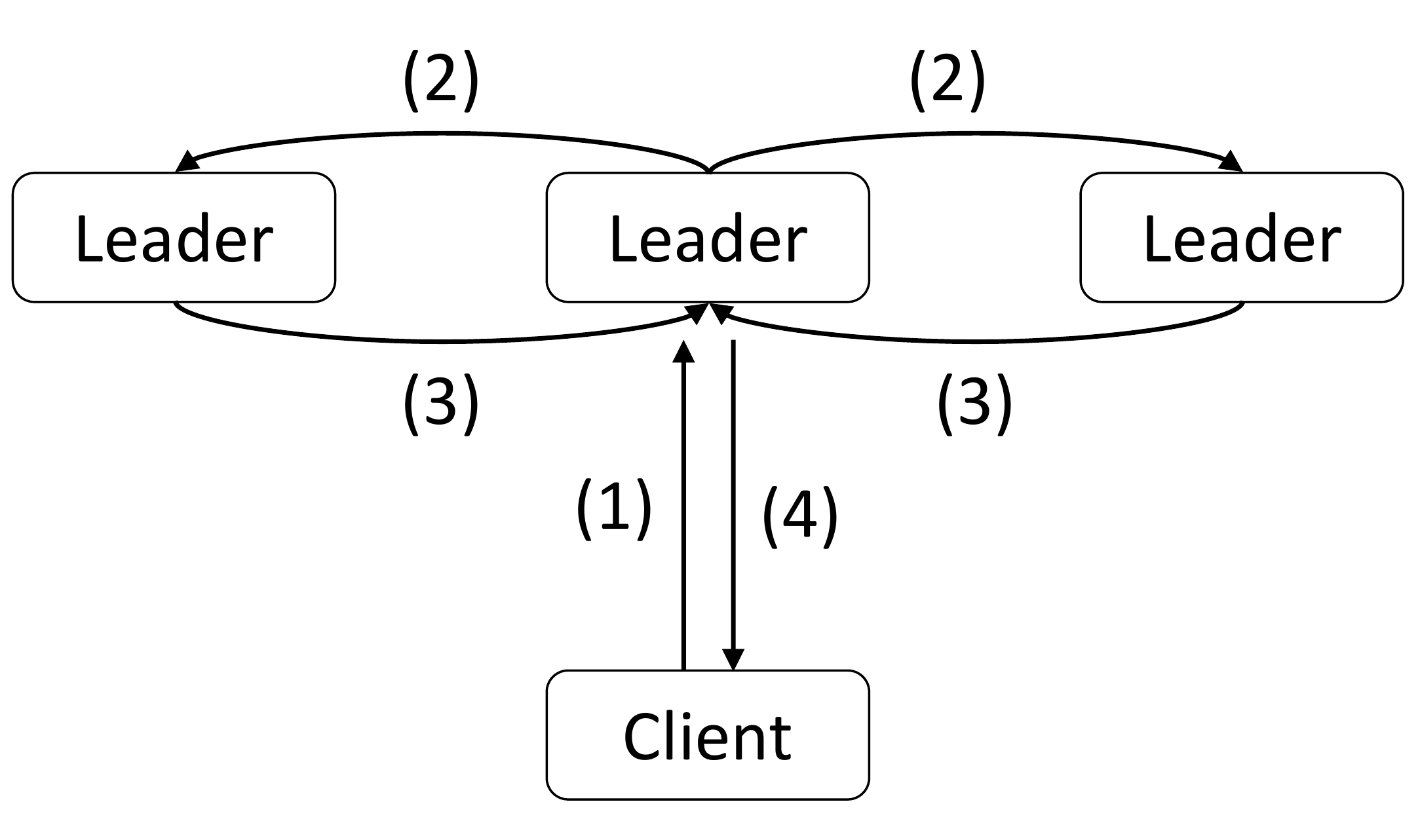}
    \caption{Message Delay of Mencius (Fast Path)}
    \label{fig:mencius-fast-md}
\end{figure}

\begin{figure}[H]
    \centering
    \includegraphics[width=6cm]{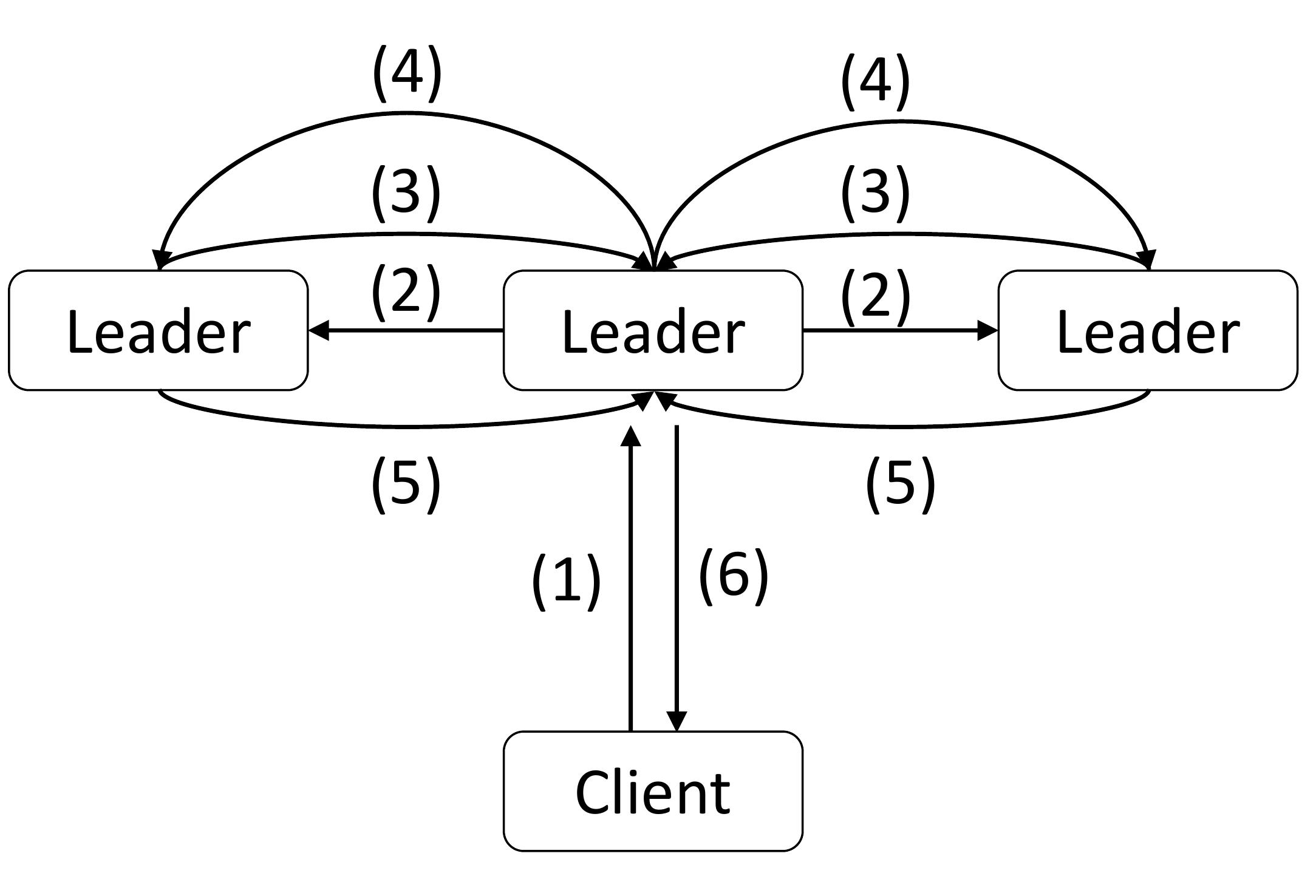}
    \caption{Message Delay of Mencius (Slow Path)}
    \label{fig:mencius-slow-md}
\end{figure}

Mencius requires 4 message delays to commit a request in the fast path (Figure~\ref{fig:mencius-fast-md}) and 6 message delays in the slow path (Figure~\ref{fig:mencius-slow-md}), because the leader who is responsible to commit the request needs to take one extra round (2 message delays) to coordinate with the other leader replicas.

\begin{figure}[H]
    \centering
    \includegraphics[width=6cm]{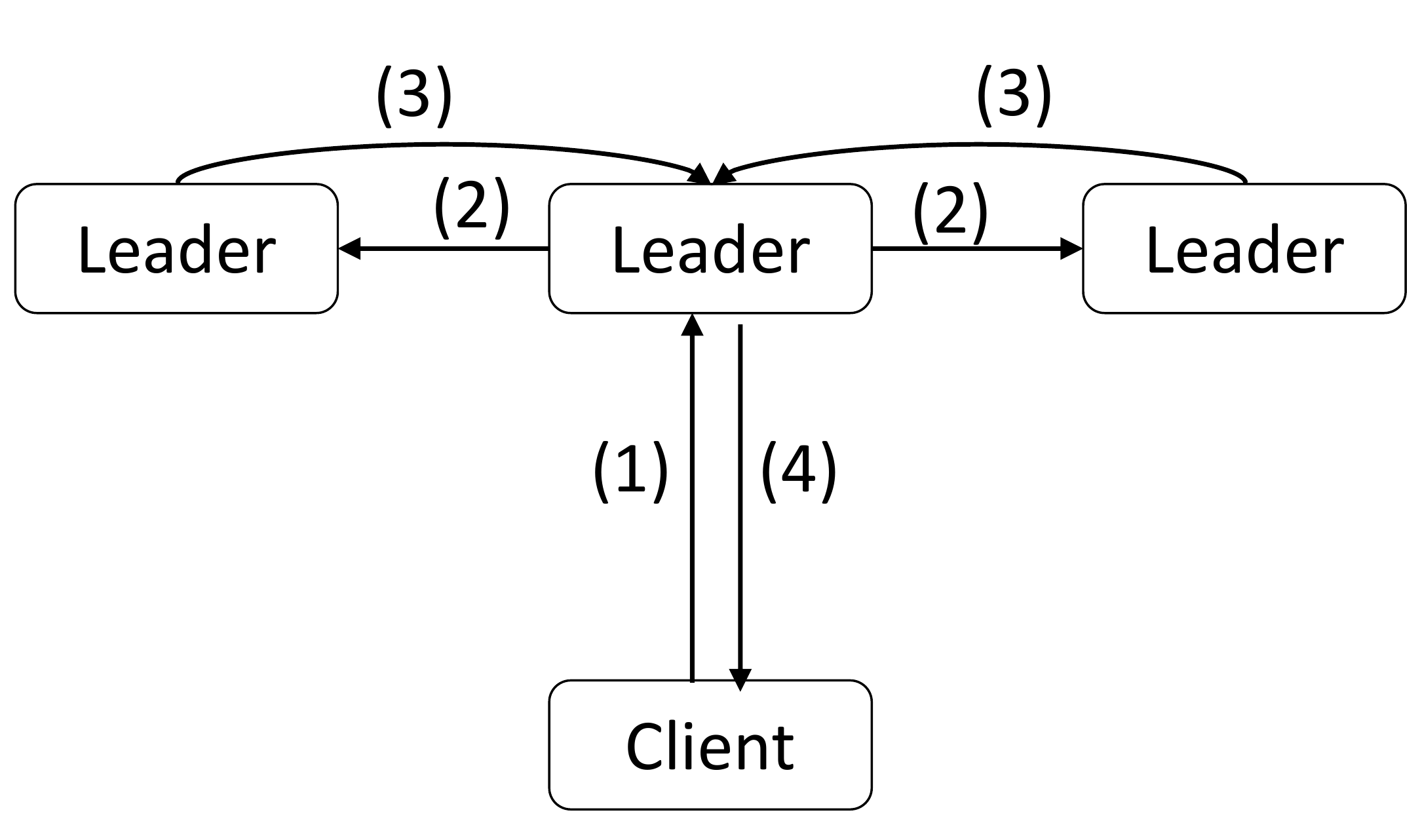}
    \caption{Message Delay of EPaxos (Fast Path)}
    \label{fig:epaxos-fast-md}
\end{figure}

\begin{figure}[H]
    \centering
    \includegraphics[width=6cm]{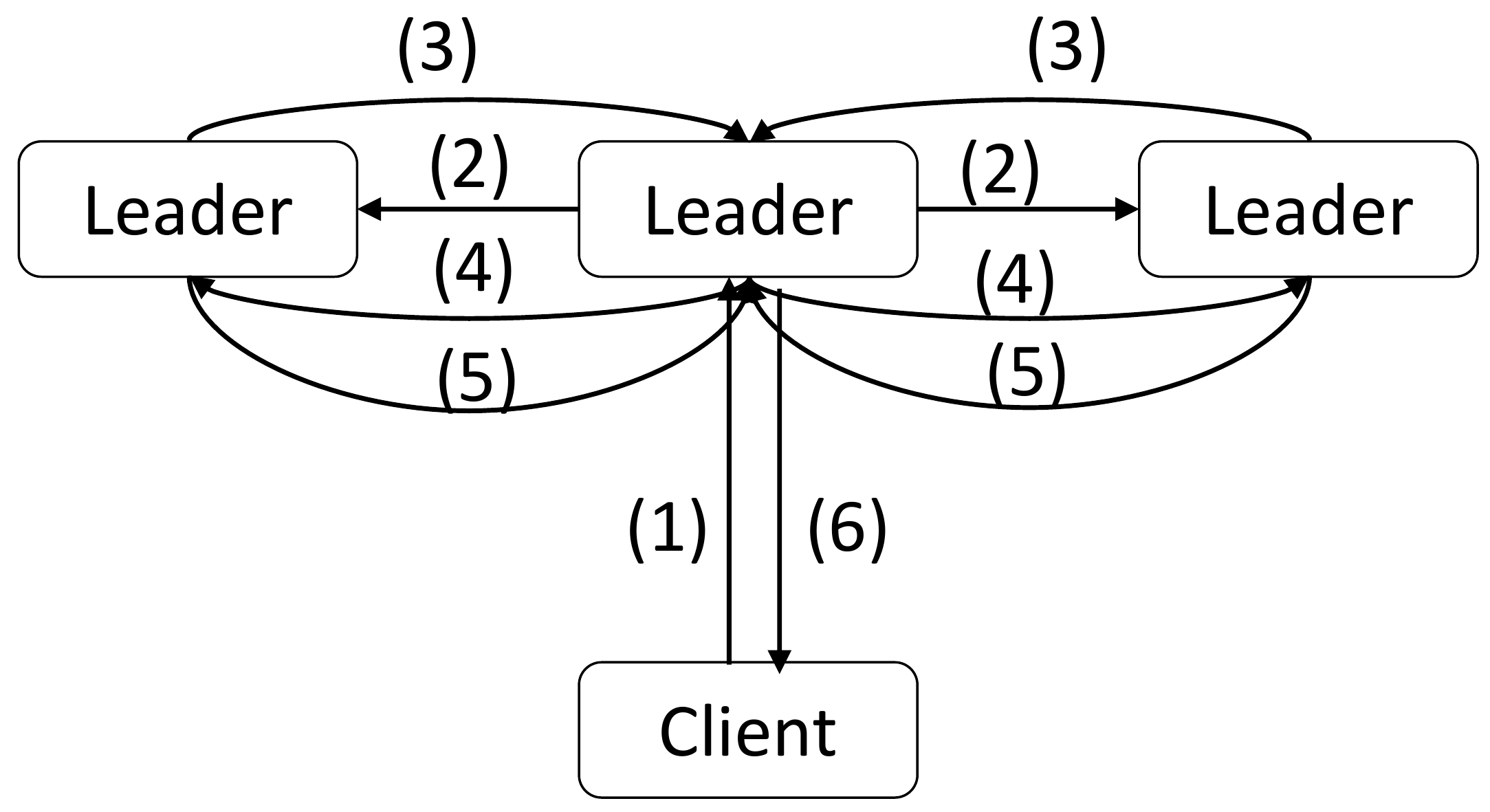}
    \caption{Message Delay of EPaxos (Slow Path)}
    \label{fig:epaxos-slow-md}
\end{figure}

EPaxos requires 4 message delays to commit a request in the fast path (Figure~\ref{fig:epaxos-fast-md}) and 6 message delays in the slow path (Figure~\ref{fig:epaxos-slow-md}).

\begin{figure}[H]
    \centering
    \includegraphics[width=6cm]{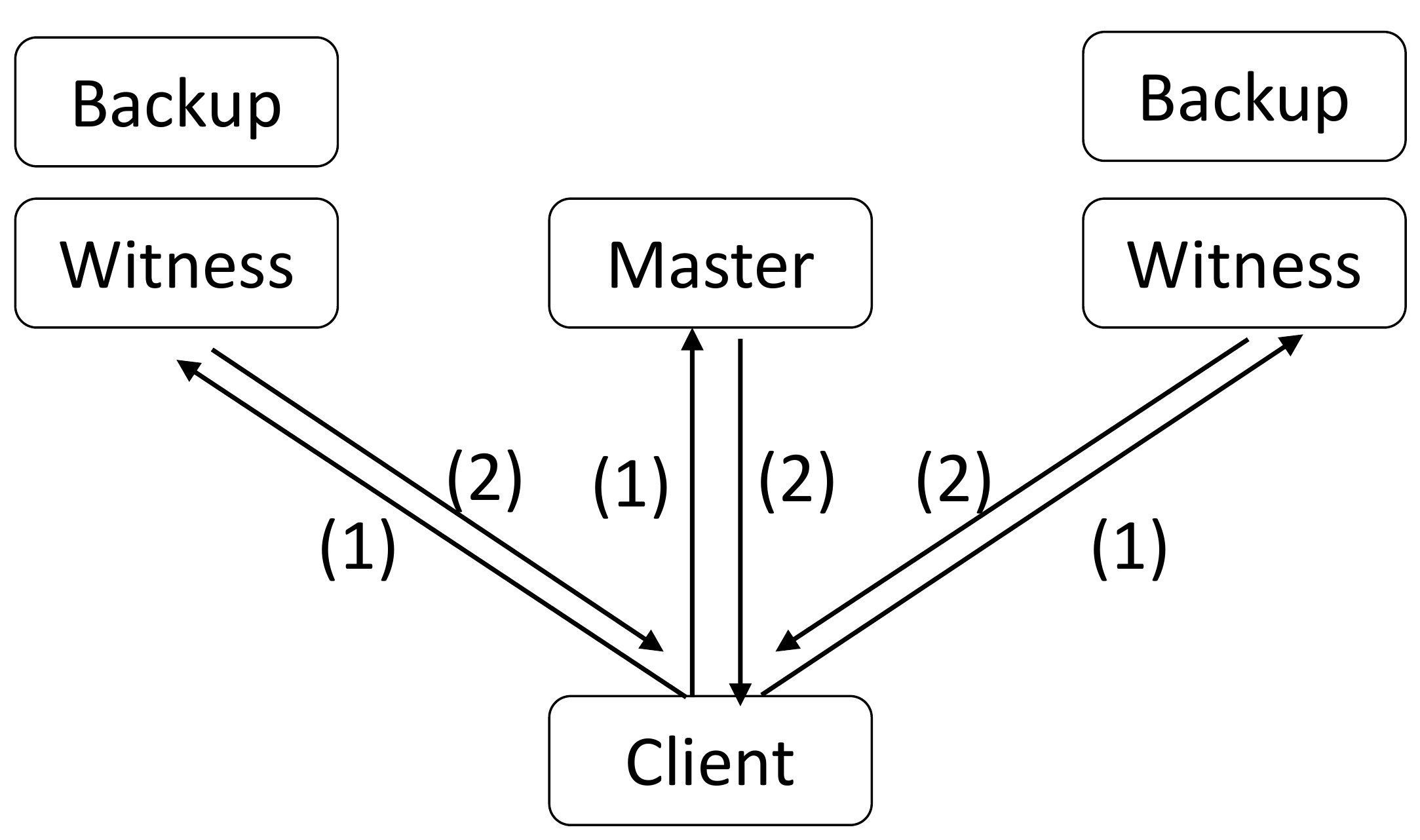}
    \caption{Message Delay of CURP (Fast Path)}
    \label{fig:curp-fast-md}
\end{figure}

\begin{figure}[H]
    \centering
    \includegraphics[width=6cm]{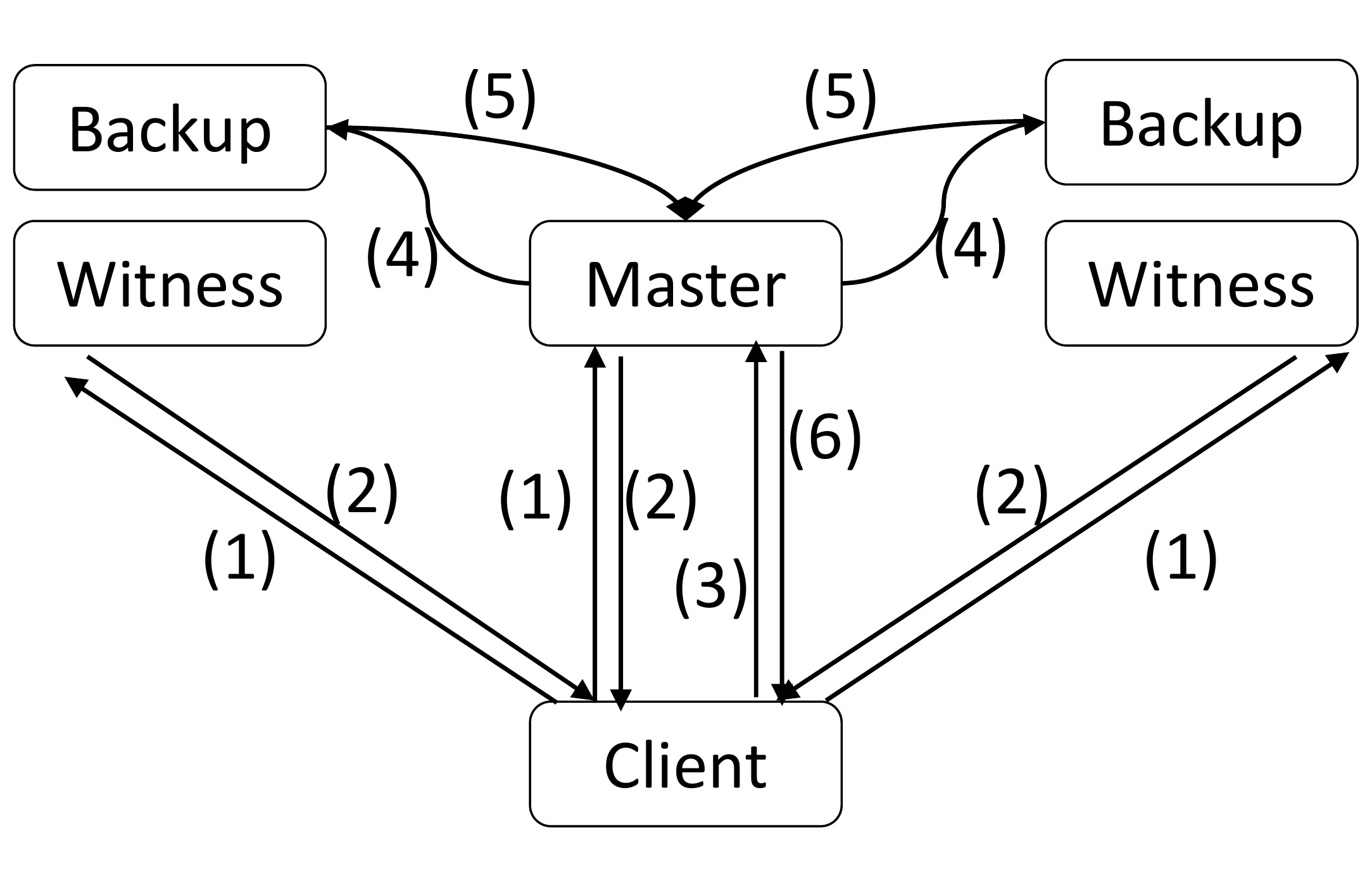}
    \caption{Message Delay of CURP (Slow Path)}
    \label{fig:curp-slow-md}
\end{figure}

CURP requires 2 message delays to commit a request in the fast path (Figure~\ref{fig:curp-fast-md}) and 4 message delays in the slow path (Figure~\ref{fig:curp-slow-md}).

\end{document}